\documentclass[%
 reprint,
superscriptaddress,
 amsmath,amssymb,
 aps,
prb,
]{revtex4-1}

\usepackage{graphicx}
\usepackage{dcolumn}
\usepackage{bm}
\usepackage{mathtools}
\usepackage{bbold}

\begin{document}

\preprint{}

\title{Quasiparticle Density of States, Localization, and Distributed Disorder in the Cuprate Superconductors}

\author{Miguel Antonio Sulangi}
\author{Jan Zaanen}
\affiliation{
	Instituut-Lorentz for Theoretical Physics, Leiden University, Leiden, Netherlands 2333 CA
}

\date{\today}

\begin{abstract}
 We explore the effects of various kinds of random disorder on the quasiparticle density of states of two-dimensional $d$-wave superconductors using an exact real-space method, incorporating realistic details known about the cuprates. Random on-site energy and pointlike unitary impurity models are found to give rise to a vanishing DOS at the Fermi energy for narrow distributions and low concentrations, respectively, and lead to a finite, but suppressed, DOS at unrealistically large levels of disorder. Smooth disorder arising from impurities located away from the copper-oxide planes meanwhile gives rise to a finite DOS at realistic impurity concentrations. For the case of smooth disorder whose average potential is zero, a resonance is found at zero energy for the quasiparticle DOS at large impurity concentrations. We discuss the implications of these results on the computed low-temperature specific heat, the behavior of which we find is strongly affected by the amount of disorder present in the system. We also compute the localization length as a function of disorder strength for various types of disorder and find that intermediate- and high-energy states are quasi-extended for low disorder, and that states near the Fermi energy are strongly localized and have a localization length that exhibits an unusual dependence on the amount of disorder. We comment on the origin of disorder in the cuprates and provide constraints on these based on known results from scanning tunneling spectroscopy and specific heat experiments.
\end{abstract}

\maketitle


\section{\label{sec:level1}Introduction}

Disorder in the high-$T_c$ superconductors has motivated many key experimental and theoretical advances in the field. Scanning tunneling spectroscopy (STS) has made wide use of the phenomenon of quasiparticle interference, which results from the presence of disorder, to provide a real-space probe of the underlying electronic nature of the cuprates.\cite{hoffman2002imaging, mcelroy2003relating, kohsaka2008cooper, lee2009spectroscopic, wang2003quasiparticle, capriotti2003wave, zhu2004power, nunner2006fourier, vishik2009momentum, kreisel2015interpretation, sulangi2017revisiting} On the theory side, the $d$-wave nature of the cuprate superconductors provided the impetus for various theoretical treatments of disorder which led to a number of differing and often contradictory predictions. Early theoretical work utilized a self-consistent treatment of disorder, which was found to result in a finite quasiparticle density of states (DOS) at the Fermi energy.\cite{gorkov1985defects, hirschfeld1988consequences, lee1993localized,durst2000impurity} Later work has shown within a similar diagrammatic approach that the DOS is suppressed.\cite{yashenkin2001nesting} Other field-theoretical treatments of disorder in $d$-wave superconductivity found a vanishing DOS at $E = 0$.\cite{nersesyan1994disorder, senthil1998quasiparticle, senthil1999quasiparticle, altland2002theories}  The manner in which the DOS vanishes as $E \to 0$ varies from approach to approach, with exponents found to be either universal or disorder-dependent. 

Meanwhile, experiments performed on YBa$_2$Cu$_3$O$_{6+\delta}$ consistently show a $T$-linear term in the specific heat at zero magnetic field, which points to a nonvanishing DOS at $E = 0$.\cite{moler1994magnetic, moler1997specific, riggs2011heat} How this nonzero DOS arises has been the subject of much speculation. According to standard self-consistent $T$-matrix theory, which assumes that impurities are located within the copper-oxide planes, this contribution is expected. It is interesting to note, however, that this $T$-linear term in YBCO persists even with very clean samples, prompting a number of exotic explanations, such as loop-current order coexisting with $d$-wave superconductivity\cite{berg2008stability, allais2012loop, kivelson2012fermi, wang2013quantum}, which give rise to a finite DOS without invoking disorder. For Bi$_2$Sr$_2$CaCu$_2$O$_{8+\delta}$, the story is a bit more complicated: it appears that no definitive evidence in favor of or against a zero $T$-linear coefficient exists, and what is present instead is considerable variation in the measured values of this coefficient . For BSCCO-2212 at low temperatures, it was found that that the coefficient is small but finite and measurable.\cite{collocott1990specific, junod1994specific} However, other experiments, performed at higher temperatures, find no discernible evidence in BSCCO-2212 for a coefficient on the same order as found in YBCO.\cite{urbach1989low} The results for the BSCCO family suggest that the cleaner the sample is, the smaller the $T$-linear coefficient becomes, with a large degree of variation present.

Given such a wide array of evidence suggesting that high-temperature superconductors do display a finite zero-energy quasiparticle DOS and the lack of any confirmation of alternative explanations, it is worth revisiting the effect of disorder, especially when incorporating inhomogeneities in the cuprates that do not fall under the random-site-energy or multiple-point-impurity categories. Previous numerical work has extensively focused on pointlike impurities and random on-site energies. In particular, Atkinson \emph{et al.} found that for realistic models (\emph{i.e.}, without a particle-hole symmetric band) with these two forms of disorder, the quasiparticle DOS becomes suppressed near $E = 0$.\cite{atkinson2000details} They point out that a \emph{constant} DOS, as seen in experiment, cannot arise from either of these two disorder models.

In any case, what is known about the cuprates makes it difficult to argue that pointlike disorder is a possible origin of the finite DOS at the Fermi energy. The consensus regarding the CuO$_2$ planes is that they are generally clean. Pointlike disorder necessarily takes the form of dopants \emph{within} the CuO$_2$ plane. Such substitutions will give rise to strong pointlike potentials. The most dramatic case of this is zinc-doped Bi$_2$Sr$_2$CaCu$_2$O$_{8+\delta}$, in which a small number of zinc atoms take the place of copper ones; STS studies of Zn-doped BSCCO show that the zinc impurities show behavior consistent with that of unitary scatterers.\cite{pan2000imaging} In contrast, STS studies of clean cuprates do not show such strong local impurities, and the conductance maps obtained from such materials are more consistent with far weaker forms of disorder.\cite{hoffman2002imaging, mcelroy2005atomic, schmidt2011electronic} More reasonable is the expectation that impurities lie in the buffer layers adjacent to the CuO$_2$ planes.\cite{eisaki2004effect, nunner2005dopant, nunner2005microwave} As they are located in an insulating layer some distance from the CuO$_2$ plane, they act as a source of an electrostatic potential which, in contrast to local pointlike potentials, is \emph{smooth}. These smooth potentials lead to small-momentum scattering processes. It is then worth examining the imprint of such smooth forms of disorder on the DOS.

In this paper, we obtain the quasiparticle DOS of a two-dimensional $d$-wave superconductor subject to various kinds of disorder: pointlike disorder, random on-site disorder, and smooth disorder. We utilize an exact real-space numerical method that allows for the evaluation of the local density of states of a disordered system with very large system sizes (a typical calculation involves 100,000 sites). The same geometry of the system also enables the \emph{direct} calculation of the localization length, which is a quantity that is difficult to extract from the exact diagonalization of small systems, given the large length scales over which localization occurs. An important feature of this work is its use of realistic band-structure and pairing parameters. As our method faces no difficulties with large system sizes, we do not need to resort to making the $d$-wave gap articially large in order to sidestep finite-size effects in related methods like exact diagonalization, and we can thus make the parameters of our lattice $d$-wave superconductor as close as possible to the real-world properties of the cuprates. 

For pointlike and random-site-energy models, we find that weak disorder---whether in the form of a low concentration of strong scatterers or a narrow distribution of on-site energies---leads to a vanishing DOS at the Fermi energy. It is only when unrealistic levels of disorder are reached that a finite DOS is generated, and even then there is an observed suppression at $E = 0$. We observe that the manner in which the $d$-wave gap ``fills'' differs depending on whether one has random-potential or unitary-scatterer disorder. With smooth disorder, however, a finite DOS at the Fermi energy is generated at fairly realistic concentrations (around $10$-$20\%$) and, strikingly, the overall structure of the $d$-wave DOS is preserved for all energies even at high dopings.

We also perform an exact calculation of the localization length $\lambda$ and its dependence on the strength of disorder for the three different kinds of disorder we consider. We find that states near the Fermi energy are strongly localized for all three models---even for weak disorder---and that at intermediate and high energies \emph{within} the $d$-wave gap the localization length is generally found to be very large for low disorder. It is worth noting that even with a high concentration of smooth scatterers, the localization length at intermediate and high energies is still very large and comparable to that seen in much lower levels of disorder in the random-potential and unitary-scatterer case, indicating that localization effects due to smooth disorder are far weaker than in the case of pointlike disorder. Unitary scatterers in turn have a weaker effect on the localization length than random-potential disorder does. 

 Finally, we comment on the nature of disorder in the cuprates based on what is known from specific heat experiments, scanning tunneling spectroscopy, and numerical simulations. We caution the reader that a major limitation of our study is that the gap is not computed self-consistently, so we cannot ascertain with any definiteness whether the effects of disorder that we detail here are preempted by the destruction of $d$-wave superconductivity once some level of disorder is reached. Incorporating full self-consistency in the real-space numerical method we use is technically difficult, especially when the system size is large. This difficulty is a part of a tradeoff we make in order to access large system sizes. That said, exact-diagonalization studies on $d$-wave superconductors with unitary scatterers, using small system sizes, find that the superfluid density of the uniform-gap case and that of the self-consistent-gap case behave very similarly to each other, except when the concentration is sufficiently large.\cite{franz1997critical} $T_c$ in turn was found to be much \emph{less} suppressed in the self-consistent case than in the uniform-gap case. It was found that while in the uniform-gap case $p \approx 8.0\%$ almost completely suppresses $T_c$, in the self-consistent case such suppression occurs at nearly twice that level of disorder. This means that the uniform-gap picture in fact \emph{overstates} the impact of disorder on the suppression of $T_c$ and the superfluid density. This is augmented by the fact that, in other exact diagonalization studies, self-consistency does not fundamentally alter the structure of the DOS of the random-potential and unitary-scatterer cases.\cite{atkinson2000gap, atkinson2000details, zhu2000quasiparticle} For certain parameter regimes it appears that the DOS for self-consistent and non-self-consistent order parameters are identical. In other regimes, the DOS is smoother and features more pronounced suppression near the Fermi energy in the self-consistent case than in the non-self-consistent one, while remaining similar to each other in other energy ranges. All of this suggests that what we find from our uniform-gap systems provides a good baseline for ascertaining the effects of site disorder on the cuprates, and very likely overestimates the pair-breaking effects of disorder. We defer a fully self-consistent treatment of these three kinds of disorder and their pair-breaking effects to a future publication.

\section{Methods} \label{methods}

We start with a tight-binding Hamiltonian describing electrons hopping on a square lattice with $d$-wave pairing:
\begin{equation}
H = -\sum_{\langle i, j \rangle}\sum_{\sigma} t_{ij}c_{i\sigma}^{\dagger}c_{j\sigma} + \sum_{\langle i, j \rangle}\Delta_{ij}^{\ast}c_{i \uparrow}c_{j \downarrow} + \sum_{\langle i, j \rangle}\Delta_{ij}c_{i \uparrow}^{\dagger}c_{j \downarrow}^{\dagger}.
\end{equation}
Nearest-neighbor and next-nearest-neighbor hoppings are both present, as is $d$-wave pairing, implemented by choosing the pairing amplitude to have the form $\Delta_{ij} = \pm\Delta_0$, where the positive (negative) value applies to pairs of nearest-neighbor sites along the $x$- ($y$-) direction. From the Hamiltonian, the Green's function takes the following expression:
\begin{equation}
G^{-1}(\omega)= \omega\mathbb{1}- H.
\label{G}
\end{equation}
Note that $H$ and $G$ are $2N_xN_y \times 2N_xN_y$ matrices written in Nambu-space form, where $N_x$ and $N_y$ are the number of lattice sites in the $x$- and $y$-directions, respectively. From $G(\omega)$, various quantities can be obtained. We will focus on the quasiparticle density of states and the localization length. 

\subsection{Quasiparticle Density of States}

The quasiparticle DOS at energy $E$ is  
\begin{equation}
\rho(E) = -\frac{1}{\pi N_x N_y}\text{Im}\text{Tr}G(E + i0^+).
\label{eq:DOS}
\end{equation}
Periodic and open boundary conditions are implemented in the $y$- and $x$-directions, respectively.  To compute $G$, we first rewrite $G^{-1}$ in the following block tridiagonal form:
\begin{equation}
\mathbf{G}^{-1} = 
\left( \begin{array}{lllllll}
\mathbf{P}_1 & \mathbf{Q}_{1} &   & \hdots &  &  & \mathbf{0}\\
\mathbf{Q}^{\dagger}_{1} & \mathbf{P}_2 & \mathbf{Q}_{2}  &  &  &  & \\
  & \ddots & \ddots &  \ddots & &  & \\ 
 \vdots &  & \mathbf{Q}^{\dagger}_{j-1} & \mathbf{P}_{j} & \mathbf{Q}_{j} & & \vdots \\
 &   &  & \ddots &\ddots & \ddots &  \\
  &  &  &  & \mathbf{Q}^{\dagger}_{N_x - 2} &\mathbf{P}_{N_x - 1} & \mathbf{Q}_{N_x  -1} \\
\mathbf{0} &  &  & \hdots  &  & \mathbf{Q}^{\dagger}_{N_x - 1}& \mathbf{P}_{N_x}
\end{array} \right).
\end{equation}
The $\mathbf{P}_i$ blocks are $2N_y\times2N_y$ submatrices and contain in their diagonal elements the frequency $\omega$ and the on-site energies at sites located on the $i$th slice of the system, where $i$ runs from $1$ to $N_x$, in addition to hopping and pairing amplitudes between sites within the $i$th slice. The $\mathbf{Q}_i$ blocks---also $2N_y\times2N_y$ submatrices---meanwhile contain hopping and pairing amplitudes from the $i$th slice to its nearest-neighbor slices. Note that the Nambu-space structure of the Green's function has been transferred to the $\mathbf{P}_i$ and $\mathbf{Q}_i$ blocks.

Because all we need is the trace of $G$ to obtain the DOS, it suffices to obtain the diagonal blocks of $G$. For this purpose we use a block-by-block matrix-inversion algorithm that applies to block tridiagonal matrices.\cite{godfrin1991method, hod2006first, reuter2012efficient}. We first define auxilliary matrices $\mathbf{R}_i$ and $\mathbf{S}_i$ in the following way:
\begin{eqnarray}
\mathbf{R}_i = 
\begin{cases}
\mathbf{Q}_{i}(\mathbf{P}_{i+1} - \mathbf{R}_{i+1})^{-1}\mathbf{Q}^{\dagger}_{i} & \text{if} \ 1 \leq i < N_x \\
\mathbf{0} & \text{if} \ i = N_x
\end{cases}
\end{eqnarray}
and
\begin{eqnarray}
\mathbf{S}_i = 
\begin{cases}
\mathbf{0} & \text{if} \ i = 1 \\
\mathbf{Q}^{\dagger}_{i-1}( \mathbf{P}_{i-1} - \mathbf{S}_{i-1})^{-1}\mathbf{Q}_{i-1} & \text{if} \ 1 < i \leq N_x.
\end{cases}
\end{eqnarray}
Once $\mathbf{R}_i$ and $\mathbf{S}_i$ have been computed, the $i$th diagonal block of $G$ can be obtained straightforwardly from the following expression:
\begin{equation}
\mathbf{G}^{ii} = (\mathbf{P}_i - \mathbf{R}_i - \mathbf{S}_i)^{-1}.
\end{equation}
We note that this procedure is exact and relies on no approximations. We set $N_x = 1000$ and $N_y = 100$ in all calculations. 

To ensure the applicability of our numerical results to the cuprates, we use a band structure that is consistent with the details known about the normal-state Fermi surface of such materials: $t = 1$, $t' = -0.3$, and $\mu = -0.8$, where $t$, $t'$, and $\mu$ are the nearest-neighbor hopping, next-nearest-neighbor hopping, and the chemical potential, respectively. We note that our parametrization of the Fermi surface is limited as higher-order hopping amplitudes are not included, but this simple form of the band structure still captures the important general features of the Fermi surface of the cuprates. We choose the pairing amplitude to be $\Delta_0 = 0.08$; this choice gives $v_F/v_{\Delta} \approx 11$, in good agreement with experiment.\cite{vishik2010doping} (All energies are expressed in units where $t = 1$.) An inverse quasiparticle lifetime given by $\eta = 0.001$ is used throughout this work. This smears out the Dirac delta function peaks $\delta(E - E_n)$, where $E_n$ is an eigenvalue of $H$, into a Lorentzian, $\frac{1}{\pi} \frac{\eta}{(E - E_n)^2 + \eta^2}$, whose full width at half maximum is $2\eta$. Because the DOS of a clean $d$-wave superconductor with this particular band structure is nonzero up to energies $E \approx \pm 6t$, this choice of broadening roughly corresponds to introducing $\mathcal{O}(10^3)$ bins for the entire energy range. As there are $2\times10^5$ eigenvalues of the Hamiltonian, this provides more than adequate resolution for the examination of the DOS as a function of energy. Note that this value of $\eta$ is parametrically much smaller than the energy resolution seen in scanning tunneling experiments (which are typically found to be 2 meV).\cite{zhu2004power} Such values of the broadening already incorporate the effects of disorder, so in order to tease out the impact of disorder on the DOS we need to pick a much smaller value of $\eta$ than seen in experiment.

The advantage of this particular method of obtaining the exact DOS, as opposed to similar methods such as the exact diagonalization of the Bogoliubov-de Gennes Hamiltonian, is threefold. First, this method is much faster in obtaining the DOS than exact diagonalization. As the DOS involves taking the trace of the Green's function, only the diagonal elements of $G$ are needed, which are precisely the quantities outputted by the algorithm in use here. Second, this method can be extended to very large system sizes. The computational complexity depends only linearly on $N_x$, and consequently the size of that dimension can increased without much trouble. Importantly, the large sizes that are accessible mean that the need to average over different disorder configurations is largely obviated---a single realization of disorder results in $10^5$ values of the local density of states to be averaged over---and hence for the most part we will focus only on a single realization of disorder for each of the cases we will consider. This makes much sense from a modeling viewpoint, especially as in experiment only \emph{one} realization of disorder is present for a measurement. Finally, as finite-size effects are minimal, we are free to set the hopping and pairing parameters to correspond closely to those known from experiment. In exact diagonalization, the smallness of the system sizes typically used means that in order to visualize the spectrum fully one is occasionally faced with the need to make $\Delta_0$ artificially large, so that within-gap physics are seen with the energy resolution available. In the method we use no such workarounds are necessary. 

The only disadvantage of this method is that self-consistency is very difficult to implement in an efficient manner. In a fully self-consistent treatment the order parameter is iteratively determined via an integral of the anomalous Green's function over a range of energies. Consequently, in energy space the Green's function needs to be evaluated over a finely spaced array of points over the full bandwidth for the numerical integral to be accurate, and this process has to be repeated for an unspecified number of times until self-consistency is achieved. The full bandwidth is several times larger than the $d$-wave gap; hence the amount of computational effort required to perform this self-consistent calculation for even one realization of disorder becomes very large and uncontrollable. (This has to be contrasted with exact diagonalization, from which one obtains all the eigenvalues and eigenvectors of the Hamiltonian at once. The gap can then be computed in terms of the eigenvectors once one diagonalization has been completed. While this method is restricted to very small geometries, it is nonlocal in energy space, and thus implementing self-consistency is much easier.)  As we have noted in the Introduction, evidence from previous numerical studies of lattice $d$-wave superconductors with strongly pair-breaking unitary scatterers suggests that self-consistent and non-self-consistent results are not drastically different from one another. We will thus take the results from our uniform-gap systems to provide a reasonable account of the effects of disorder on the various quantities of interest to us.

 It is also easy to obtain the \emph{local} quasiparticle density of states (LDOS) from $G$. Because $G$ is written in a real-space basis, the LDOS $\rho(\mathbf{r}, E)$ is simply
 \begin{equation}
 \begin{aligned}
 \rho(\mathbf{r}, E) = &-\frac{1}{\pi}\text{Im}\big(G_{11}(\mathbf{r},\mathbf{r}, E + i0^+) \\
 &+ G_{22}(\mathbf{r},\mathbf{r}, E + i0^+)\big),
 \end{aligned}
 \label{eq:LDOS}
 \end{equation}
where $G_{11}$ and $G_{22}$ are the particle and hole parts, respectively, of the Nambu-space Green's function. At this point it is worth emphasizing the fact that, from the way we have defined them, these maps are \emph{not} the same as the local density of states maps obtained from STS studies. The conductance maps obtained in STS experiments are proportional to the local \emph{electron} density of states, which are taken solely from the electron part of the Green's function: $\rho_{tunn}(\mathbf{r}, E) = -\frac{1}{\pi}\text{Im}G_{11}(\mathbf{r},\mathbf{r}, E + i0^+)$. In contrast, the quasiparticle DOS at energy $E$, as defined in Eq.~\ref{eq:LDOS}, includes contributions from both the electron and hole Green's function. We will frequently show these maps to visualize the extent to which disorder affects the degree of inhomogeneity in the quasiparticle wavefunctions at a particular energy $E$.

We also calculate, for completeness, the quasiparticle DOS of a \emph{clean} $d$-wave superconductor in order to provide a baseline from which one can examine the impact of disorder. Unlike the disordered case, we perform this calculation in momentum space. We use the formula
\begin{equation}
\rho(E) = \sum_{\mathbf{k} \in \text{BZ}}  \delta(E - E_{\mathbf{k}}),
\label{eq:kDOS}
\end{equation}
where $E_{\mathbf{k}}$ are the eigenvalues of the clean Hamiltonian, given for positive energies by
\begin{equation}
 E_{\mathbf{k}} = \sqrt{\epsilon^2_{\mathbf{k}} + \Delta^2_{\mathbf{k}}}.
\end{equation}
Here $\epsilon_{\mathbf{k}} = -2t(\cos k_x + \cos k_y) - 4t'\cos k_x\cos k_y - \mu$ and $\Delta_{\mathbf{k}} = 2\Delta_0(\cos k_x - \cos k_y)$ are the normal-state dispersion and the gap function in momentum space, respectively. Only positive energies need to be considered because of particle-hole symmetry. For consistency with the real-space calculations of the disordered cases, we also broaden the delta functions that enter Eq.~\ref{eq:kDOS} into a Lorentzian with broadening $\eta = 0.001$. In our momentum-space calculations we discretize the first Brillouin zone into a grid with $4000 \times 4000$ points. This choice results in a smooth DOS as a function of $E$ which is free from finite-size effects.

\subsection{Specific Heat}

The quasiparticle contribution to the specific heat $C$ is easily derived from the density of states by means of the following equation,\cite{hirschfeld1988consequences}
\begin{equation}
C = 2\times \frac{\partial}{\partial T}\int_{0}^{\infty} dE \rho(E) E \frac{1}{e^{E/k_B T} + 1},
\label{eq:SH}
\end{equation}
where the factor of two arises from the two spin species present. We are interested in $C$ in the low-temperature regime, so we can neglect the dependence of $\rho(E)$ on $T$, and because $T \ll 4 \Delta_0$ (the $d$-wave gap edge, which itself is much bigger than $T_c$) we can impose a cutoff $E_c \approx 4\Delta_0$ so that only energies within the $d$-wave gap are integrated over. As such, Eq.~\ref{eq:SH} becomes 
\begin{equation}
C = 2\times \frac{1}{k_B T^2}\int_{0}^{E_c} dE \rho(E) E^2 \frac{e^{E/k_B T}}{(e^{E/k_B T} + 1)^2}.
\end{equation}
It can further be shown that the contribution of $\rho(E = 0)$ to the specific heat is
\begin{equation}
C_0 = \gamma_0 T = \frac{1}{3} \pi^2 \rho(E = 0) k^2_B  T.
\label{eq:sommerfeld}
\end{equation}
When $C_0/T$ is plotted versus $T$, the plot is flat, and the $y$-intercept of this plot is equal to $\gamma_0$. In our numerical results we will typically set $k_B = 1$ and measure the temperature $T$ in units of the hopping energy $t$ ($t \approx 0.150$ eV $\approx 1700$ K).

Note that the scaling of $C$ with $T$ is dependent on how $\rho$ scales with $E$. At low energies the DOS of a clean $d$-wave superconductor is a linear function of $E$; thus the quasiparticles of a clean $d$-wave superconductor contribute a $T^2$-dependent term to $C$. When this coexists with a finite quasiparticle DOS at $E = 0$, the most general scaling of $C$ due to the $d$-wave quasiparticles is
\begin{equation}
C \approx \gamma_0 T + \alpha T^2,
\end{equation}
and a $C/T$-versus-$T$ plot would have a slope equal to $\alpha$ and a $y$-intercept equal to $\gamma_0$. In the most general disordered case we should not expect this form of scaling to arise, as disorder can lead to a nonlinear dependence of $\rho$ on $E$. However, a finite value of $\gamma_0$ is a feature that unambiguously suggests the presence of a finite DOS at the Fermi energy.

\subsection{Localization Length}

The geometry of our system is particularly amenable to exact calculations of the localization length $\lambda$, owing to the fact that $N_x$ can be made very large relative to $N_y$, allowing us to measure the localization length even when it is much bigger than the transverse dimension. This calculation is all but impossible using exact diagonalization, as that method is restricted to fairly small system sizes whose linear dimension is much smaller than typical localization lengths.

We will use the following definition of $\lambda$:\cite{mackinnon1981one, buŀka1985mobility, kramer1993localization, xiang1995effect}
\begin{equation}
\lambda^{-1} = -\frac{1}{2(N_x - 1)}\ln\frac{\sum_{ij\sigma\sigma'}|G^{N_x1}_{ij\sigma\sigma'}( E)|^2}{\sum_{ij\sigma\sigma'}|G^{11}_{ij\sigma\sigma'}(E)|^2}.
\label{eq:loclength}
\end{equation}
The $\frac{\sum_{ij\sigma\sigma'}|G^{N_x1}_{ij\sigma\sigma'}( E)|^2}{\sum_{ij\sigma\sigma'}|G^{11}_{ij\sigma\sigma'}(E)|^2}$ factor measures the transmission probability from the left end of the system (the $1$st slice) to the right end (the $N_x$th slice); the denominator in the aforementioned factor is for normalization. The sums are performed over all sites and spin indices within the relevant block. The off-diagonal block $G^{N_x1}(E)$ can be recursively computed from the diagonal block $G^{11}( E)$ by an algorithm that applies to block tridiagonal matrices.\cite{godfrin1991method, hod2006first, reuter2012efficient} Using the $\mathbf{P}_i$, $\mathbf{Q}_i$, $\mathbf{R}_i$, $\mathbf{S}_i$, and $\mathbf{G}^{ii}$ matrices obtained earlier, any off-diagonal blocks of $G$ can be computed using this formula:
\begin{eqnarray}
\mathbf{G}^{i j} = 
\begin{cases}
-(\mathbf{P}_i - \mathbf{R}_i)^{-1}\mathbf{Q}^{\dagger}_{i-1}\mathbf{G}^{i-1,j} & \text{if} \ i > j, \\
-(\mathbf{P}_i - \mathbf{S}_i)^{-1}\mathbf{Q}_{i}\mathbf{G}^{i+1,j} & \text{if} \ i < j.
\label{eq:offdiag}
\end{cases}
\end{eqnarray}

We calculate the localization length only for fixed values of $N_x$ and $N_y$. We do not extract the actual localization length via finite-size analysis. We thus provide the necessary caveat that the values of $\lambda$ that we cite here are meaningful only in comparison with systems with \emph{identical} system sizes. That is, a direct comparison is possible between $\lambda$'s computed with the same $N_x$ and $N_y$ but for different disorder types and strengths, but \emph{not so} when these system-size parameters are altered relative to one another. 

\section{Models of Disorder}

In this paper we will focus on three distinct models of disorder. Many of these forms of disorder have been discussed in the older literature on the subject, and in particular some of them can be treated, on some level, analytically in either the Born approximation or the $T$-matrix approximation. Here we will make use of the ability to simulate systems with very large system sizes to cover regimes where the approximations that enable analytical treatments of disorder fail. Below we will enumerate these models of disorder, their properties, and the degree to which these describe the actual disorder present in the cuprates.

\subsection{Random-Potential Disorder}

The first model is random and \emph{spatially uncorrelated} on-site energies. We assume that the potential at each lattice site consists of two parts: the uniform chemical potential and a normally distributed random component $V$ with zero mean and variance $\sigma^2$:
\begin{eqnarray}
&\langle V(\mathbf{r})  \rangle = 0,\\
&\langle V(\mathbf{r}_1) V(\mathbf{r}_2)  \rangle = \sigma^2\delta_{\mathbf{r}_1 \mathbf{r}_2}.
\label{eq:deltacorr}
\end{eqnarray}
From the perspective of diagrammatic perturbation theory, this is a particularly tractable model of disorder: given the above conditions, the Fourier transform of the two-point \emph{averaged} correlation function of the disorder potential is a constant in momentum space:
\begin{equation}
\begin{aligned}
W({\mathbf{k}}) &= \sum_{\mathbf{r}} \langle V(\mathbf{r}) V(\mathbf{0})  \rangle e^{-i\mathbf{k \cdot r}}\\
&= \sum_{\mathbf{r}}  \sigma^2\delta_{\mathbf{r0}} e^{-i\mathbf{k \cdot r}}\\
&= \sigma^2.
\label{eq:deltacorrft}
\end{aligned}
\end{equation}
This property of the model allows one to analytically obtain the self-energy easily using the Born approximation in the limit that $\sigma$ is small.\cite{lee1993localized} Physically this model can be obtained from the multiple point-impurity model when one takes the strength of these impurities to be very weak and the spacing between impurities very small.

 A related version of this disorder potential was studied numerically by Atkinson \emph{et al.}; however they utilized box disorder instead of Gaussian distributions.\cite{atkinson2000details} We on the other hand will focus exclusively on normally-distributed on-site energies. This form of disorder is physically realistic, as recent work has shown that narrowly-distributed Gaussian disorder of this sort could give rise to quasiparticle scattering interference (QPI) patterns in $d$-wave superconductors that are in reasonably good agreement with those seen in experiments on BSCCO.\cite{sulangi2017revisiting}

\subsection{Multiple Unitary Scatterers}

The second model we will discuss is another paradigmatic form of disorder in the cuprates: unitary pointlike scatterers situated \emph{within} the copper-oxide plane. Unitary scatterers in $d$-wave superconductors have been extensively studied experimentally and theoretically. Zinc dopants within the CuO$_2$ planes of BSCCO are the most well-known studied form of unitary scatterers in the cuprates, and in fact their resonances have been directly imaged in STS experiments.\cite{pan2000imaging} Unitary scatterers also arise in the cuprates in the form of vacancies within the CuO$_2$ plane. Like the Gaussian random-disorder case discussed earlier, unitary scatterers, which induce scattering phase shifts equal to $\delta_0 = \pi/2$, are quite tractable to model in practice: the $T$-matrix for a single pointlike impurity is momentum-independent, allowing one to obtain the full Green's function, including the impurity and its effects, in an exact manner. This can then be extended to the many-impurity case in the dilute limit (\emph{i.e.}, at low concentrations $p$) in the form of a multiple-scattering $T$-matrix.\cite{hirschfeld1988consequences} (Note that if one takes the strength of the impurities to be small, the phase shift is $\delta_0 \approx 0$, and the corresponding $T$-matrix problem becomes identical to the Born-scattering limit of the Gaussian random-potential case discussed previously.\cite{hirschfeld1988consequences, lee1993localized})

We will eschew the $T$-matrix approach and instead obtain the full Green's function and the DOS exactly using the methods described in Section~\ref{methods}. This will allow us to examine cases where the concentration $p$ is large enough that the system enters the strong-disorder regime. We will vary $p$ to cover small, intermediate, and large concentrations; the strength of the impurity is fixed at $V_{u} = 10$, and we will make this potential attractive, to mimic the effect of zinc impurities, which are attractive potential scatterers.\cite{martin2002impurity,kreisel2015interpretation} These impurities are distributed randomly over the entire system, with each lattice site having a $p$ chance of hosting a unitary impurity and a $1 - p$ probability of not having one. Our choice of $V_{u} = 10$ gives a resonance energy at around $E \approx - 0.06$---the negative-bias peak in the bare electron LDOS at the sites adjacent to an isolated impurity is far more prominent than the positive-bias one---which is near, but not at, the Fermi energy. (To perform a sanity check, we checked the case of an isolated impurity with $V_{u} = 100$, which yielded a resonance energy of $E \approx -0.045$. Increasing the impurity potential tenfold indeed pushed the resonance closer to the Fermi energy, but only by a small amount. In fact, if we do a single-impurity (\emph{i.e.}, non-self-consistent) $T$-matrix calculation,\cite{zhu2003two} assuming unitary scatterers with $V_{u} \to \infty$ and using the same band-structure and pairing details as in our exact numerical calculations, we find that the resonance is at $E \approx -0.04$. For generic band structures and arbitrary but strong $V_{u}$ the resonance due to a strong, attractive scatterer is located close to, but not at, the Fermi energy, although for the purposes of our paper its precise location is not very important.) Note that the effect of unitary scatterers on the DOS of $d$-wave superconductors has been studied by Atkinson \emph{et al.},\cite{atkinson2000gap, atkinson2000details} but we will go beyond their work by varying $p$ such that both dilute and strong-disorder limits are covered, and by delving deep into the statistics of the DOS at the Fermi energy in considerable detail.

\subsection{Smooth Disorder}

The third and final form of disorder that we will discuss is off-plane disorder. As we have noted earlier, for the cuprates, disorder due to doping is generally due to dopants that are located some distance away from the CuO$_2$ planes. Doping in the cuprates is accomplished using oxygen atoms, and these oxygens are in general not found within the conducting planes. For BSCCO, the BiO planes host the excess oxygens arising from doping. In the case of YBCO, the doped oxygens are found in the one-dimensional CuO chains some distance away from the CuO$_2$ planes. YBCO is a particularly interesting case to consider because the amount of doping, and hence disorder, can be controlled rather precisely: very clean samples have been synthesized. Thermal conductivity experiments on clean YBCO find that transport does not resemble either Born or unitary scattering (\emph{i.e.}, the previous two models at low levels of disorder).\cite{hill2004transport} Thus it is an interesting theoretical puzzle as to why precisely a finite DOS at the chemical potential is consistently found in specific heat studies of YBCO, even with clean samples.

We will attempt to revisit the effects of off-plane disorder on the quasiparticle DOS of a $d$-wave superconductor. Off-plane dopants will produce a screened Coulomb potential which affects the electrons on the CuO$_2$ plane in the form of a \emph{smooth} disorder potential.\cite{pan2001microscopic, eisaki2004effect, bobowski2010oxygen} In the absence of a more microscopic model of disorder, we will take the disorder potential from one off-plane dopant located on the $a$-$b$ plane at $\mathbf{r}_n$ to have the following reasonably general form:
\begin{equation}
V_n(\mathbf{r}) = V_0 \frac{e^{-\frac{s(\mathbf{r, r}_n) }{L}} }{s(\mathbf{r, r}_n) }.
\label{eq:vn}
\end{equation}
For brevity we have defined $s(\mathbf{r, r}_n)$ as
\begin{equation}
s(\mathbf{r, r}_n) = \sqrt{(\mathbf{r - r}_n)^2 + l^2_z },
\end{equation}
and $L$ is the screening length of the Coulomb potential, $l_z$ is the distance along the $c$-axis from the dopant to the CuO$_2$ plane, and $V_0$ quantifies the ``strength'' of the potential. For our calculations we take $L = 4$, $l_z = 2$, and $V_0 = 0.5$. Because we do not exactly know the details of this disorder potential, we will assume two different scenarios for how this form of disorder is spatially distributed. For the first scenario, we will take the general disorder potential to have the \emph{same} sign, such that the net potential, expressed as a function of the doping concentration $p$, takes the following form:
\begin{equation}
V_s(\mathbf{r}) = \sum_{n = 1}^{pN_x N_y} V_n(\mathbf{r}).
\label{eq:vs}
\end{equation}
The second scenario assumes that there is an equal number of positive- and negative-strength potentials,
\begin{equation}
V_z(\mathbf{r}) = \sum_{n = 1}^{pN_x N_y} (-1)^{a(n)} V_n(\mathbf{r}),
\label{eq:vz}
\end{equation}
where $a(n)$ is a random integer. This leads to a potential whose spatial average is zero, and whose average over disorder configurations (\emph{i.e.}, positions of the dopants, with the number of dopants held fixed) is also zero:
\begin{equation}
\langle V_z(\mathbf{r})  \rangle = 0.
\end{equation}

The second scenario relies on a finely-tuned equality between the number of positive- and negative-strength dopants, and as such we do not claim that it necessarily corresponds to a realistic disorder potential. Nevertheless, from a theoretical standpoint $V_z$ is a particularly interesting form of disorder because, like the Gaussian random-potential disorder case discussed earlier, its spatial and configuration average is zero. However $V_z(\mathbf{r})$ differs from the Gaussian case because it is \emph{not} spatially uncorrelated: its disorder-averaged two-point correlator is not a delta function. Rather, this correlator decays much more slowly than a delta function. The length scales associated with this disorder potential drastically affect the allowed scattering processes. Recall that a $d$-wave superconductor has four nodes where gapless Bogoliubov quasiparticles exist at $E = 0$, which then morph into banana-shaped contours of constant energy (CCEs) once energy is increased from zero. When one has elastic scattering off of pointlike impurities, there is no restriction on scattering processes aside from phase-space considerations: scattering has to occur between states lying on CCEs.\cite{wang2003quasiparticle, capriotti2003wave, kreisel2015interpretation, sulangi2017revisiting} With smooth disorder, however, the matrix elements of the potential vanish very quickly as momentum is increased, leading to a suppression of large-momentum scattering processes.\cite{nunner2005microwave} For this form of disorder, the dominant scattering processes occur only within one node, and to a first approximation scattering between states on different nodes can be neglected. This has been studied from the perspective of quasiparticle scattering interference, and smooth disorder potentials have been found to result in the marked suppression of large-momentum peaks in the Fourier-transformed LDOS.\cite{nunner2006fourier, sulangi2017revisiting} 

The distinction between pointlike disorder (\emph{e.g.}, random normally-distributed on-site potentials and multiple unitary impurities) and smooth disorder is rarely discussed on a theoretical level. Prominent exceptions are the pioneering and extensive work by Nunner \emph{et al}. on Coulomb-potential disorder,\cite{nunner2005dopant, nunner2005microwave, nunner2006fourier}, by Durst and Lee on extended linear scatterers,\cite{durst2002microwave} and field-theoretical work motivated by the possibility that scattering in the cuprate superconductors is primarily forward (\emph{i.e.}, small-momenta) in nature.\cite{nersesyan1994disorder} It has been argued that, from the standpoint of effective field theory, the microscopics of the disorder determine the symmetry class of the effective theory of the disordered system, and consequently pointlike and smooth disorder belong to different universality classes.\cite{altland2002theories} While this does make sense from this particular viewpoint, from a more microscopic perspective such as ours such a distinction is not as clear-cut: one can, at least in principle, continuously tune the length scales of the disorder potential to come close to the pointlike limit, so it is difficult to argue that the lattice tight-binding Hamiltonian exhibits such a sharp distinction between two different universality classes. There is also a difficulty in extending these field-theoretical results to the intermediate- and strong-disorder regimes, as these take as a starting point the presence of \emph{weak} disorder. Nevertheless, as we shall see with our numerics, smooth disorder does lead to effects that differ dramatically from either random Gaussian disorder or multiple-impurity models.

The main variable we use to manipulate the amount of disorder in the superconductor is the concentration $p$ of off-plane dopants. To be more specific, $p$ here is the number of off-plane dopants per copper site \emph{at} the CuO$_2$ plane. From what is known about LSCO, BSCCO, and YBCO, $p$ is generally a large fraction which is usually of the order of $p \approx 0.1$-$0.2$. The precise doping level of YBCO is a complicated quantity to determine because it is not at all obvious how many of the oxygen dopants go to the chains and to the planes; we will not incorporate these subtleties in our calculations, but we do note that microwave conductivity measurements on YBCO are generally found to be consistent with a concentration of defects on the CuO chains given by $p \approx 0.1$.\cite{bobowski2010oxygen} We will cover this regime of doping, as this is the most physically relevant one, although we will cover low and high concentrations as well. It is not clear \emph{a priori} whether a density of $p \approx 0.1$-$0.2$ corresponds to weak or strong disorder, so we will scan through $p$ to see precisely what regimes are covered by these impurity concentrations.
 
\section{Quasiparticle Density of States: An Overview}

\begin{figure}[t]
	\centering
	\includegraphics[width=.5\textwidth]{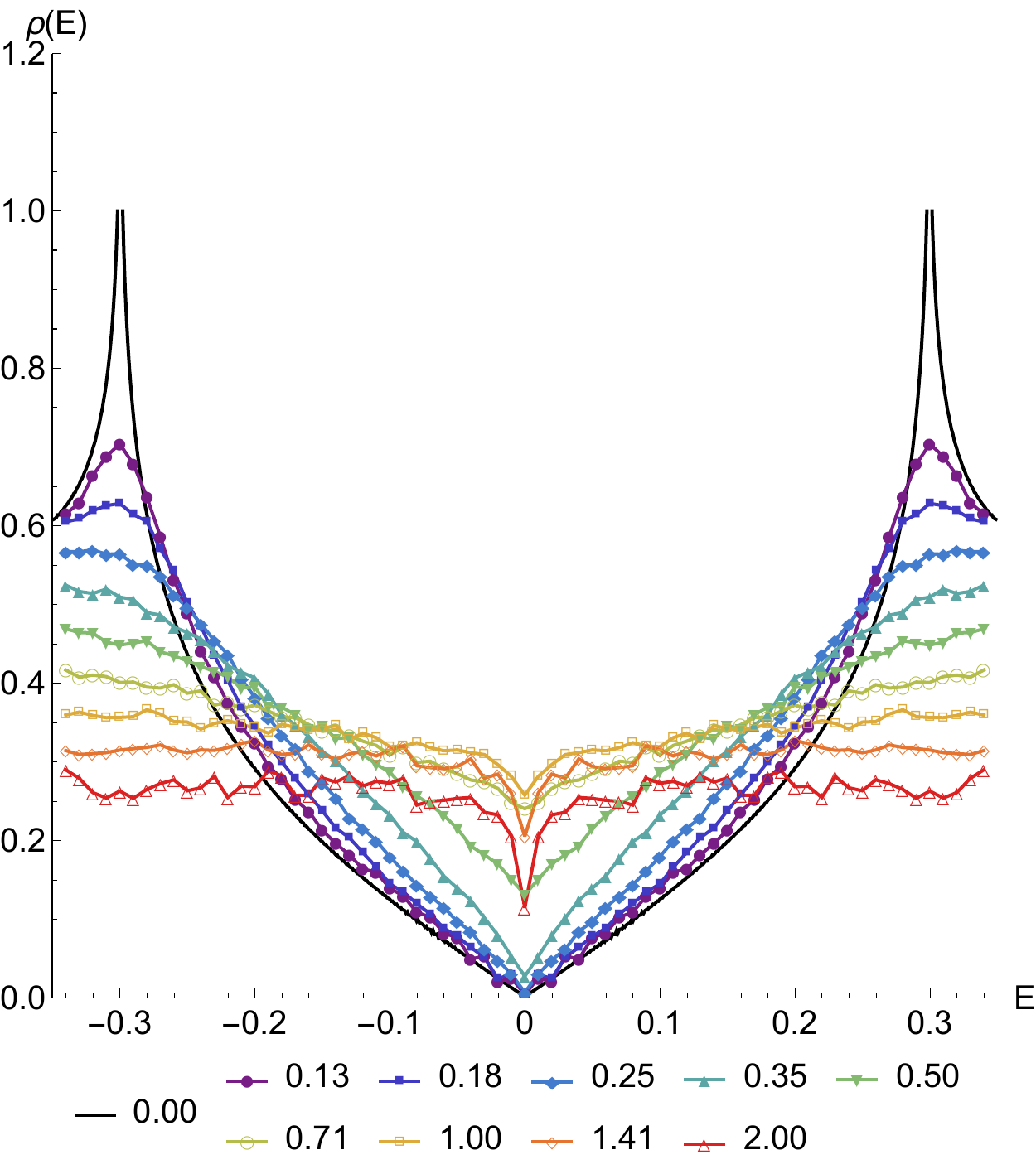}\hfill
	\caption{Plots of the quasiparticle DOS as a function of energy $E$ for the Gaussian random-potential model, for various values of $\sigma$.}
	\label{fig:rp}
\end{figure}	

\begin{figure*}
	\centering
	\includegraphics[width=.2\textwidth]{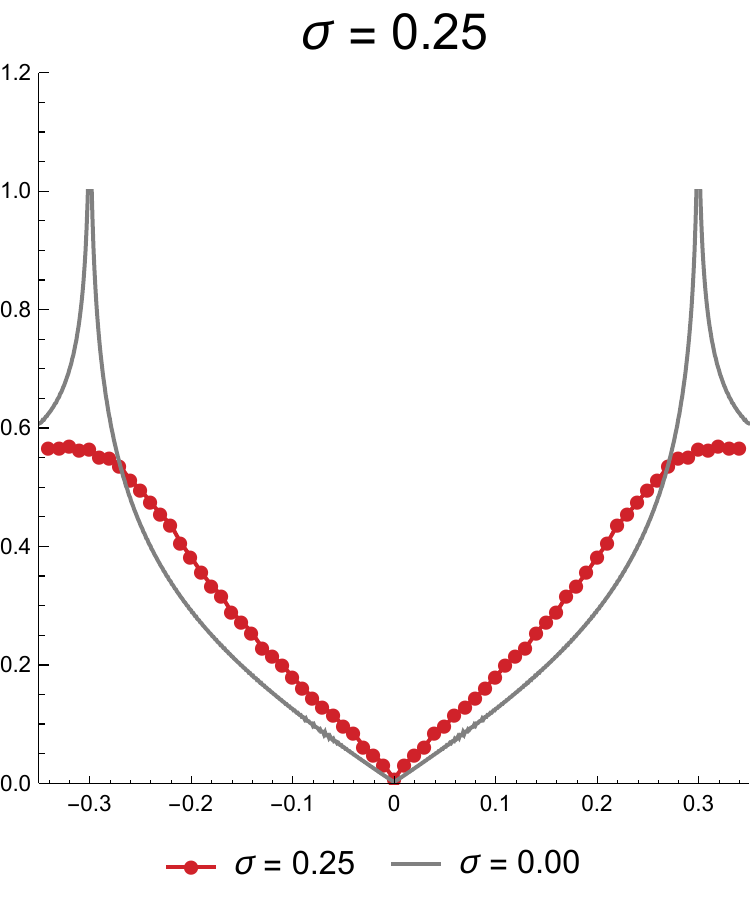}
	\includegraphics[width=.2\textwidth]{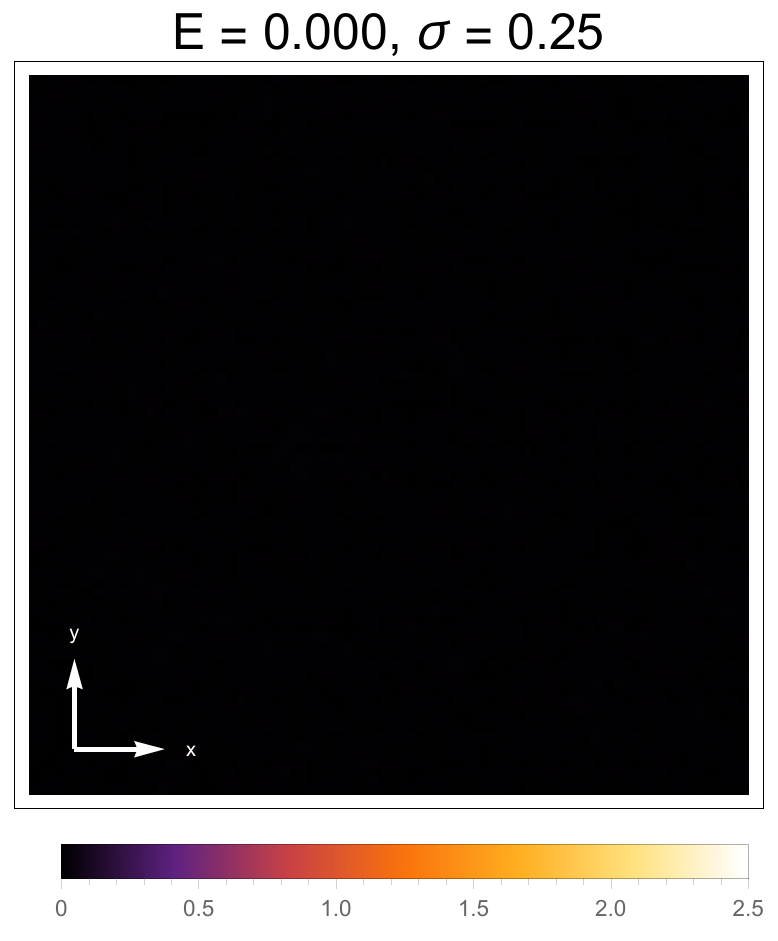}
	\includegraphics[width=.2\textwidth]{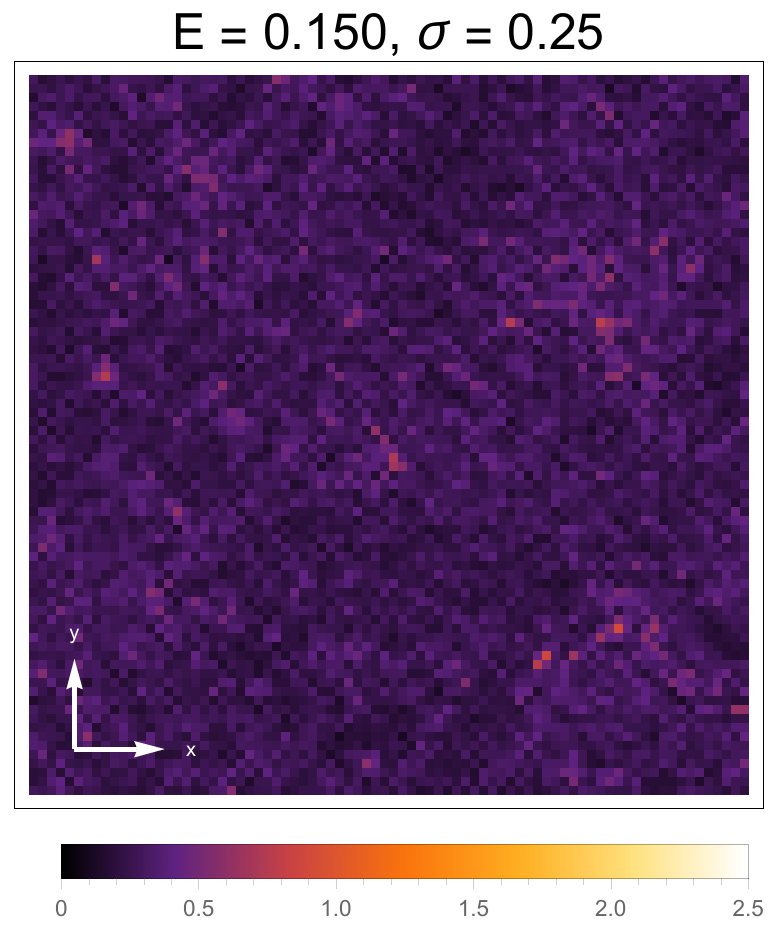}
	\includegraphics[width=.2\textwidth]{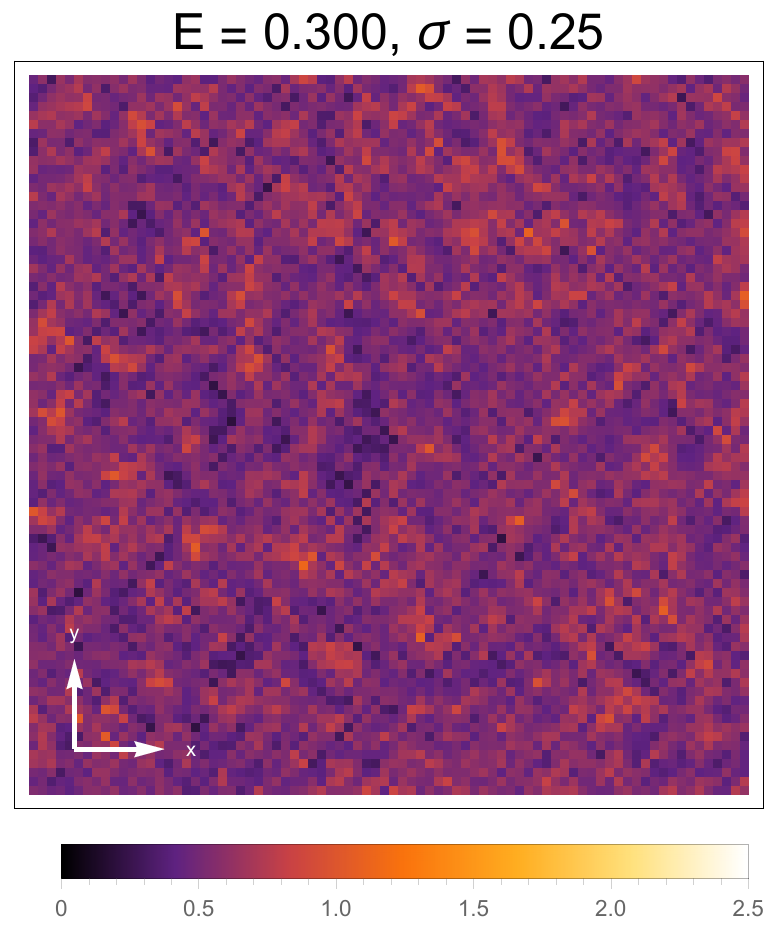} \\
	\includegraphics[width=.2\textwidth]{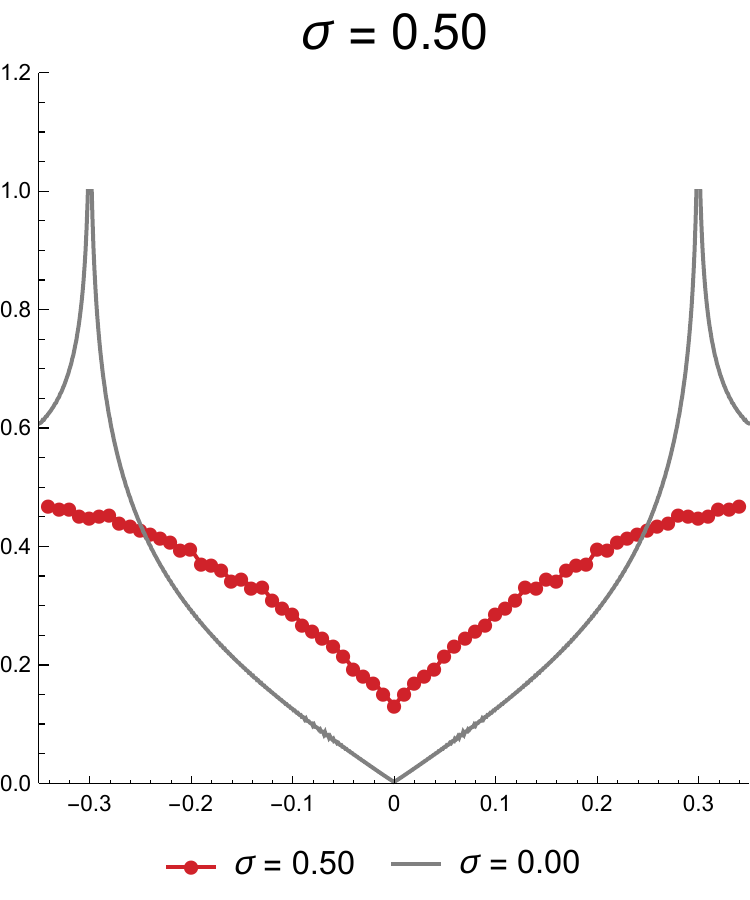}
	\includegraphics[width=.2\textwidth]{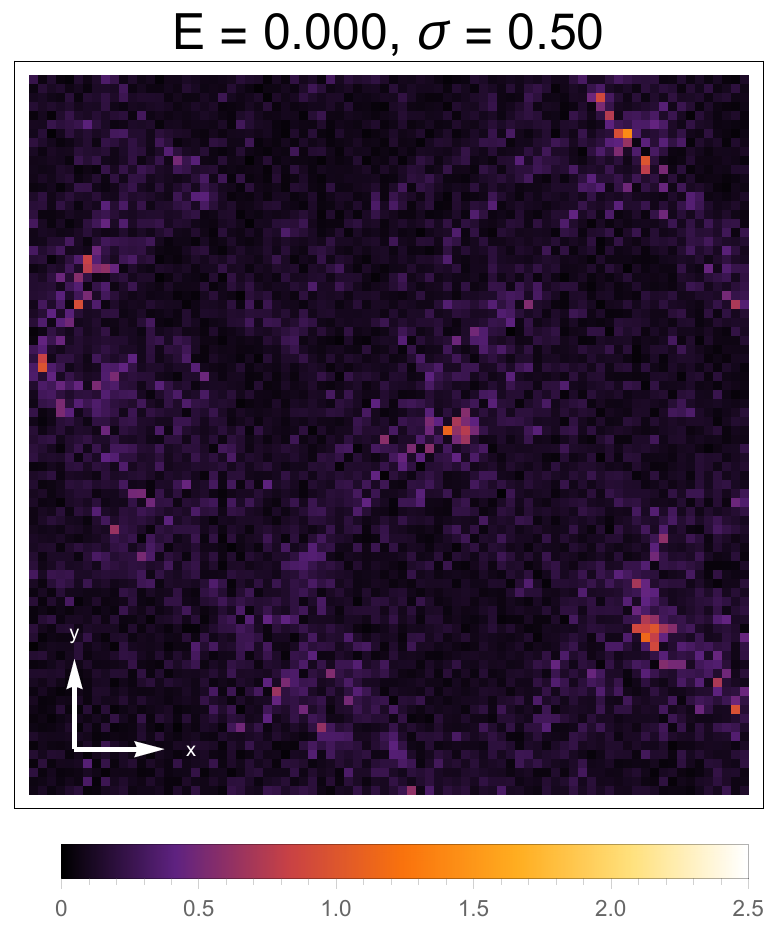}
	\includegraphics[width=.2\textwidth]{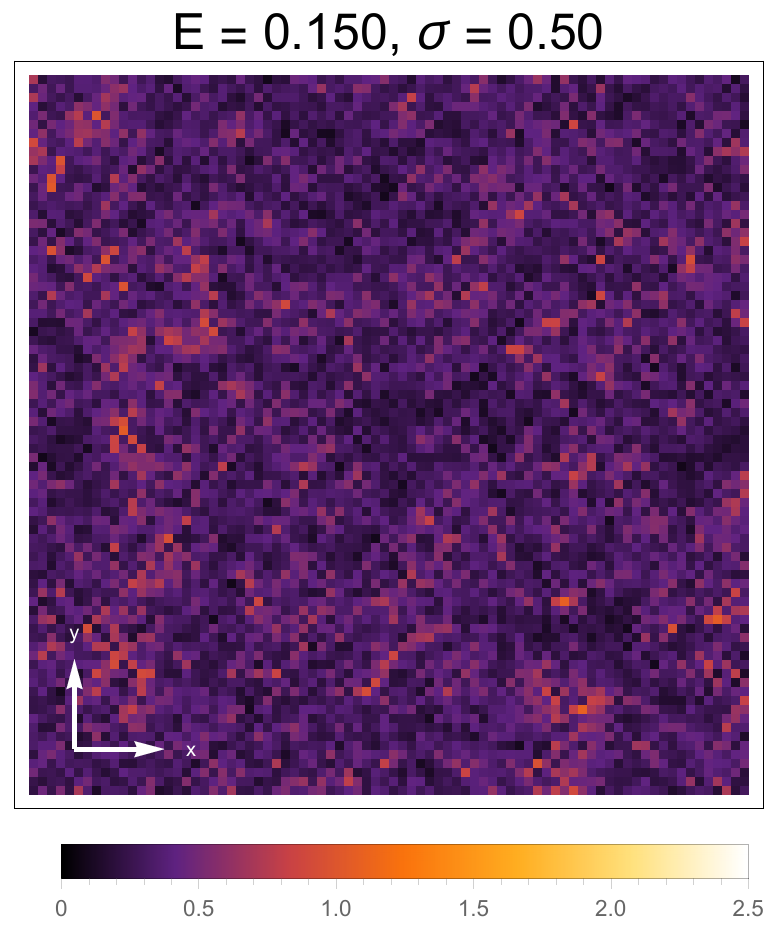}
	\includegraphics[width=.2\textwidth]{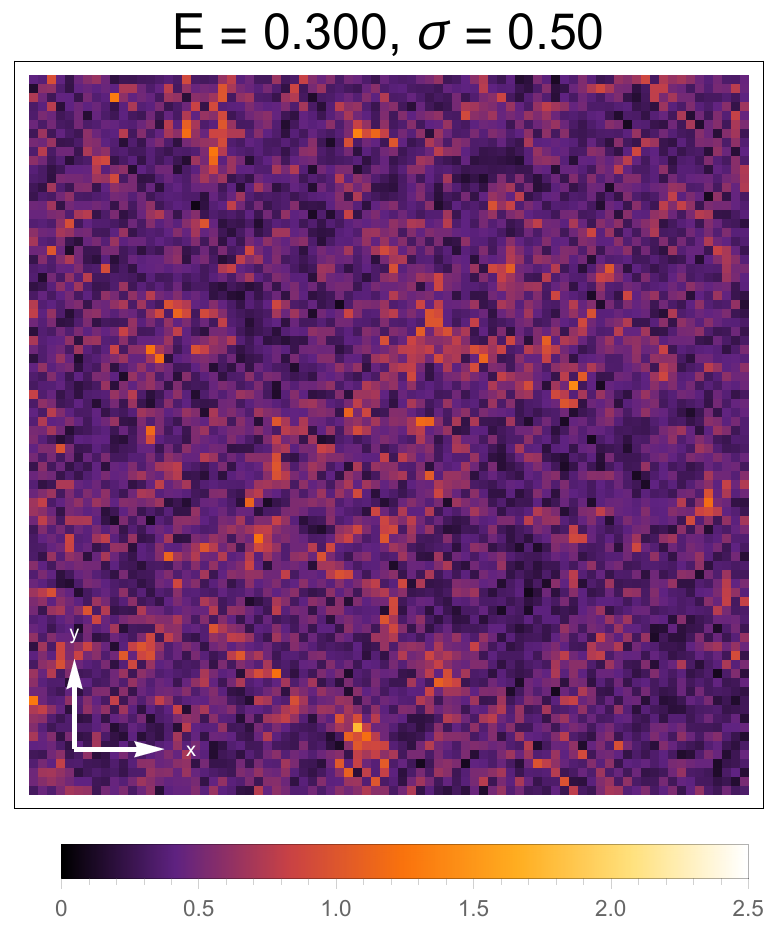} \\
	\includegraphics[width=.2\textwidth]{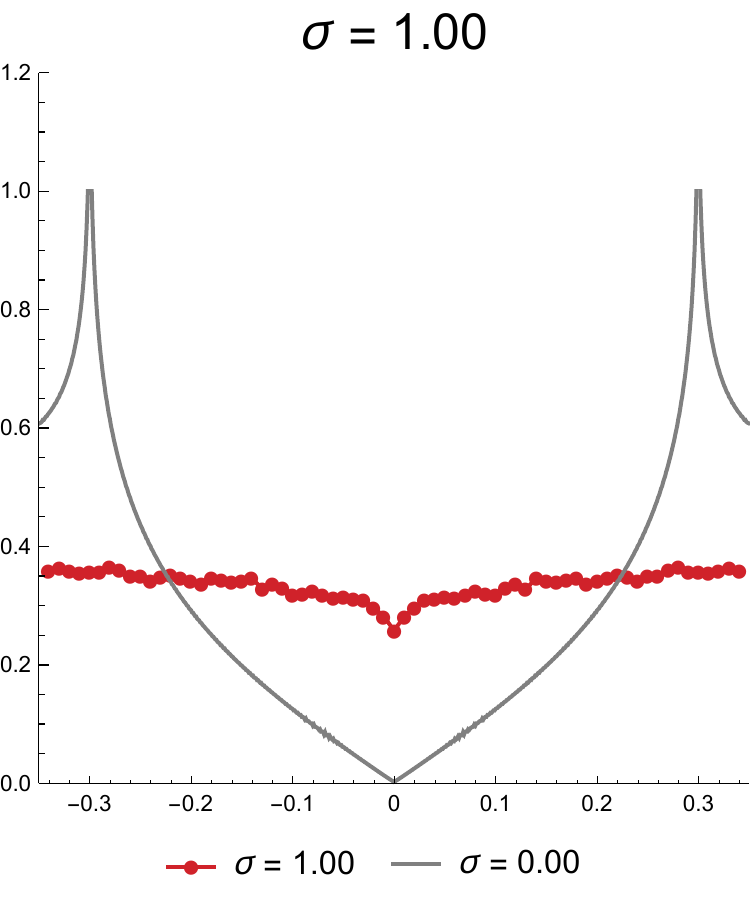}
	\includegraphics[width=.2\textwidth]{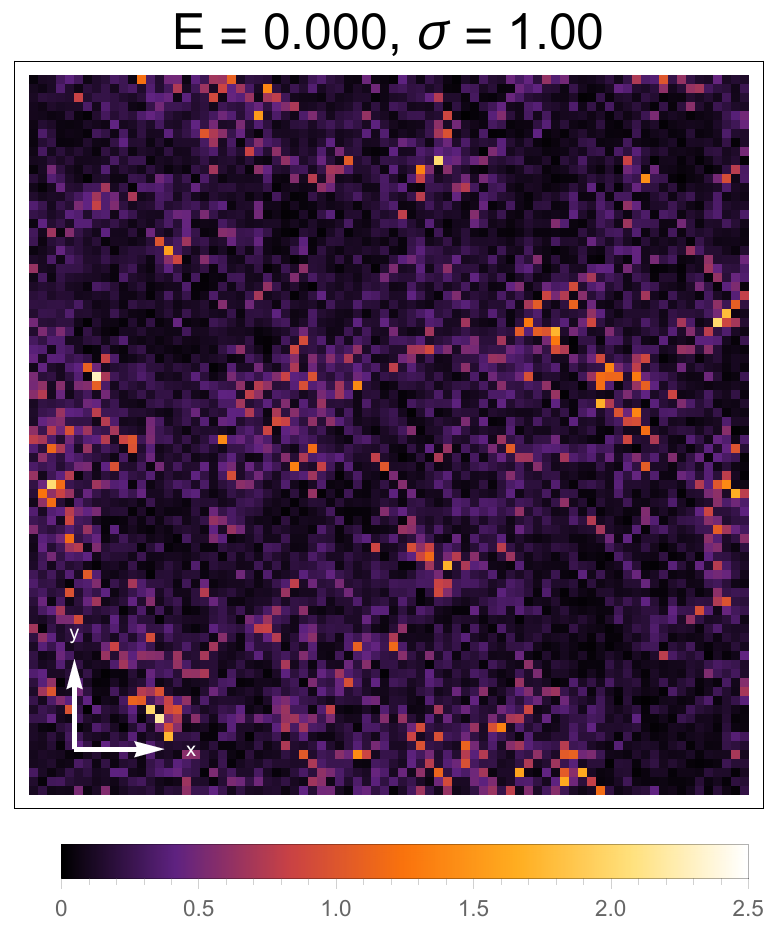}
	\includegraphics[width=.2\textwidth]{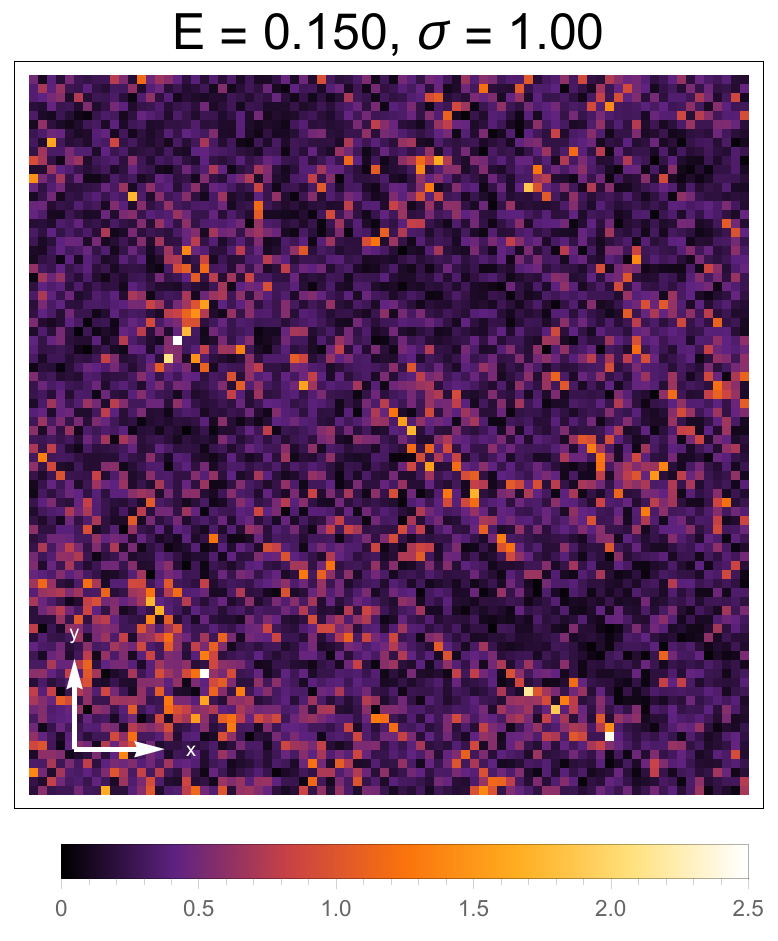}
	\includegraphics[width=.2\textwidth]{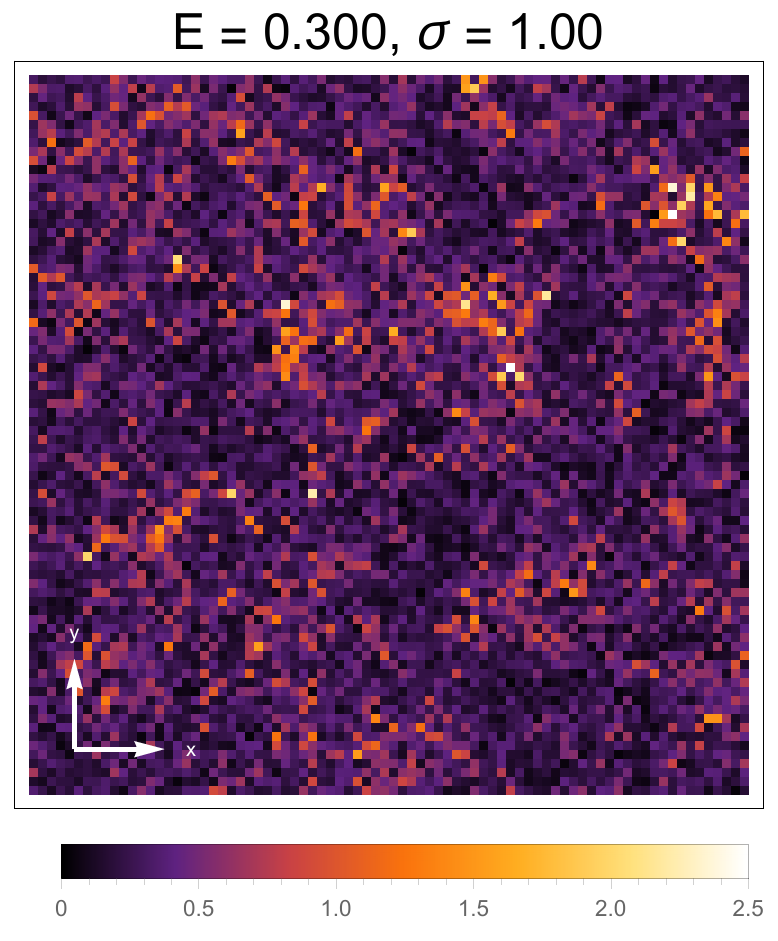}
	\caption{Snapshots of the real-space quasiparticle density of states for random Gaussian disorder with increasing standard deviation $\sigma$ (top to bottom) and energy $E$ (left to right), extracted from the middlemost $80 \times 80$ subset of the full system. The leftmost column shows plots of the DOS as a function of energy for a particular $\sigma$, along with plots of the clean case for comparison. The same disorder realizations as in Fig.~\ref{fig:rp} are used here. The color scale is the same for all plots.}
	\label{fig:rpqpdos}
\end{figure*}

 \begin{figure}[t]
 	\centering
 	\includegraphics[width=.5\textwidth]{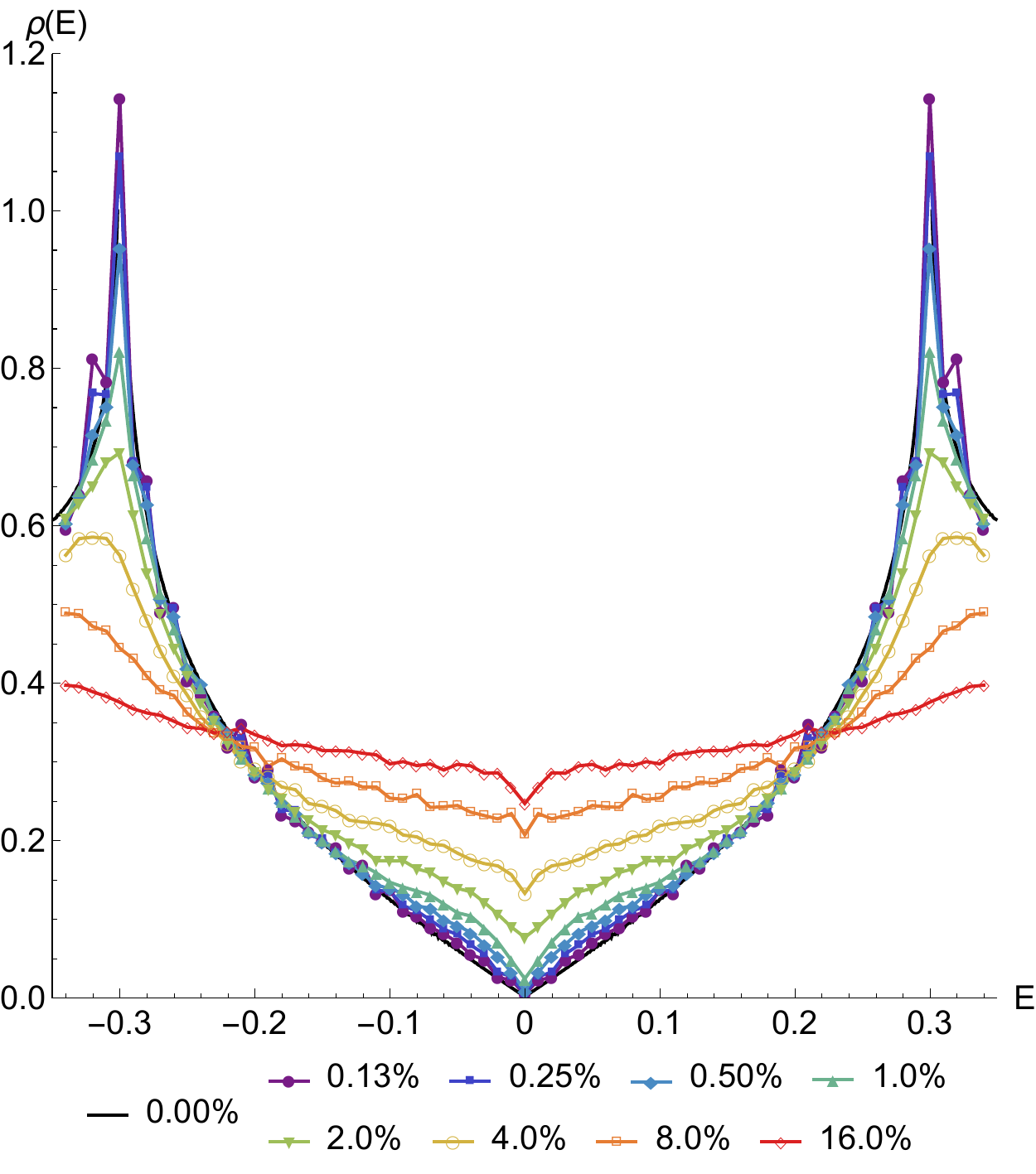}\hfill
 	\caption{Plots of the quasiparticle DOS as a function of energy $E$ for the multiple unitary-scatterer model, for various impurity concentrations.}
 	\label{fig:mp}
 \end{figure}	
 
 \begin{figure*}
 	\centering
 	\includegraphics[width=.2\textwidth]{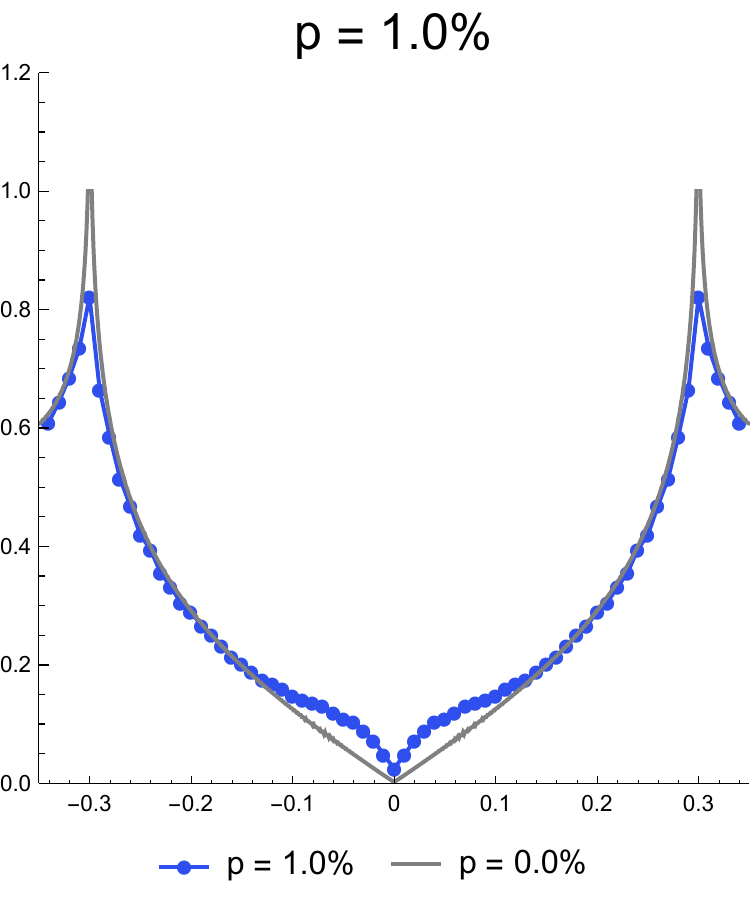}
 	\includegraphics[width=.2\textwidth]{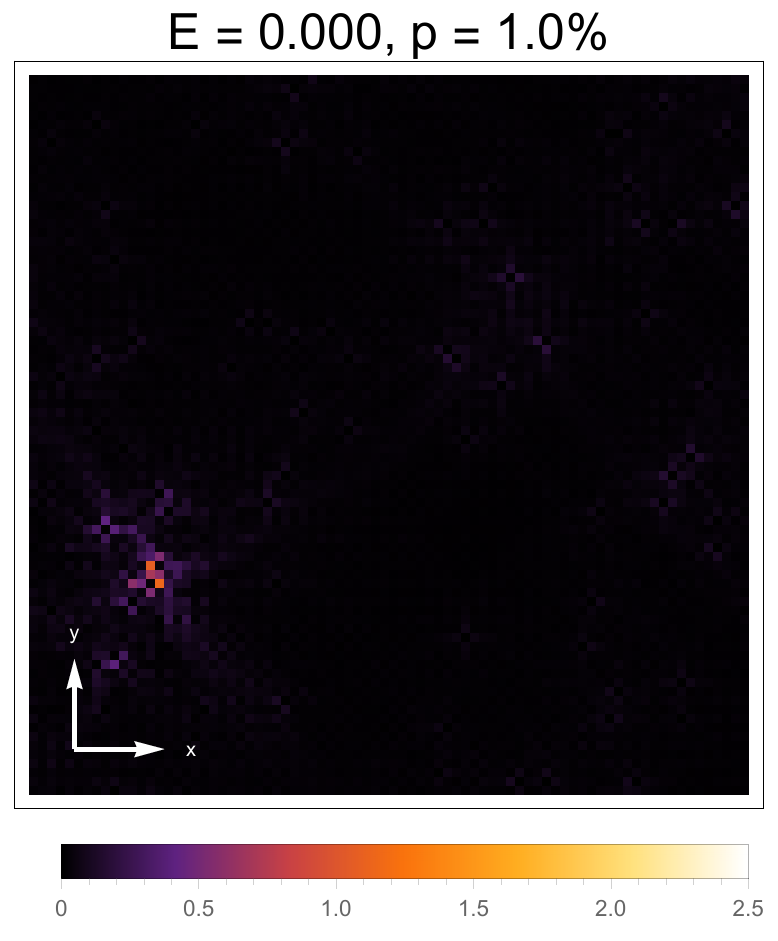}
 	\includegraphics[width=.2\textwidth]{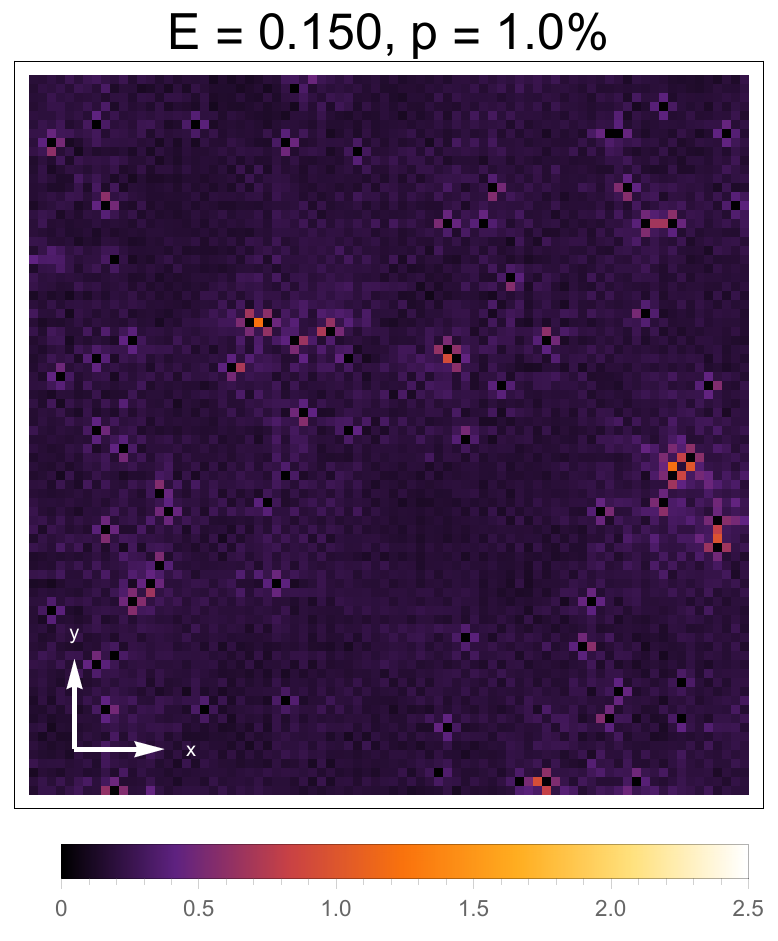}
 	\includegraphics[width=.2\textwidth]{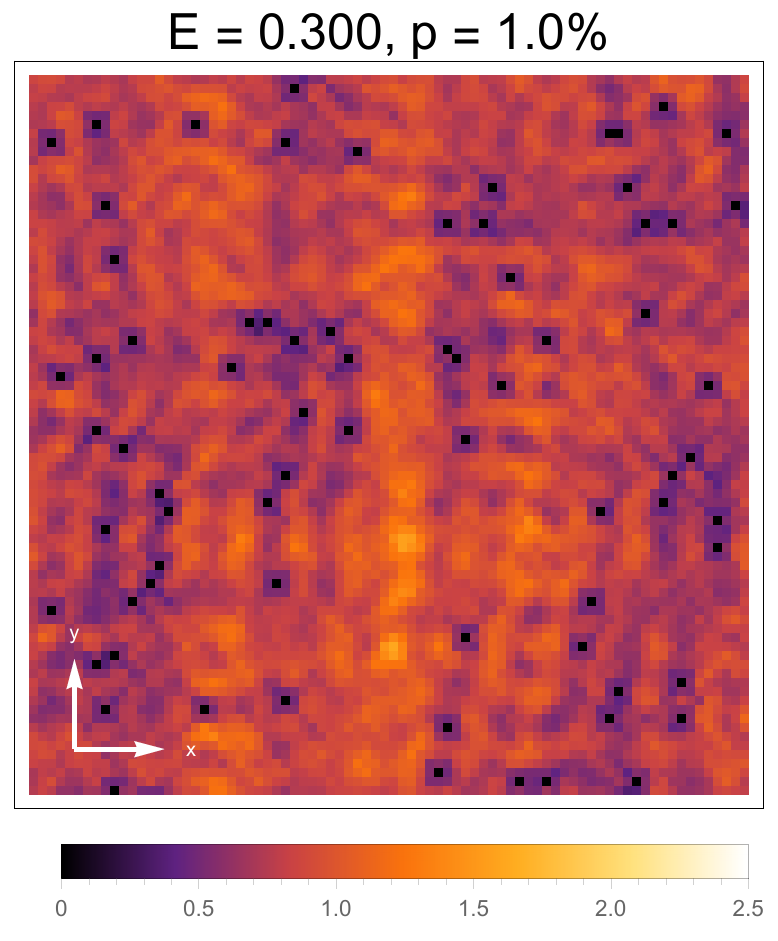} \\
 	\includegraphics[width=.2\textwidth]{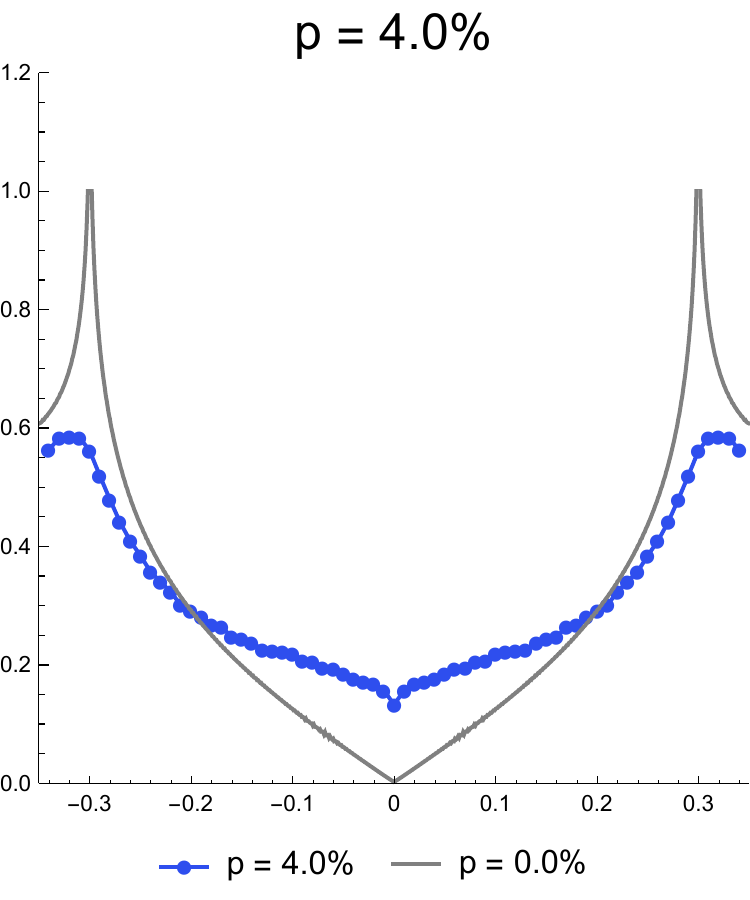}
 	\includegraphics[width=.2\textwidth]{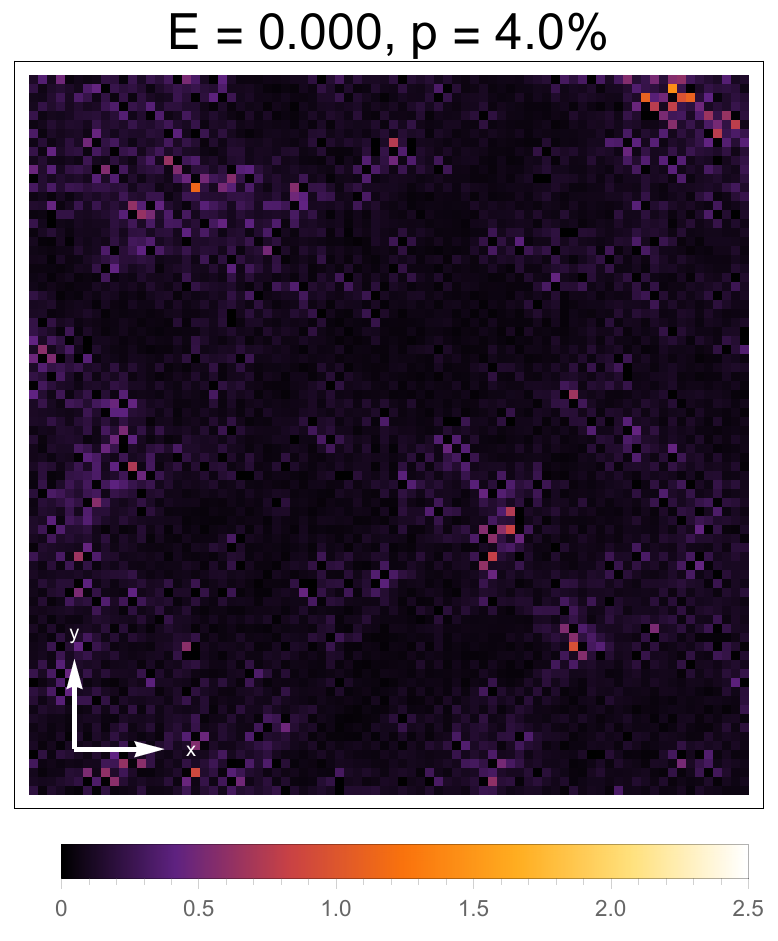} 
 	\includegraphics[width=.2\textwidth]{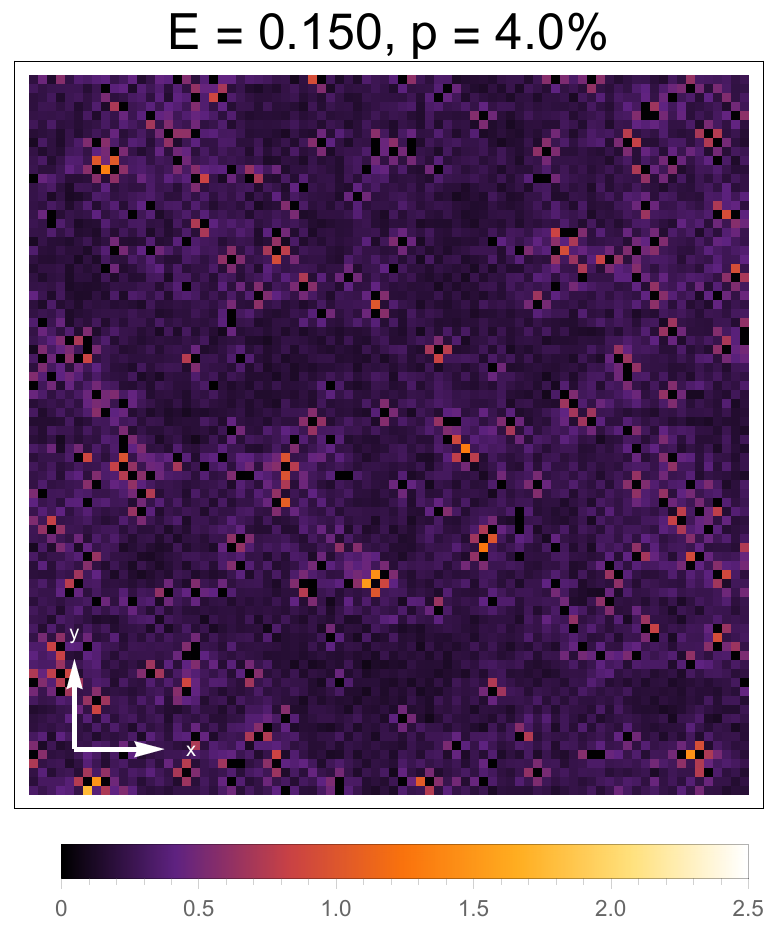}
 	\includegraphics[width=.2\textwidth]{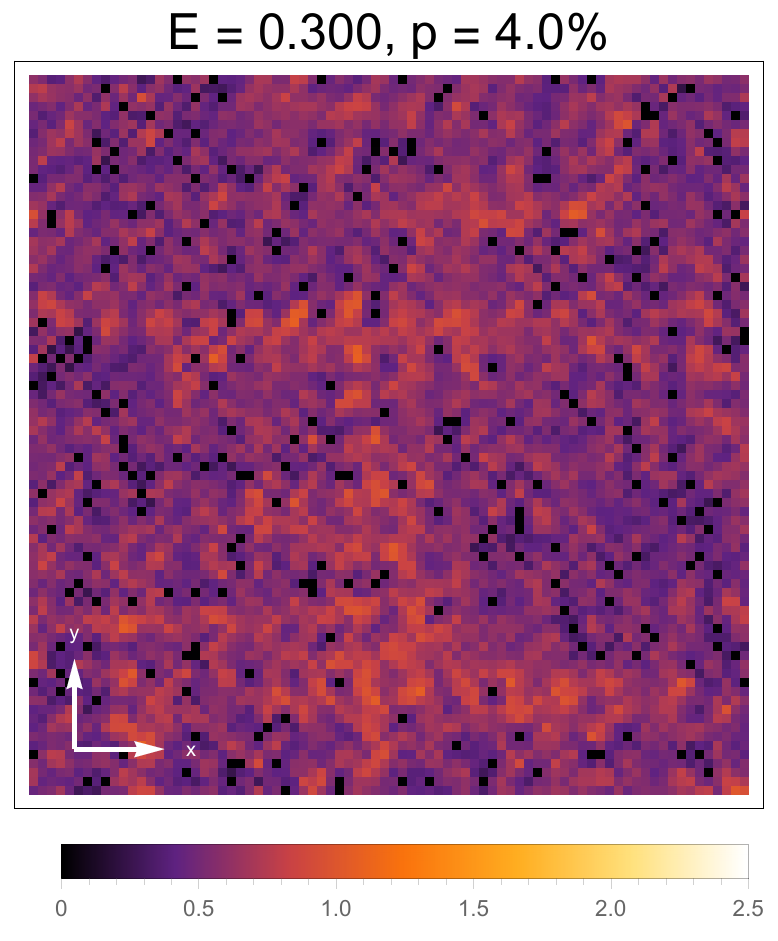} \\
 	\includegraphics[width=.2\textwidth]{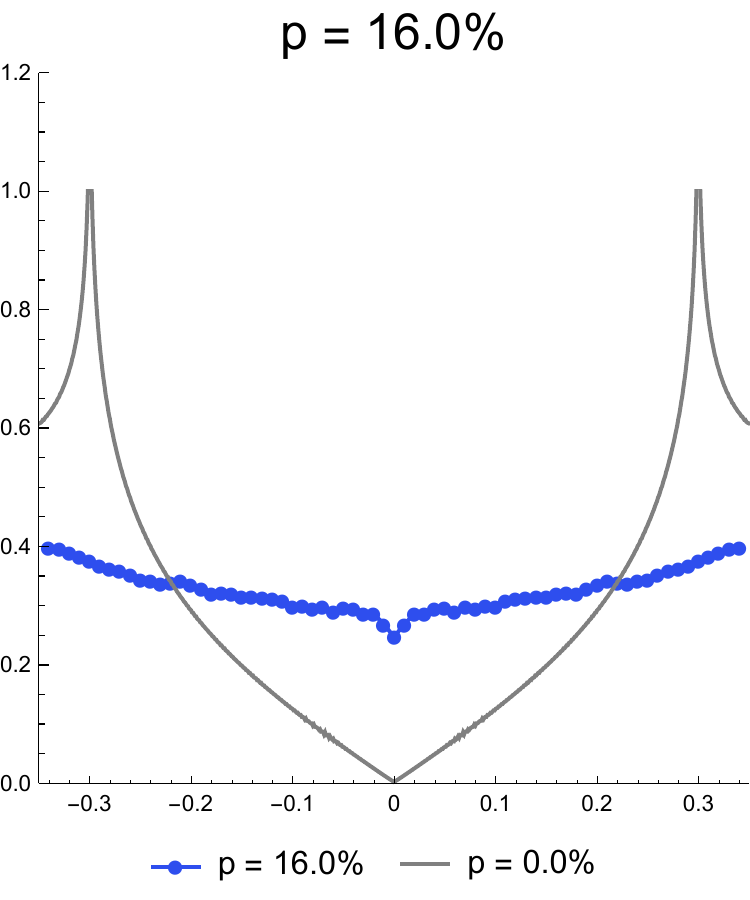}
 	\includegraphics[width=.2\textwidth]{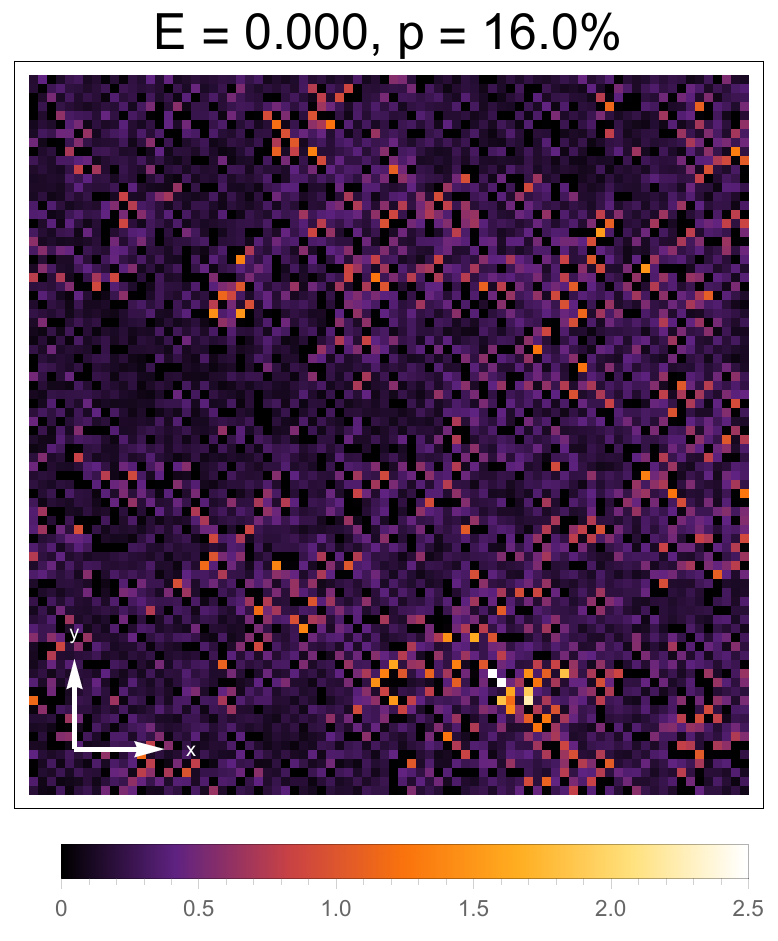}
 	\includegraphics[width=.2\textwidth]{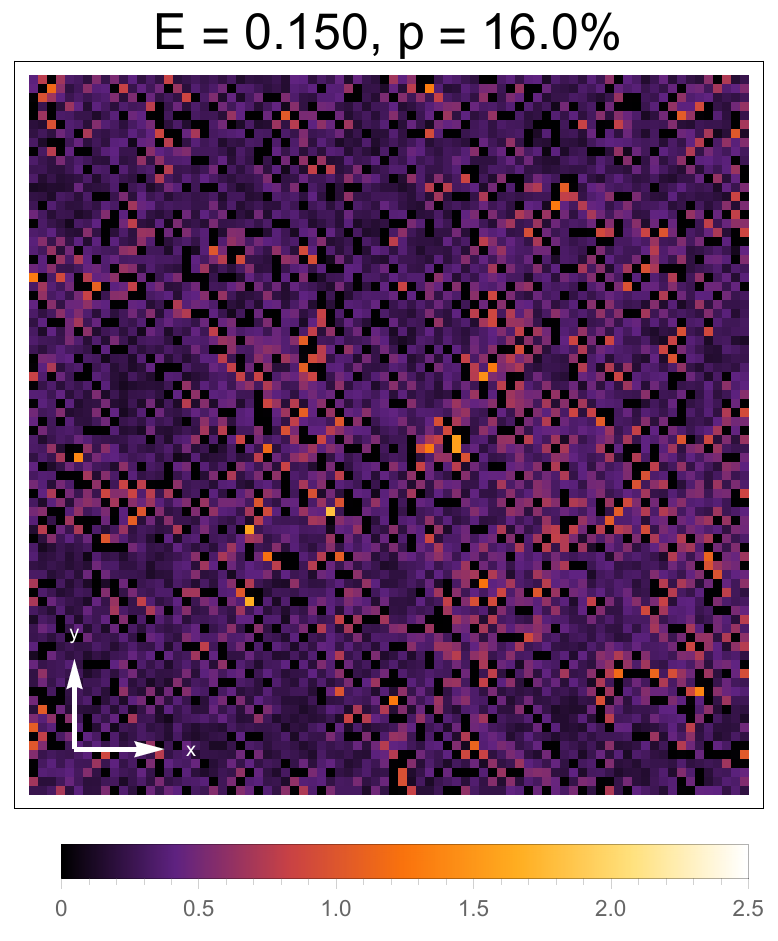}
 	\includegraphics[width=.2\textwidth]{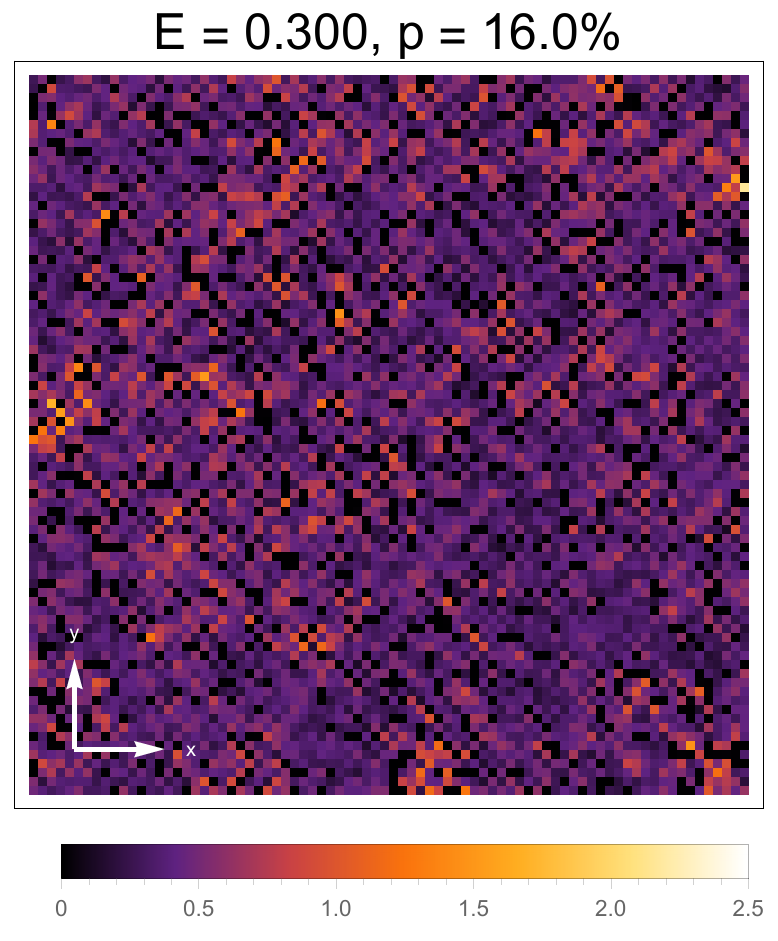}
 	\caption{Snapshots of the real-space quasiparticle density of states for an ensemble of unitary pointlike scatterers ($V_U = 10$) with increasing impurity concentration $p$ (top to bottom) and energy $E$ (left to right), extracted from the middlemost $80 \times 80$ subset of the full system. The leftmost column shows plots of the DOS as a function of energy for a particular $p$, along with plots of the clean case for comparison. The same disorder realizations as in Fig.~\ref{fig:mp} are used here. The color scale is the same for all plots.}
 	\label{fig:mpqpdos}
 \end{figure*}
 
  \begin{figure}[t]
  	\centering
  	\includegraphics[width=.5\textwidth]{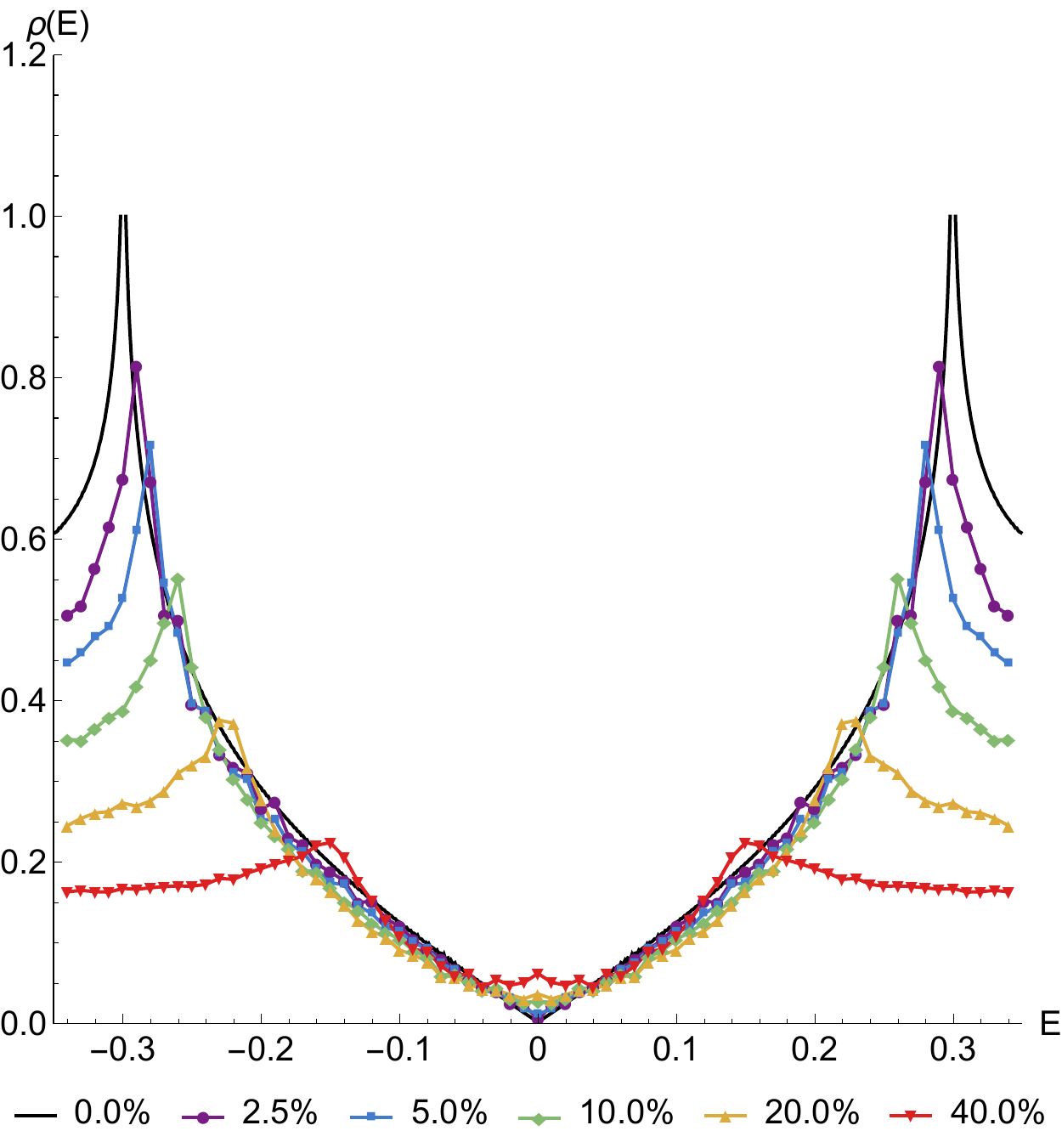}\hfill
  	\caption{Plots of the quasiparticle DOS as a function of energy $E$ for the multiple smooth-scatterer model with \emph{positive} net potential, for various impurity concentrations.}
  	\label{fig:ss}
  \end{figure}	
  
  \begin{figure*}
  	\centering
  	\includegraphics[width=.2\textwidth]{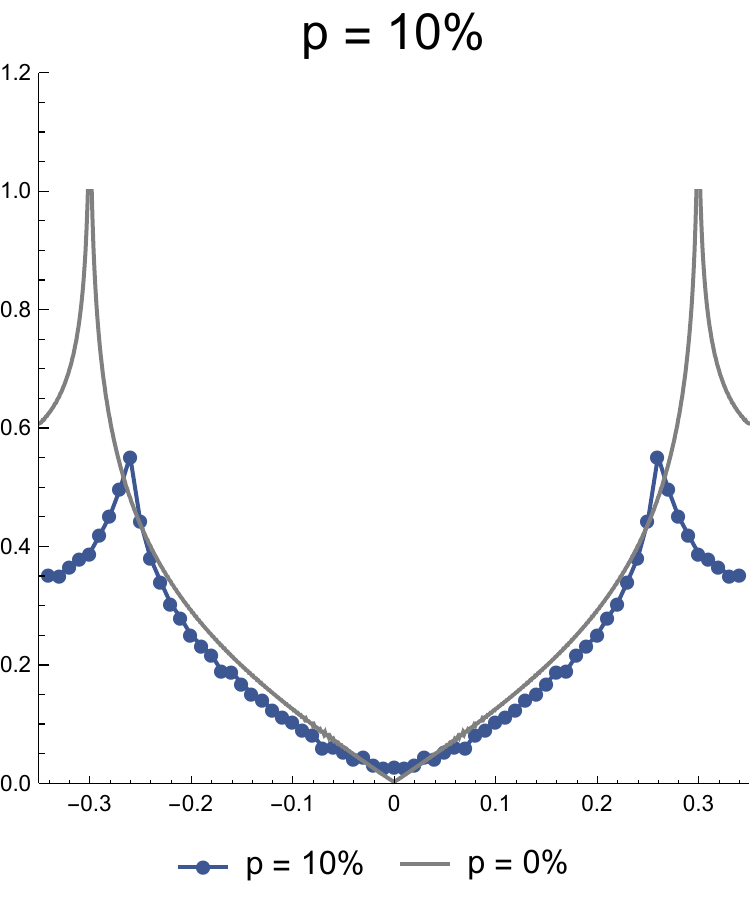}
  	\includegraphics[width=.2\textwidth]{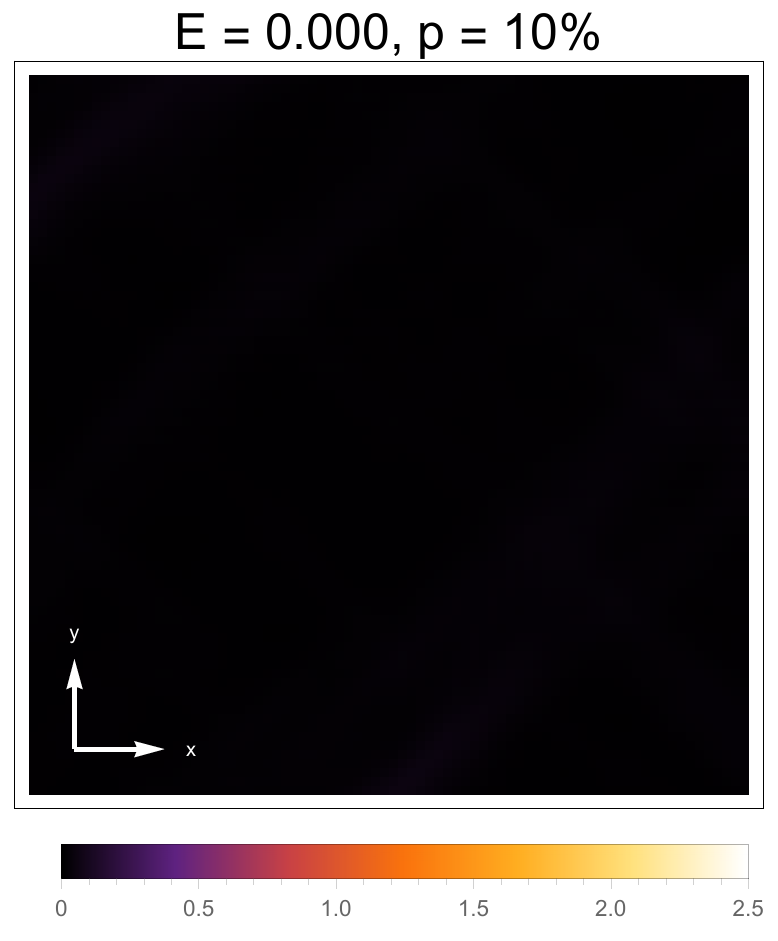}
  	\includegraphics[width=.2\textwidth]{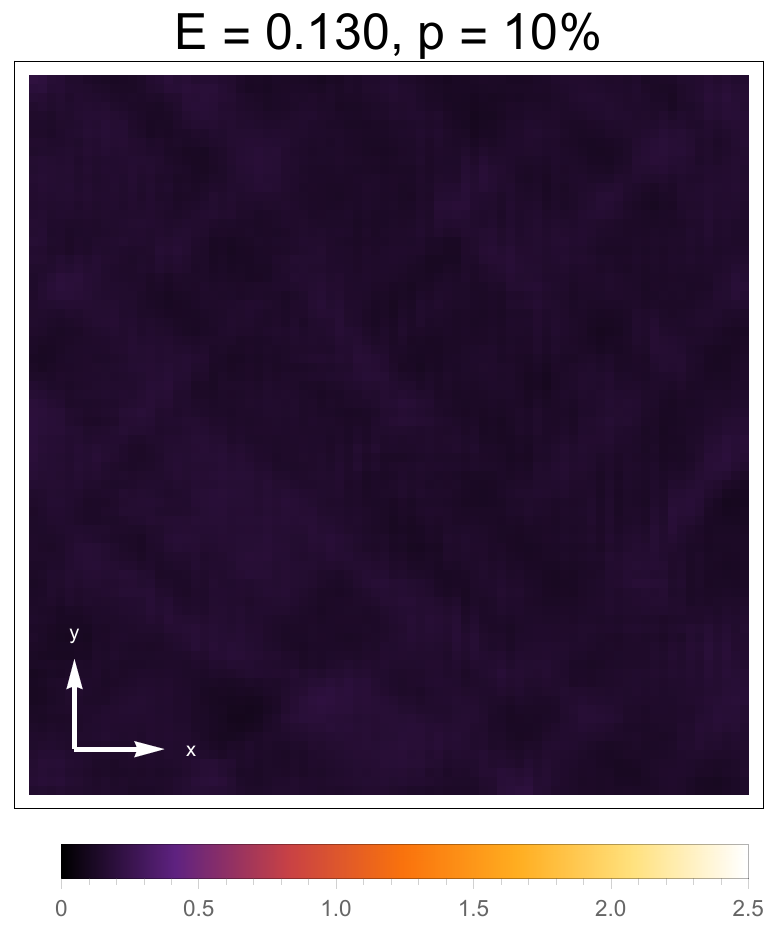}
  	\includegraphics[width=.2\textwidth]{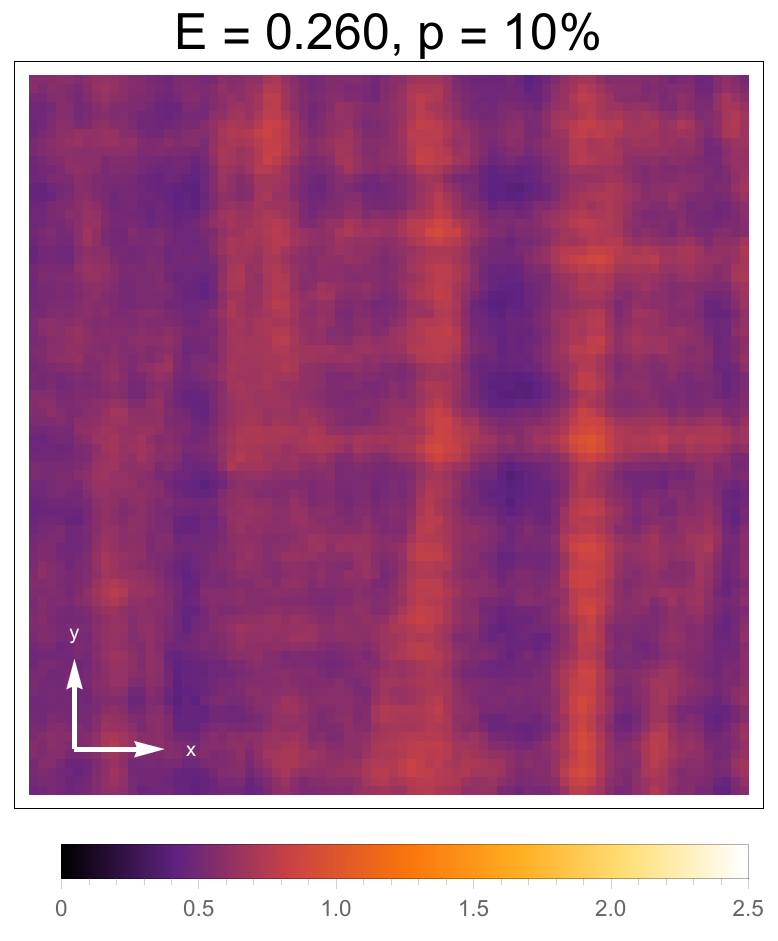} \\
  	\includegraphics[width=.2\textwidth]{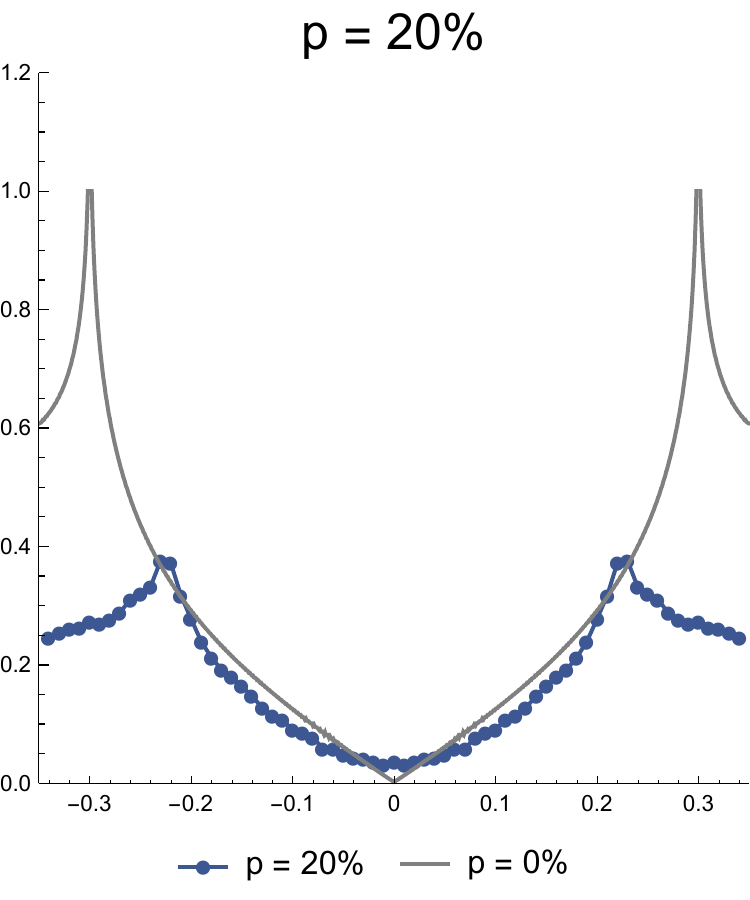}
  	\includegraphics[width=.2\textwidth]{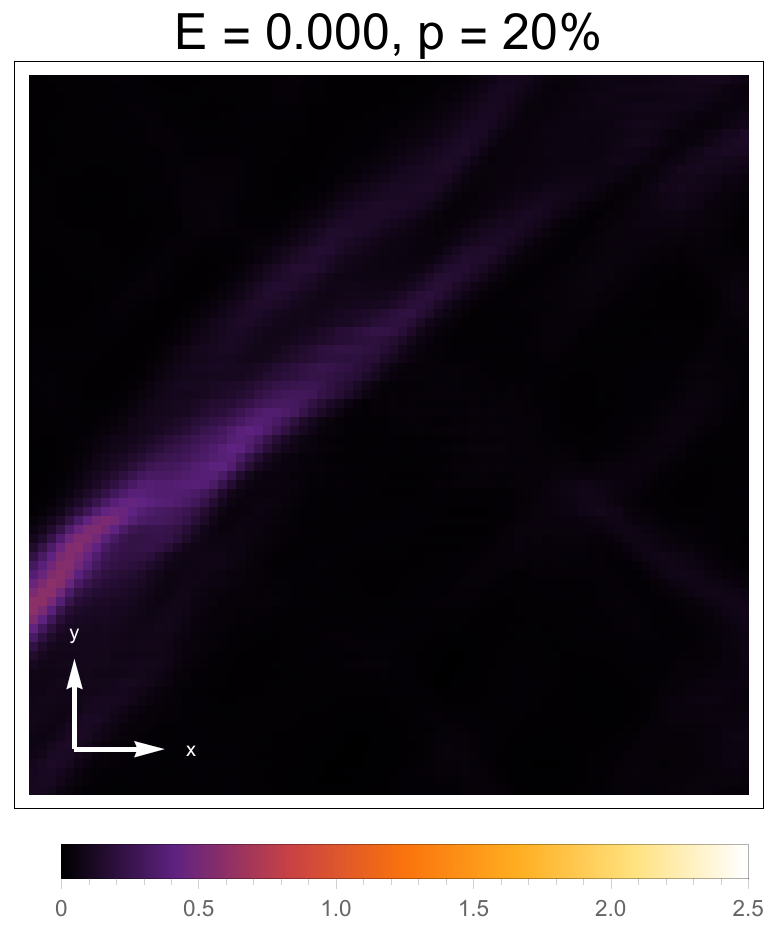}
  	\includegraphics[width=.2\textwidth]{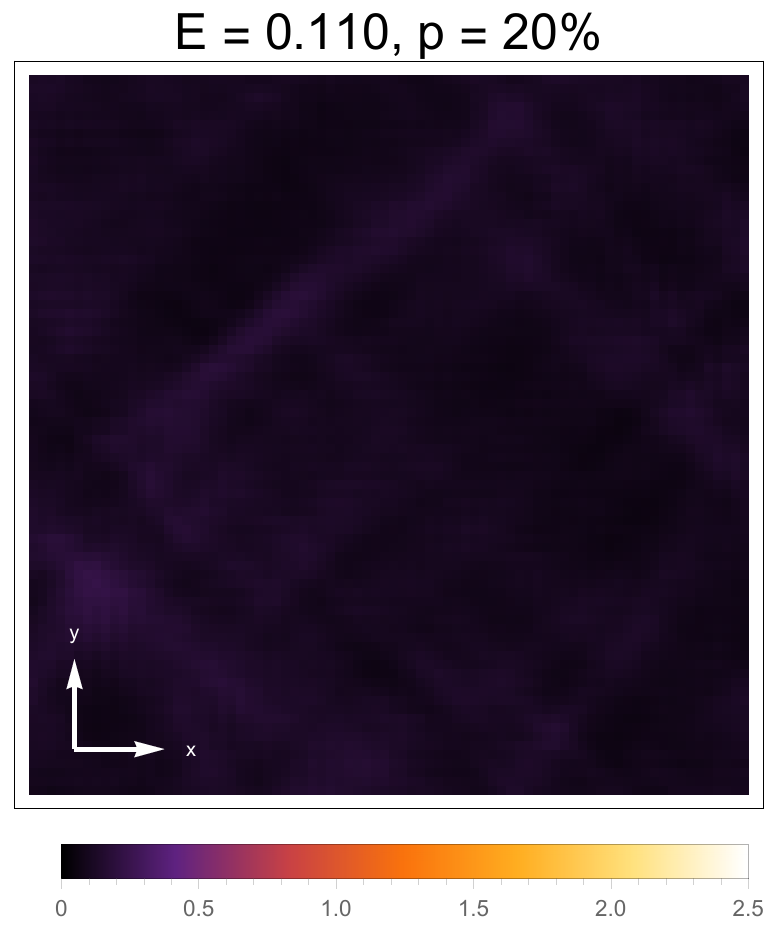}
  	\includegraphics[width=.2\textwidth]{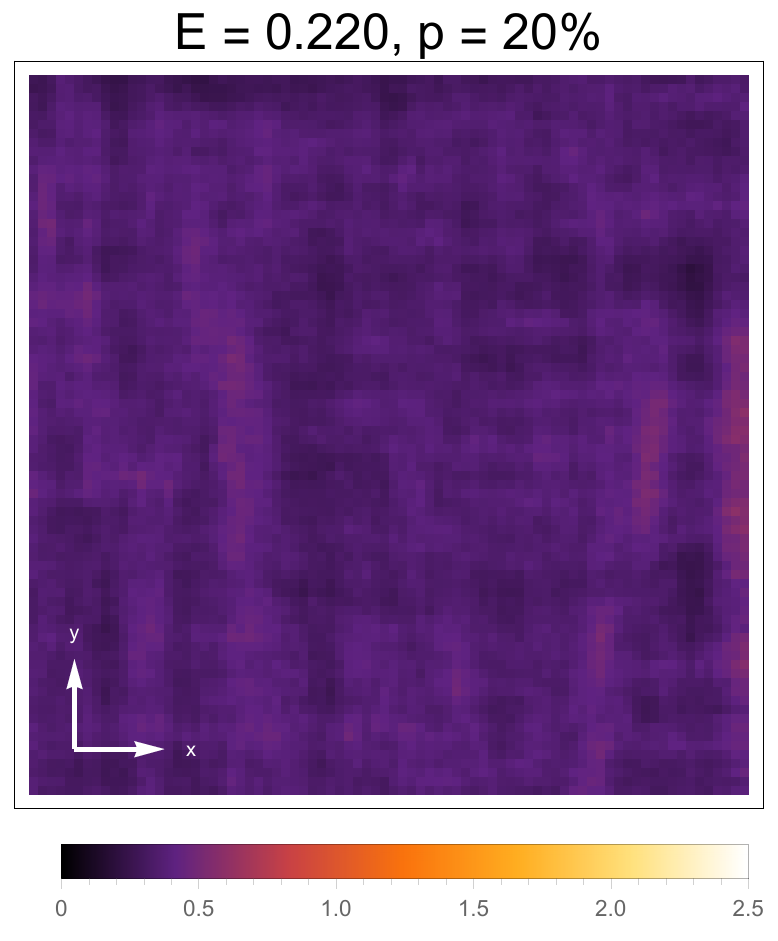} \\
  	\includegraphics[width=.2\textwidth]{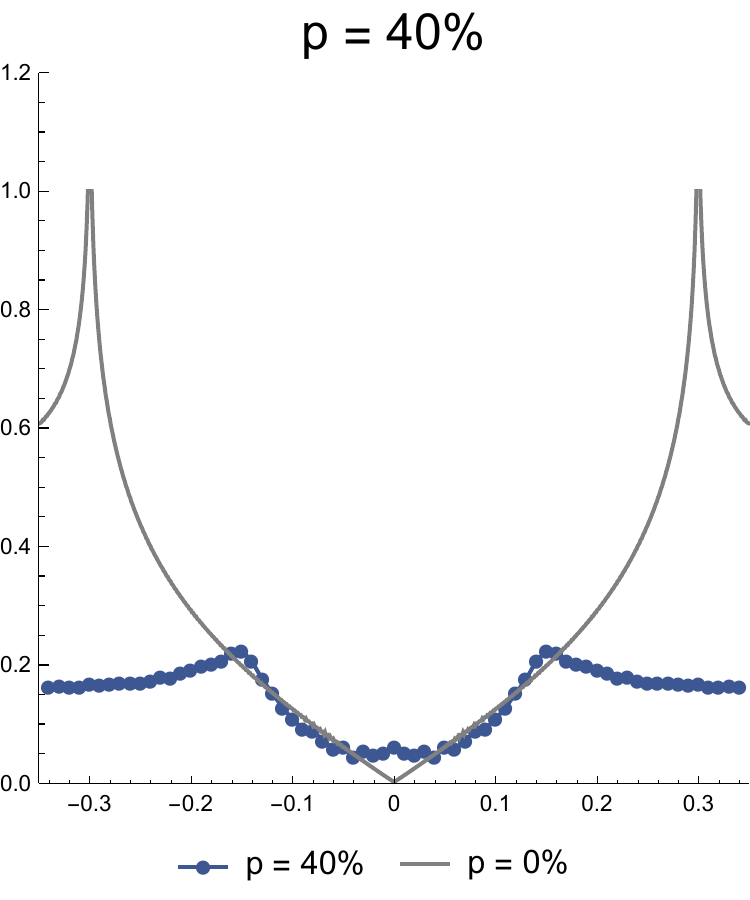}
  	\includegraphics[width=.2\textwidth]{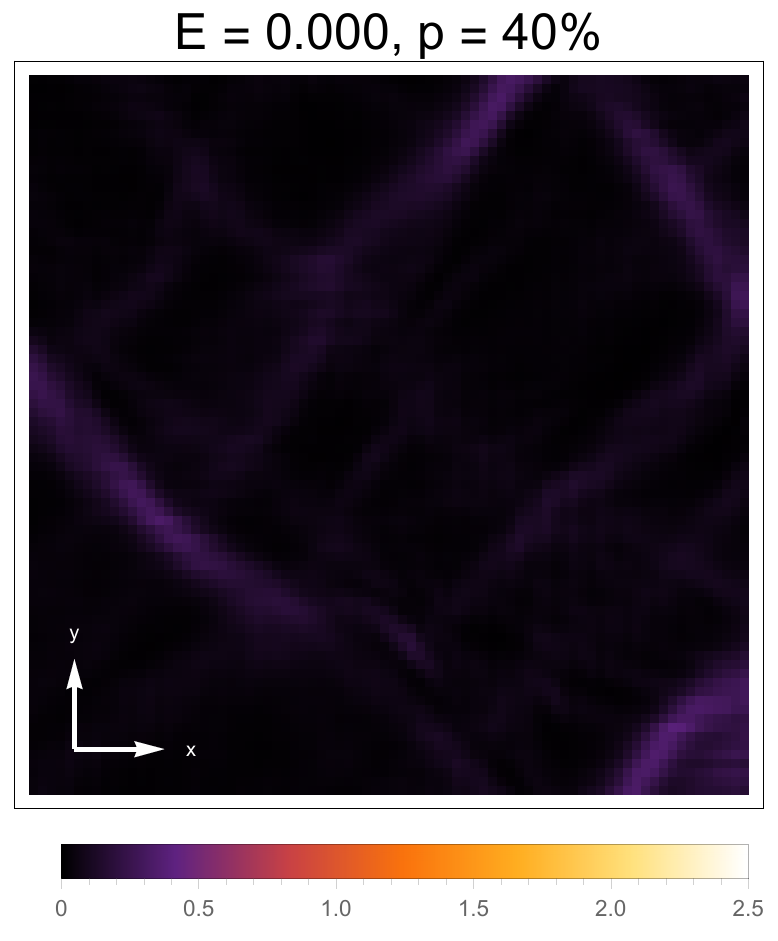}
  	\includegraphics[width=.2\textwidth]{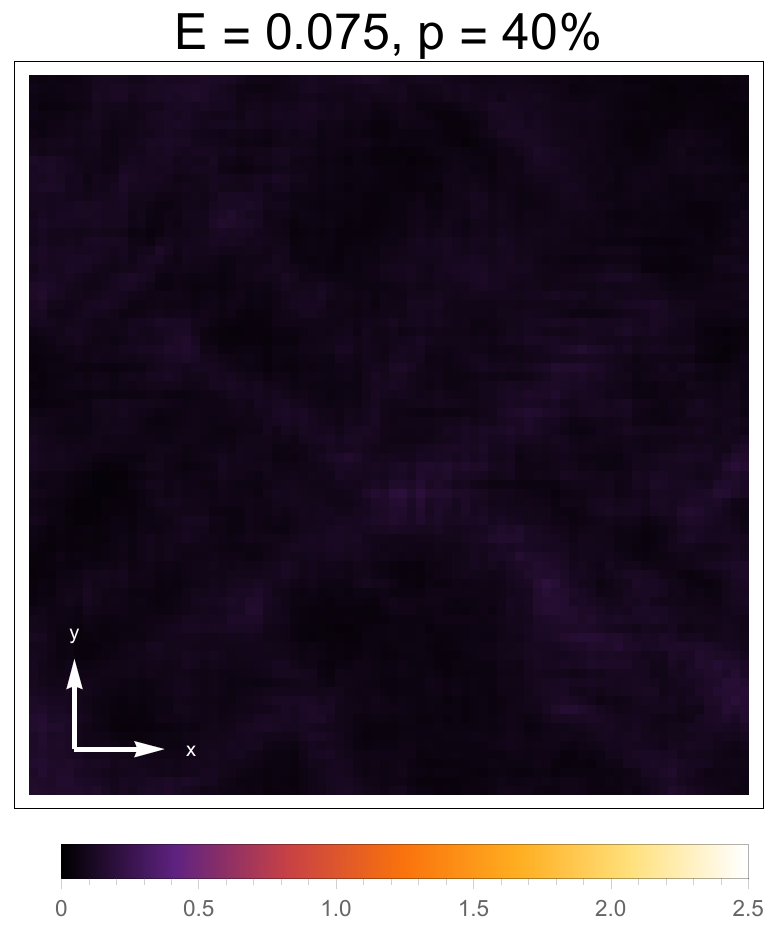}
  	\includegraphics[width=.2\textwidth]{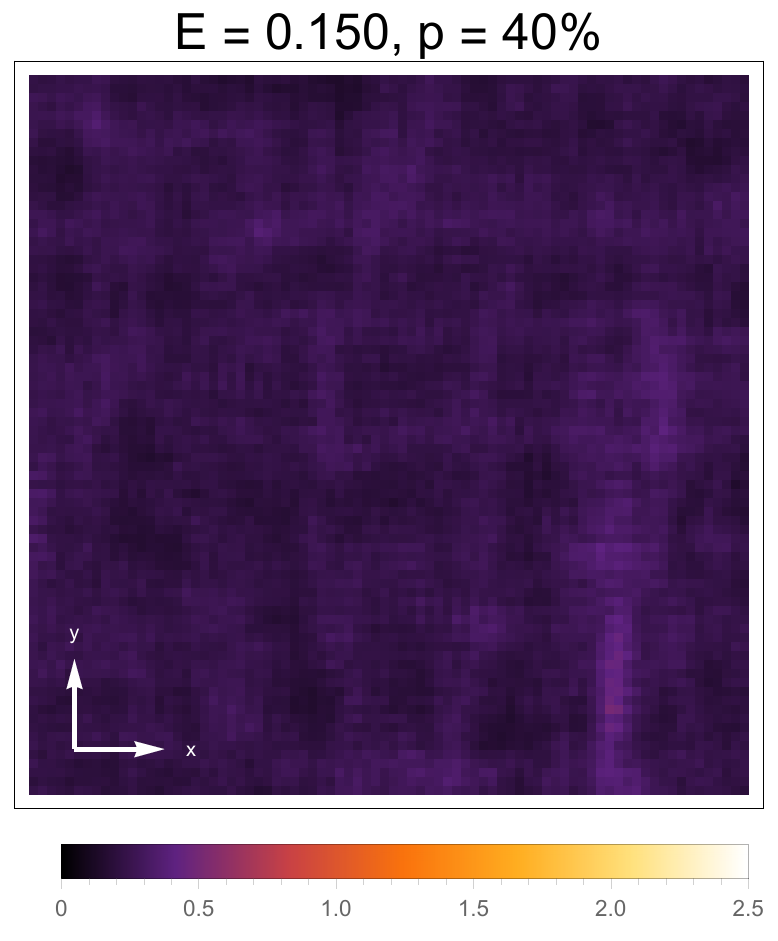}
  	\caption{Snapshots of the real-space quasiparticle density of states for smooth disorder (with \emph{positive} net potential) with increasing impurity concentration $p$ (top to bottom) and energy $E$ (left to right), extracted from the middlemost $80 \times 80$ subset of the full system. The energy at the rightmost column corresponds to the location at which the coherence peaks can be found, while the energy at the middle column is half the coherence-peak energy. The leftmost column shows plots of the DOS as a function of energy for a particular $p$, along with plots of the clean case for comparison. The same disorder realizations as in Fig.~\ref{fig:ss} are used here. The color scale is the same for all plots.}
  	\label{fig:ssqpdos}
  \end{figure*}

    \begin{figure}[t]
    	\centering
    	\includegraphics[width=.5\textwidth]{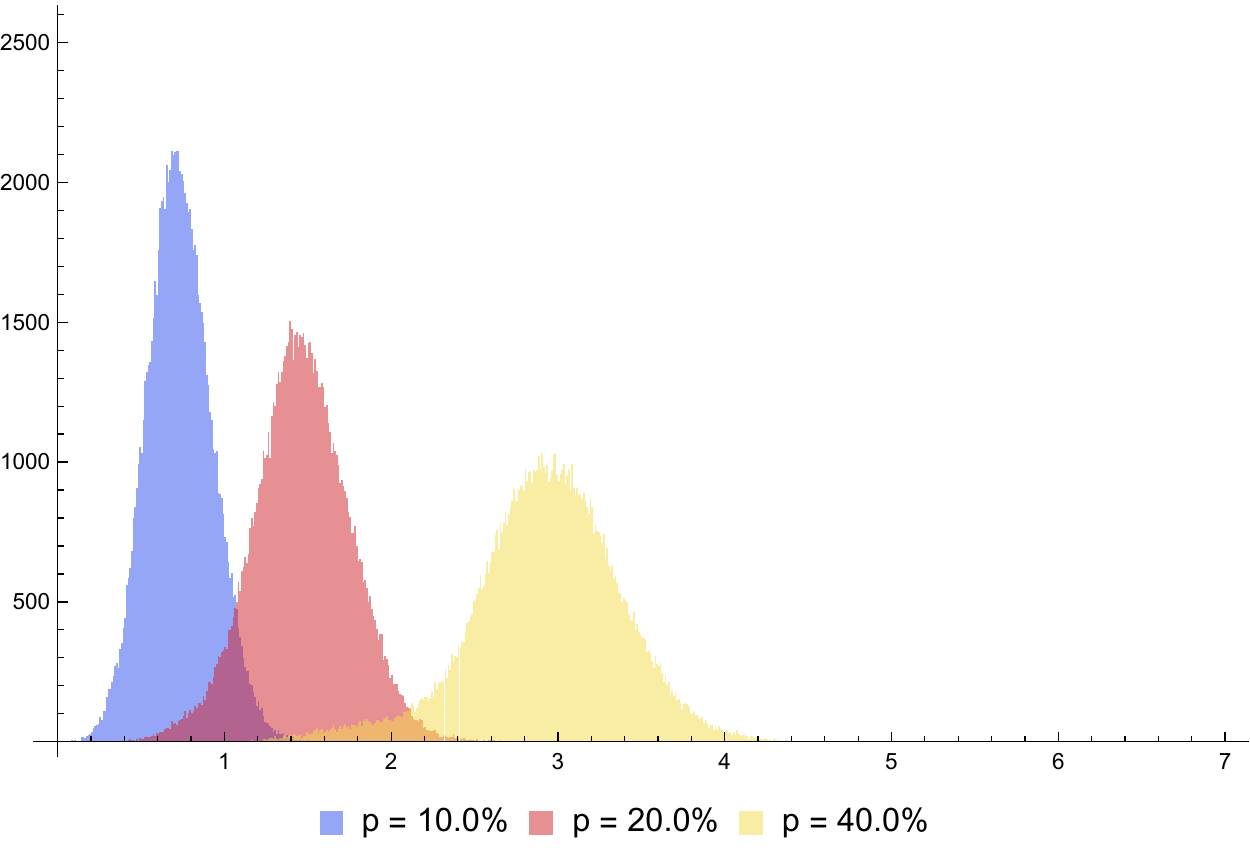}\hfill
    	\caption{Histogram of the values of the disorder potential for smooth disorder with positive net potential for three values of $p$. The width of each bin is 0.01. Notice that the mean of the disorder potential is nonzero, leading to a shift in the average chemical potential of the overall system.}
    	\label{fig:SSh}
    \end{figure}	
  
        \begin{figure}[t]
        	\centering
        	\includegraphics[width=.5\textwidth]{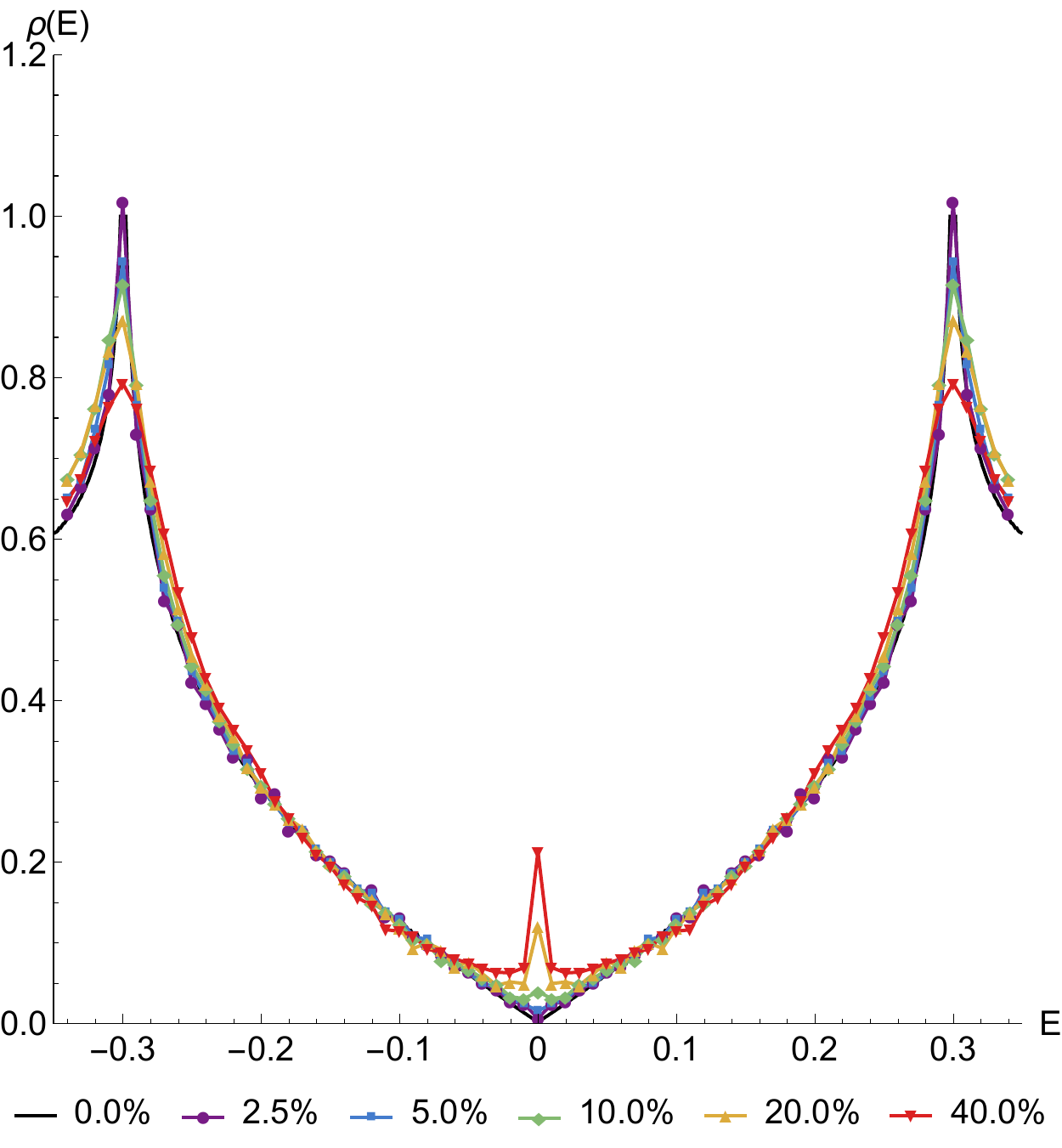}\hfill
        	\caption{Plots of the quasiparticle DOS as a function of energy $E$ for the multiple smooth-scatterer model with \emph{zero} net potential, for various impurity concentrations.}
        	\label{fig:ssn}
        \end{figure}	
        
        \begin{figure*}
        	\centering
        	\includegraphics[width=.2\textwidth]{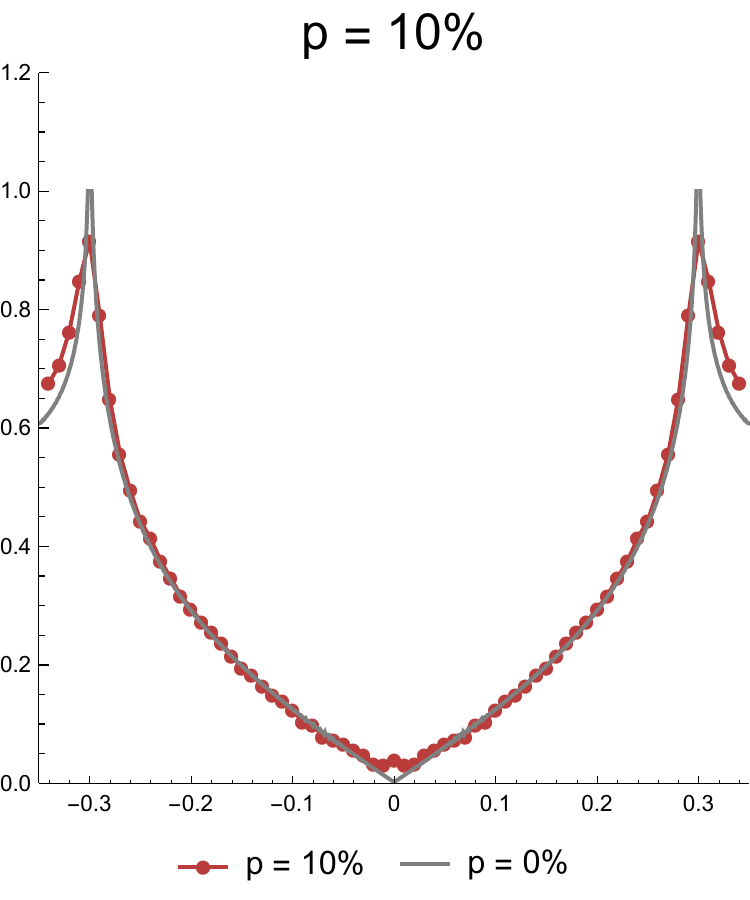}
        	\includegraphics[width=.2\textwidth]{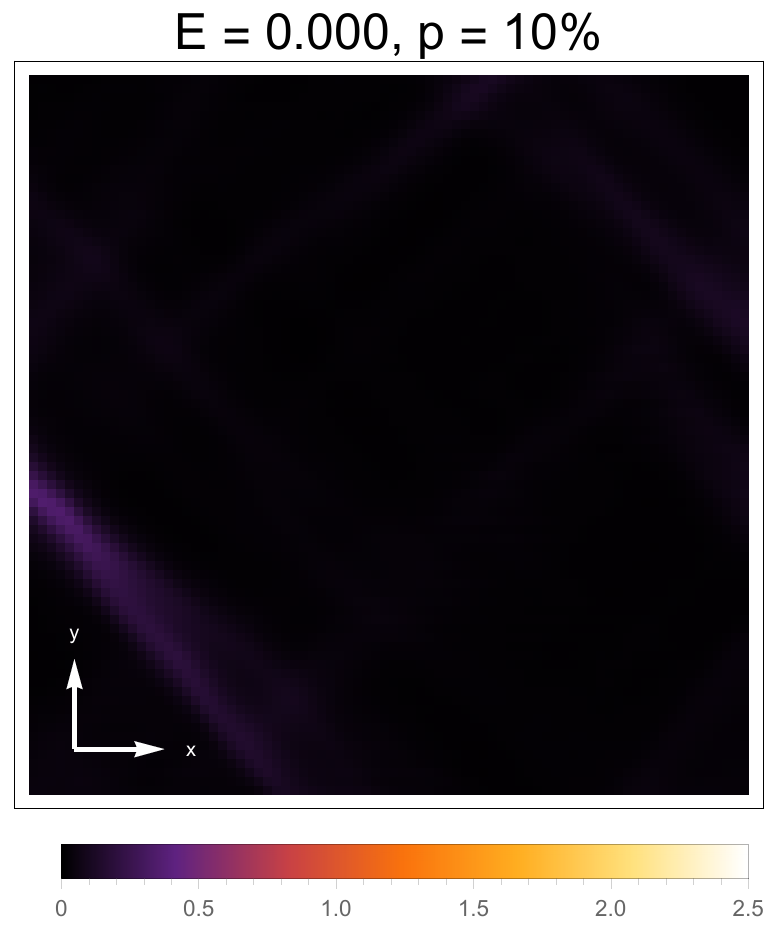}
        	\includegraphics[width=.2\textwidth]{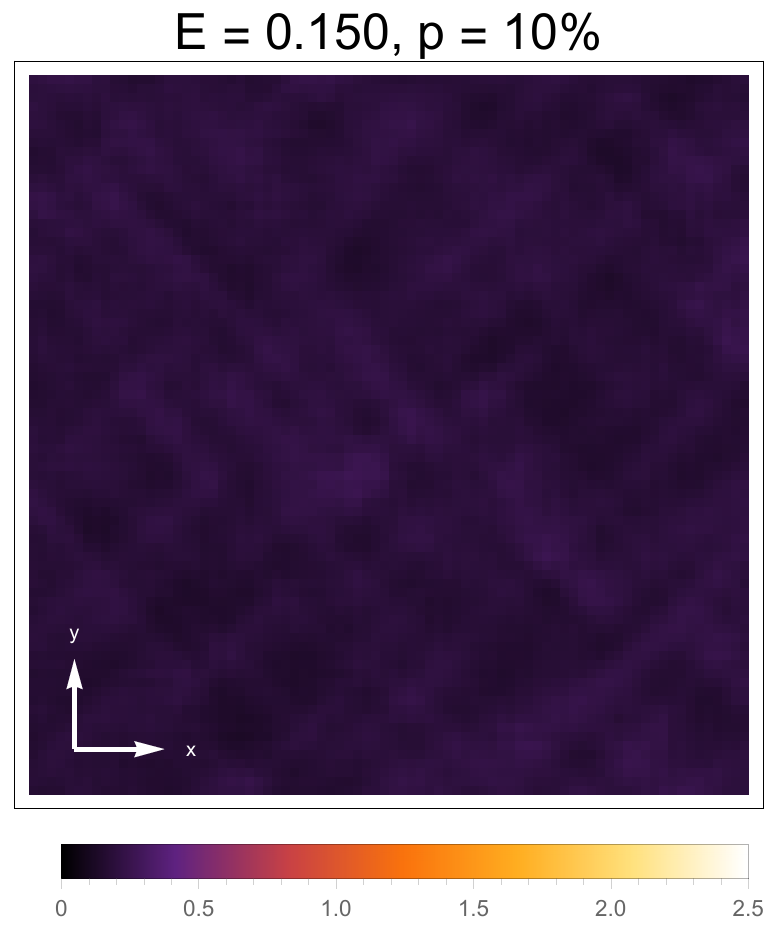}
        	\includegraphics[width=.2\textwidth]{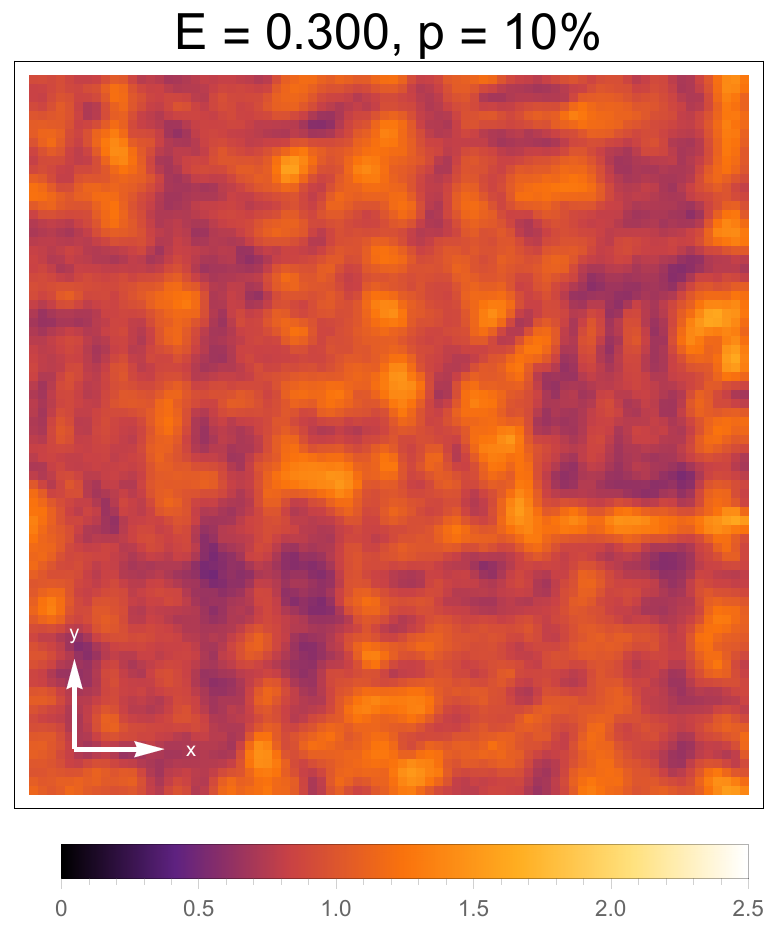} \\
        	\includegraphics[width=.2\textwidth]{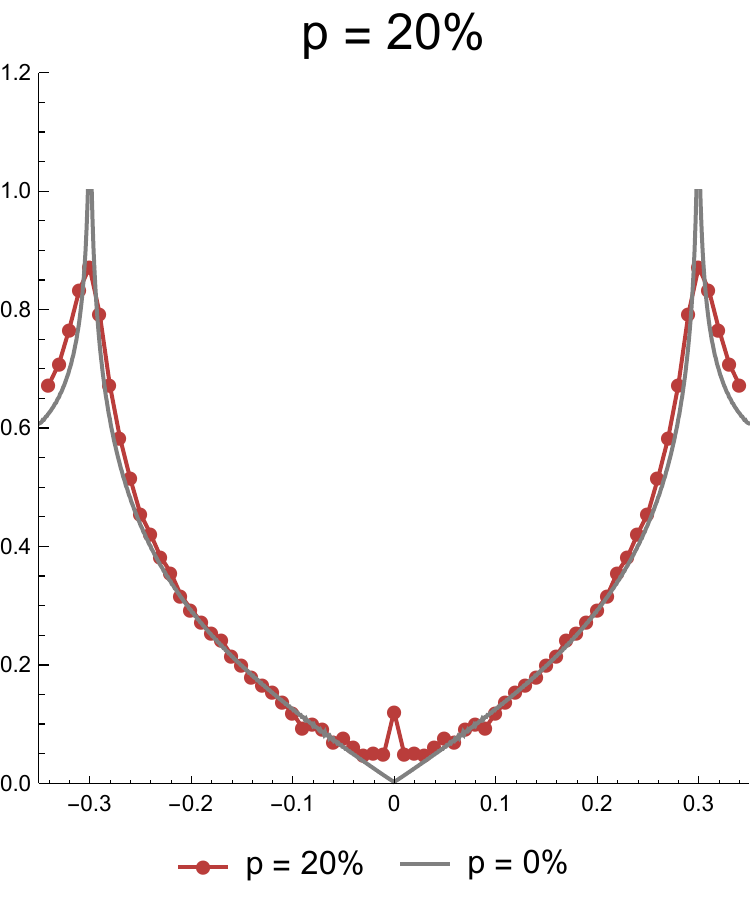}
        	\includegraphics[width=.2\textwidth]{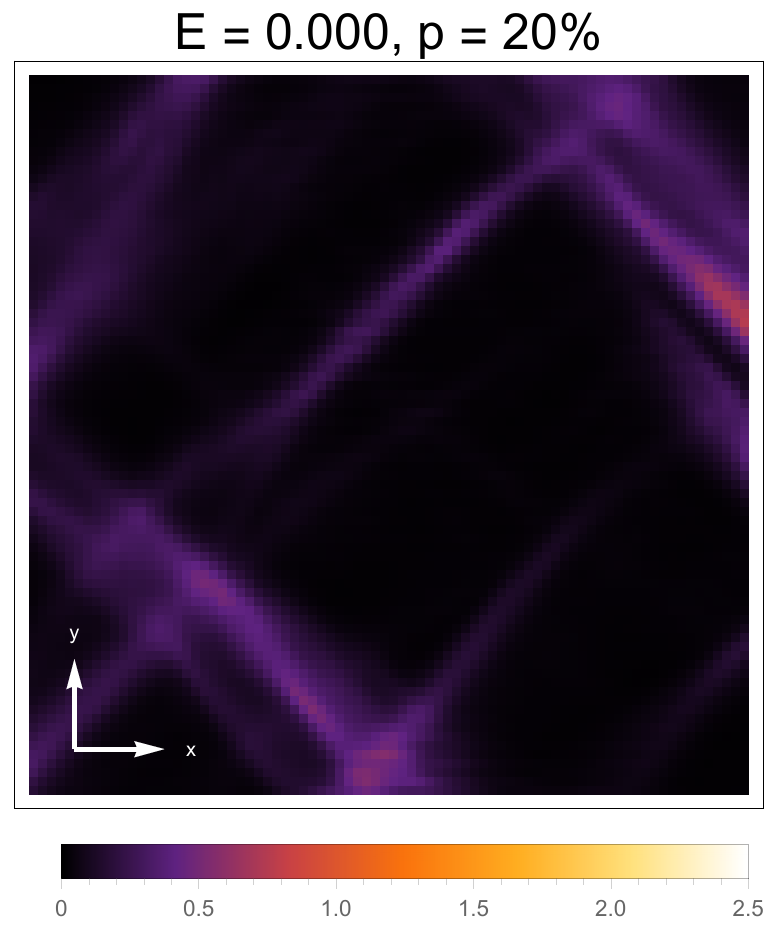}
        	\includegraphics[width=.2\textwidth]{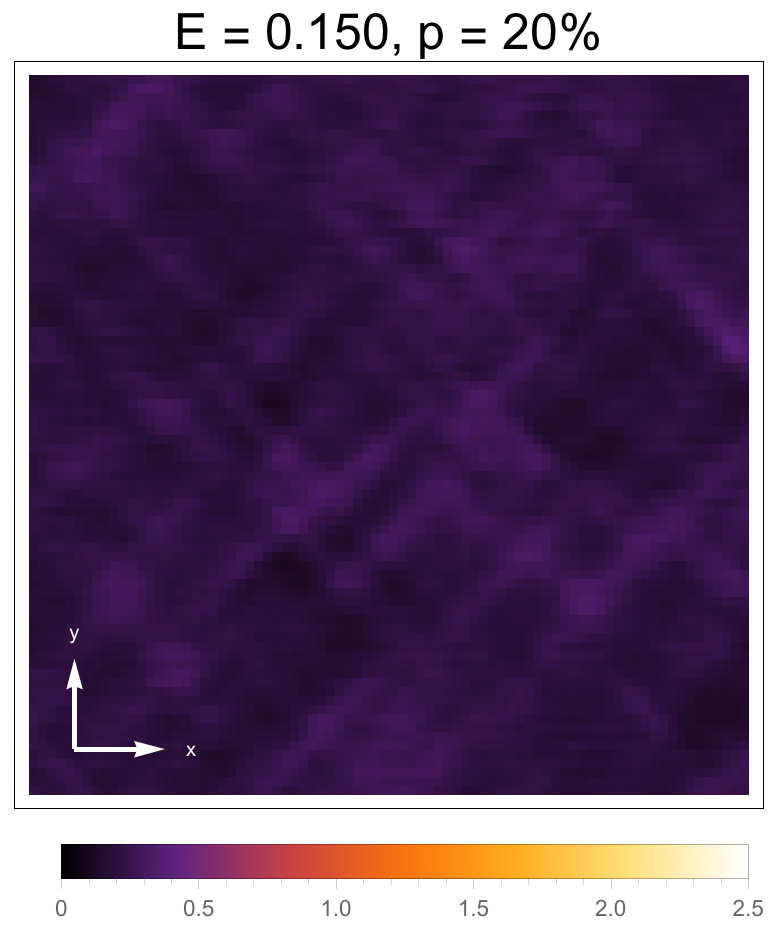}
        	\includegraphics[width=.2\textwidth]{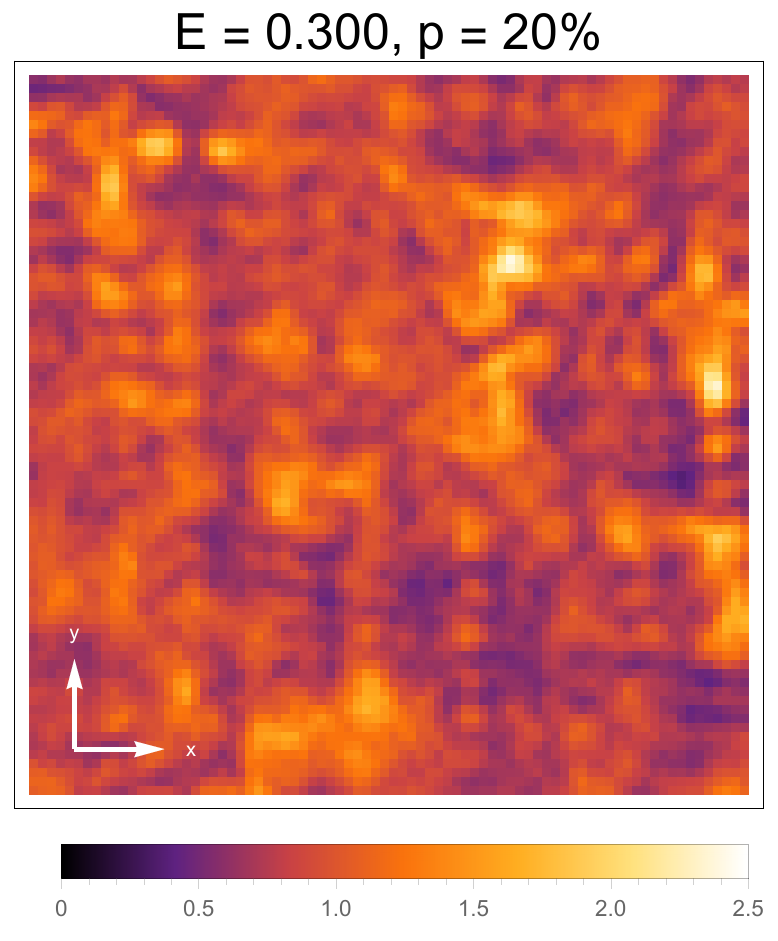} \\
        	\includegraphics[width=.2\textwidth]{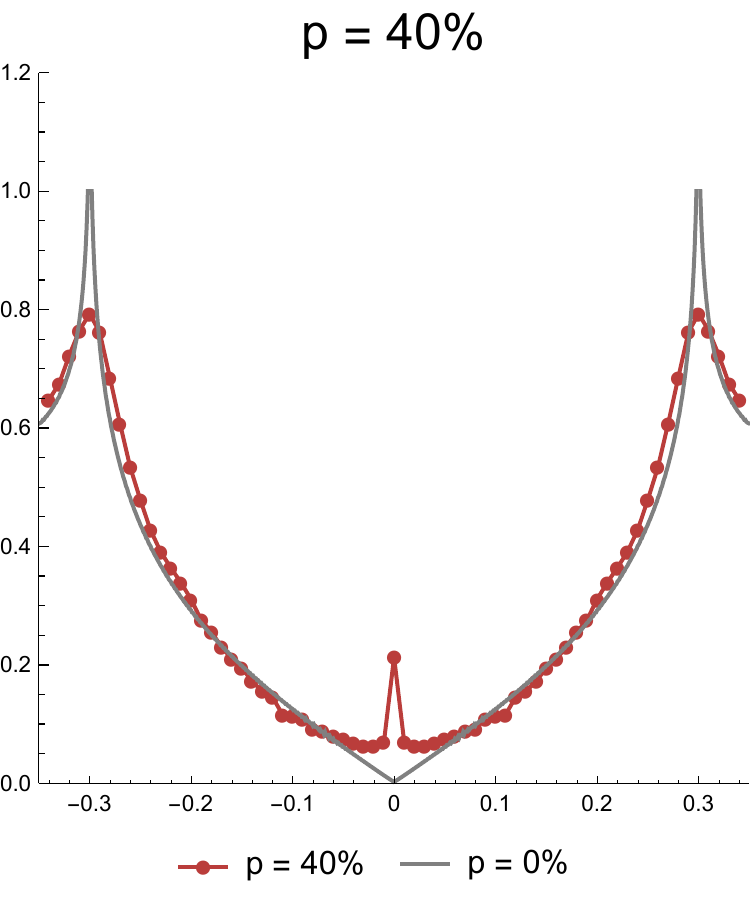}
        	\includegraphics[width=.2\textwidth]{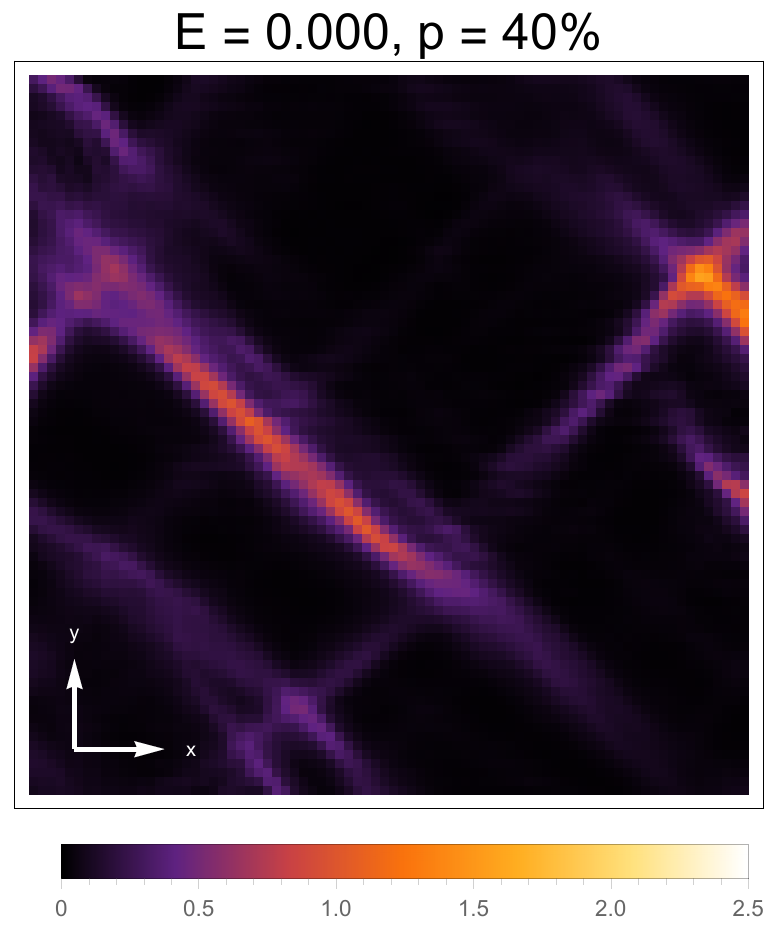}
        	\includegraphics[width=.2\textwidth]{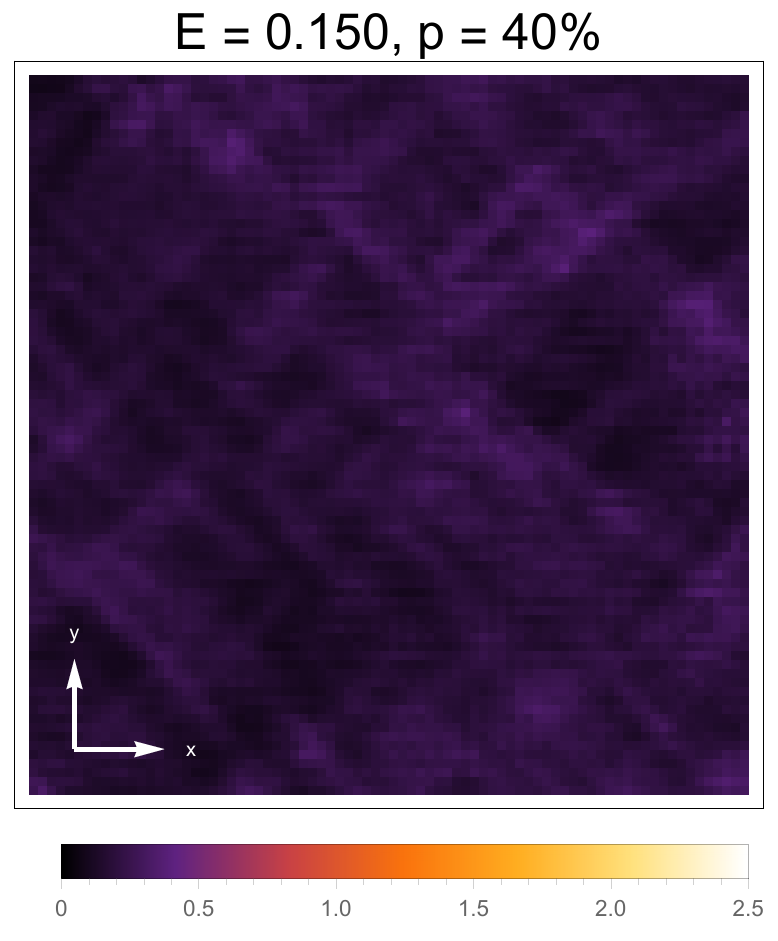}
        	\includegraphics[width=.2\textwidth]{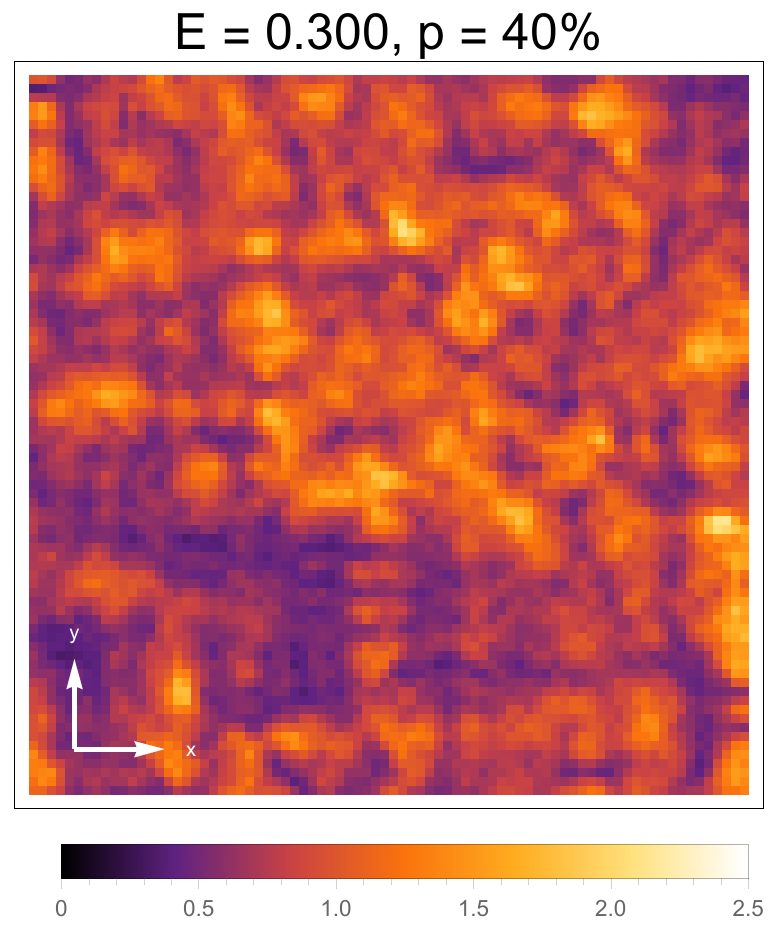}
        	\caption{Snapshots of the real-space quasiparticle density of states for smooth disorder (with \emph{zero} net potential) with increasing impurity concentration $p$ (top to bottom) and energy $E$ (left to right), extracted from the middlemost $80 \times 80$ subset of the full system. The leftmost column shows plots of the DOS as a function of energy for a particular $p$, along with plots of the clean case for comparison. The same disorder realizations as in Fig.~\ref{fig:ssn} are used here. The color scale is the same for all plots.}
        	\label{fig:ssnqpdos}
        \end{figure*}
      
         \begin{figure}[t]
         	\centering
         	\includegraphics[width=.5\textwidth]{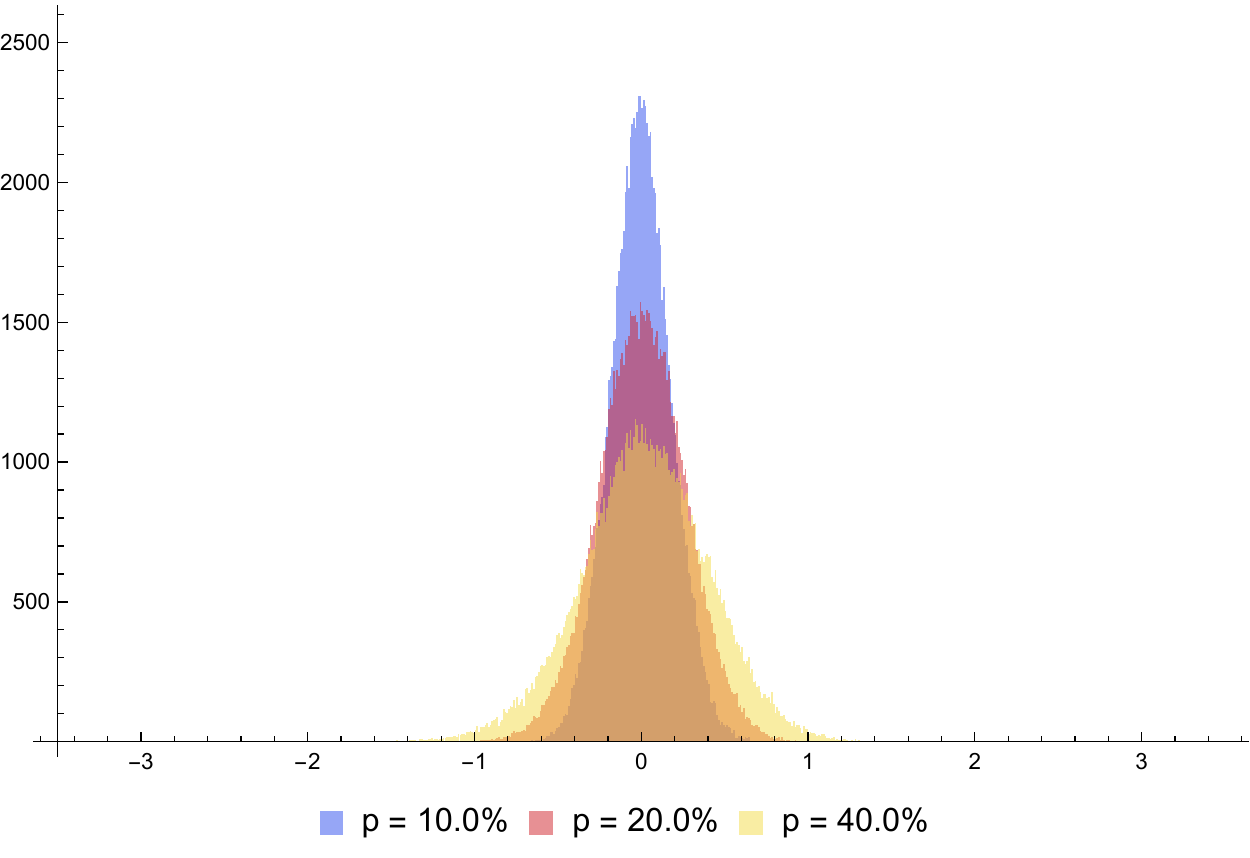}\hfill
         	\caption{Histogram of the values of the disorder potential for smooth disorder with zero net potential for three values of $p$. The width of each bin is 0.01. The mean of the disorder potential is zero, and the average chemical potential of the system as a whole is not shifted.}
         	\label{fig:SSNh}
         \end{figure}	
         
                    \begin{figure}[t]
                    	\centering
                    	\includegraphics[width=.5\textwidth]{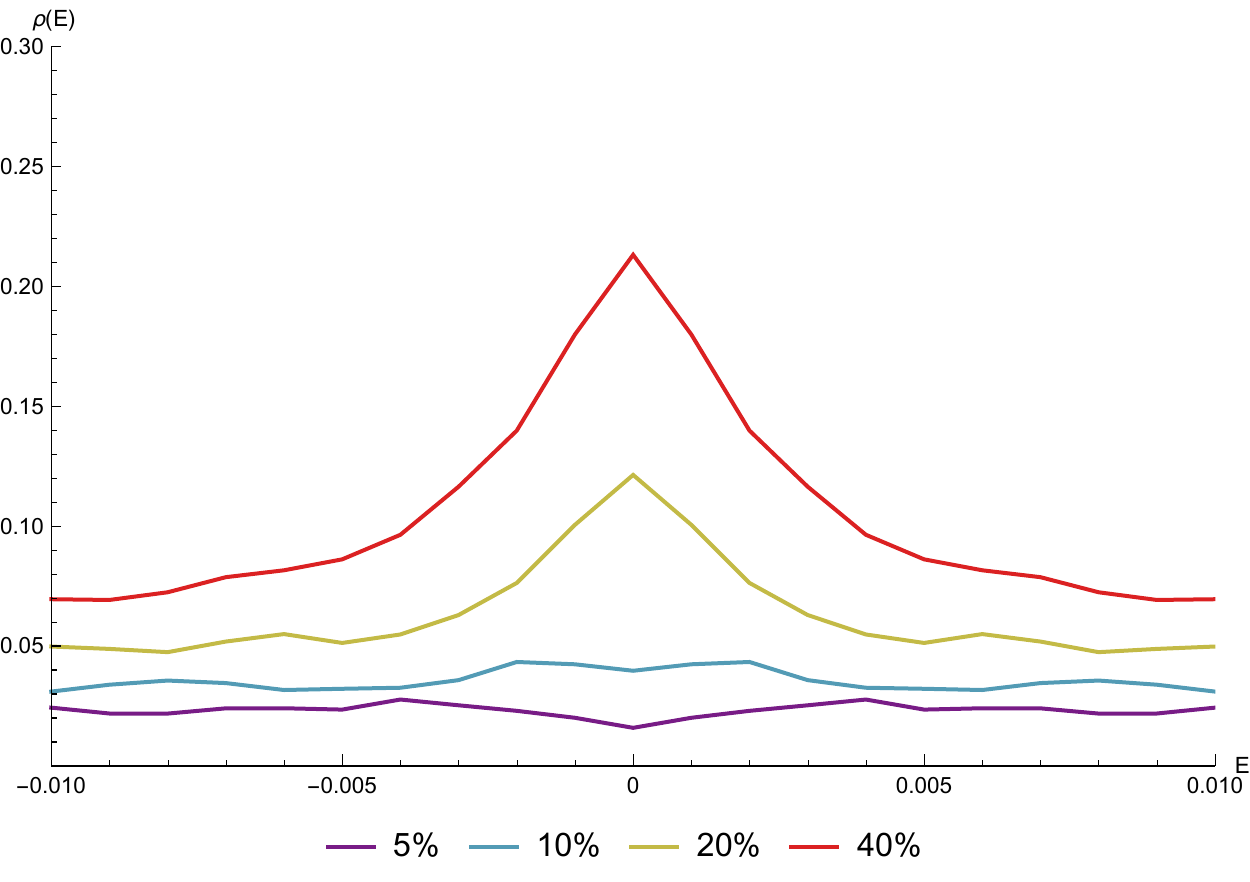}\hfill
                    	\caption{Plot of the density of states at and near $E = 0$ for the multiple-smooth-scatterer case for various impurity concentrations $p$. For the $p = 20\%$ and $p = 40\%$ the resonance is seen to have a width of approximately 0.006.}
                    	\label{fig:resonancezoom}
                    \end{figure}
           
           \begin{figure}[t]
           	\centering
           	\includegraphics[width=.5\textwidth]{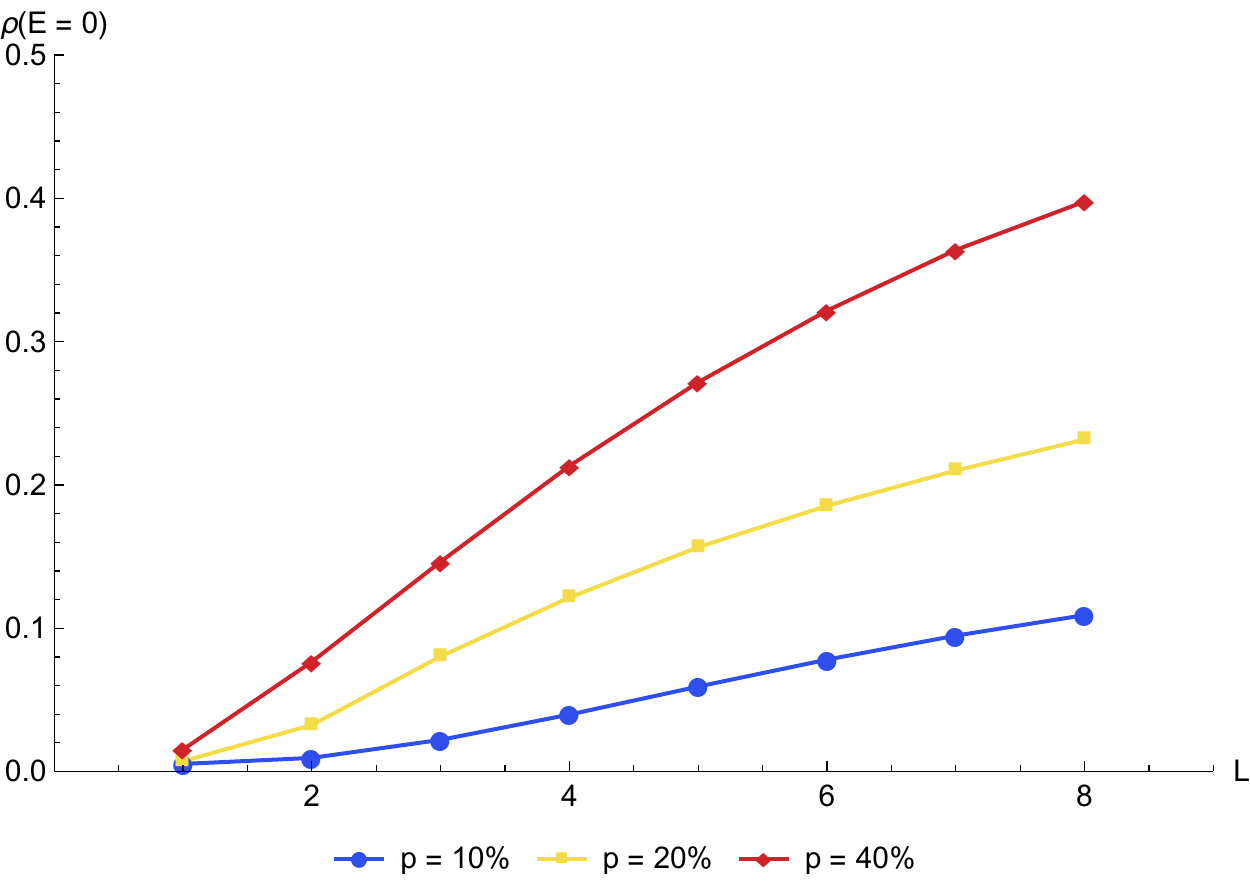}\hfill
           	\caption{Plot of the density of states at $E = 0$ for the multiple-smooth-scatterer case as a function of screening length $L$. For a given $p$, the positions of the smooth scatterers are fixed, with only the screening length and the amplitude of the disorder potential adjusted as discussed in the text. At fixed $p$ the zero-energy DOS increases monotonically with $L$. In addition, at fixed $L$ the DOS at $E = 0$ increases with increasing $p$.}
           	\label{fig:rhovsl}
           \end{figure}
           
        \begin{figure*}
        	\centering
        	\includegraphics[width=.2\textwidth]{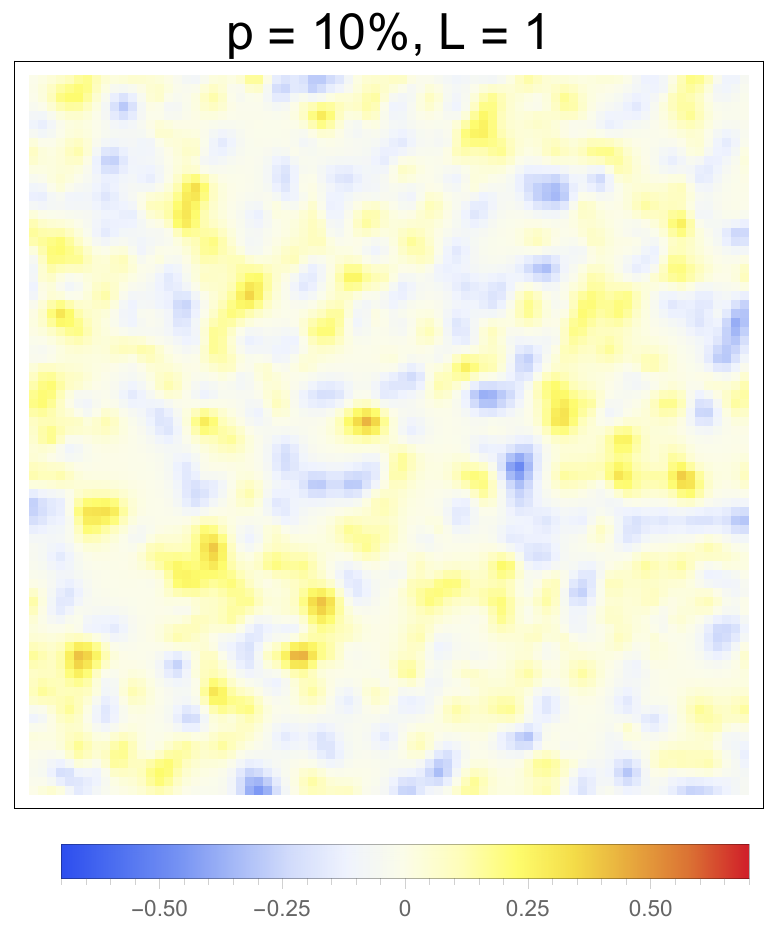}
        	\includegraphics[width=.2\textwidth]{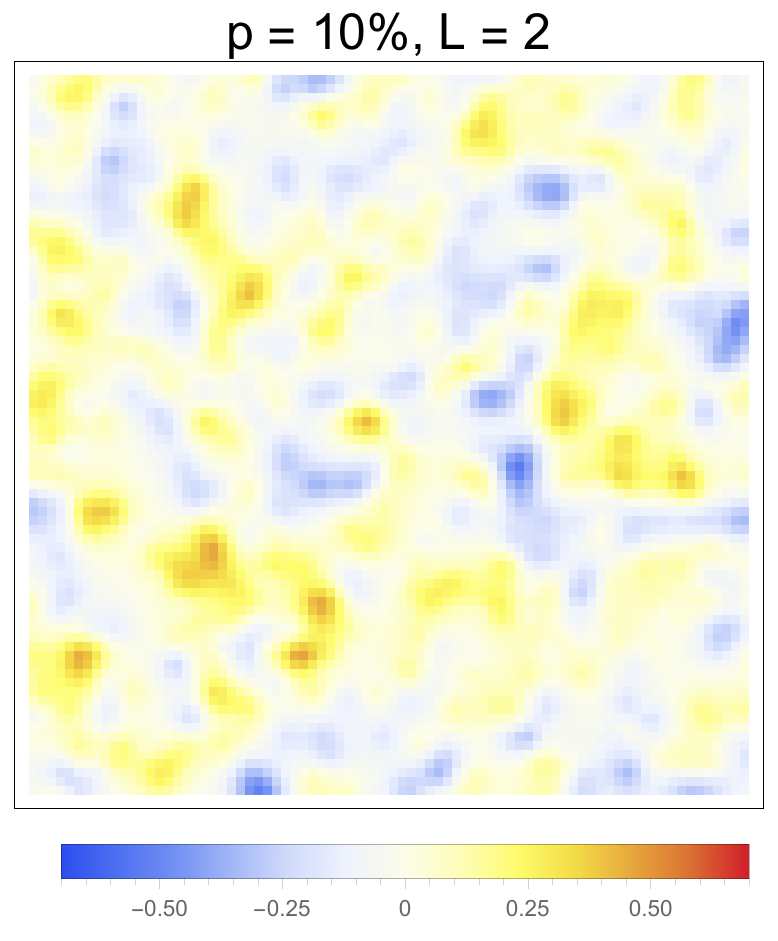}
        	\includegraphics[width=.2\textwidth]{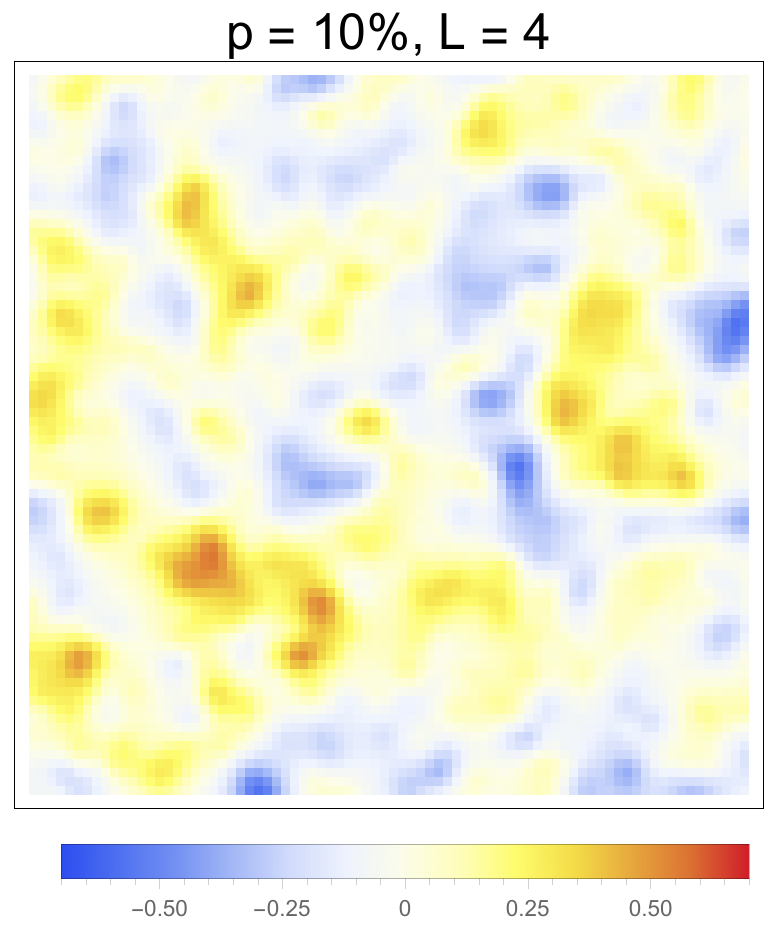}
        	\includegraphics[width=.2\textwidth]{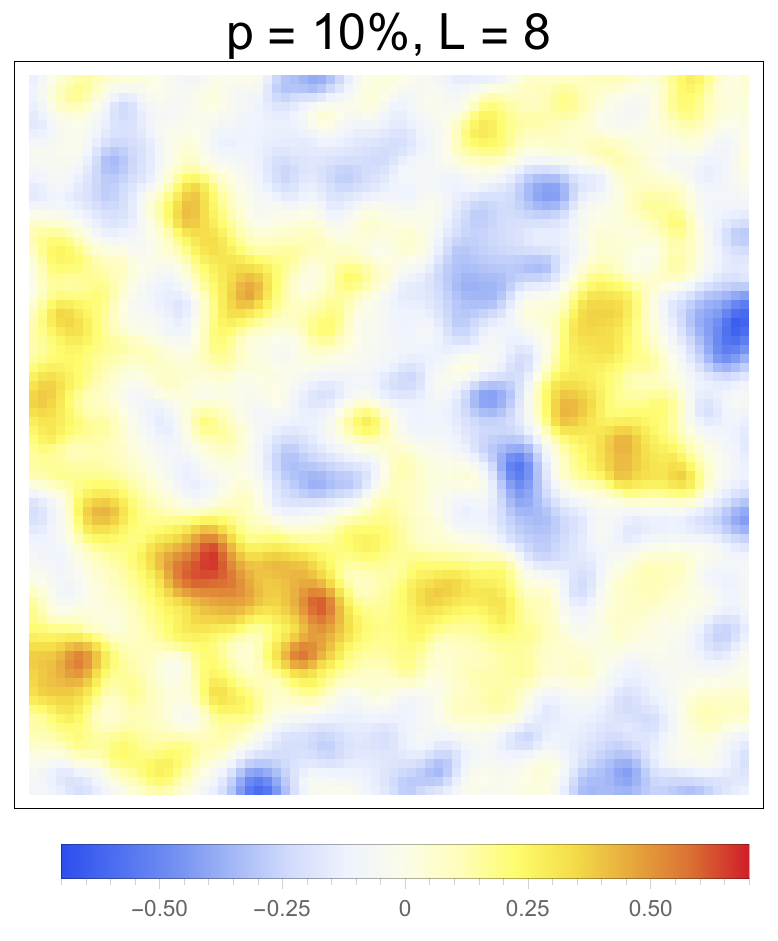} 
        	\caption{Smooth disorder potentials used in Fig.~\ref{fig:rhovsl}, with concentration $p = 10\%$, evaluated at various screening lengths $L$. Shown are $L = 1$, $L = 2$, $L = 4$, and $L = 8$. The color scale is the same for all plots. Notice that as $L$ is increased the disorder potential becomes smoother and more spatially correlated.}
        	\label{fig:smoothpotentialvsl}
        \end{figure*}
    
    We now discuss our numerical results for the quasiparticle density of states. We first focus on random-potental disorder. Fig.~\ref{fig:rp} shows the quasiparticle DOS as a function of energy for various values of $\sigma$. There are a number of interesting features in these plots that are worth mentioning. We focus first on the DOS near $E = 0$. For small values of $\sigma$ (\emph{i.e.}, $\sigma = 0.125$ and $\sigma = 0.25$), the DOS vanishes markedly at $E = 0$. For these cases the DOS scales roughly linearly with $E$ near $E = 0$. The weakest disorder distribution we consider ($\sigma = 0.125$) has a DOS curve that is concave upward between $E = 0$ and the coherence peaks. This changes for $\sigma = 0.25$, for which the DOS is almost perfectly linear from zero energy up to the coherence peaks, and from $\sigma = 0.35$ upwards the DOS curves are all \emph{concave downward}.  At $\sigma = 0.35$ and $\sigma = 0.50$, a finite DOS at $E = 0$ is generated, but despite this offset the DOS still scales approximately linearly with $E$. For higher values of $\sigma$, the DOS at the Fermi energy is still finite, but there is a very visible dip around $E = 0$ relative to nearby energies. In the strong-disorder regime, the DOS scales linearly with $E$ only within a small neighborhood of $E = 0$, then becomes dramatically concave downward as energies increase.
       
       At $E \approx 0.3$, one can see the coherence peaks becoming more rounded and decreasing in height with increasing $\sigma$.  With relatively weak disorder, the peaks retain their prominence, but as disorder becomes stronger these peaks flatten. In fact, for the strongest disorder cases we consider ($\sigma = 1.41$ and $\sigma = 2.00$) the DOS near (but not \emph{at}) $E = 0$ barely differs from the DOS at $E \approx 0.3$. For energies between $E=0$ and $E\approx0.3$, the slope of the DOS decreases with increasing $\sigma$. The overall effect of increasing disorder of this kind is to shift spectral weight away from the coherence peaks towards a broad range of low and intermediate energies, consequently filling in the $d$-wave gap.
       
       Qualitatively there are three distinct regimes that are encountered as random on-site disorder is increased. At low values of $\sigma$, the superconductor is only weakly disordered: the DOS vanishes at $E = 0$ and coherence peaks are prominent. At intermediate values of $\sigma$, a finite value of the DOS forms at the Fermi energy, but the DOS still varies linearly with $E$ over a broad energy range, and traces of the coherence peaks (now rounded and diminished in height) still remain. Finally, when $\sigma$ is large, we enter the strong-disorder regime, where the DOS is linear only within a small neighborhood of $E = 0$ and saturates very quickly to a constant value (albeit with considerable random fluctuations about that value). The DOS is suppressed at $E = 0$ relative to the value to which it eventually saturates, and in fact tends toward zero once more as disorder is increased. In this regime almost no trace of the structure of the DOS of the clean $d$-wave superconductor remains.
       
       To closely examine the origins of both the generation of a finite DOS at $E = 0$ and the smoothening of the coherence peaks, we extract real-space maps of the \emph{quasiparticle} local DOS (LDOS) for various disorder strengths and energies. We take these samples from the middle $80 \times 80$ section of the full system. These maps are shown in Fig.~\ref{fig:rpqpdos}. At $E=0$, the weak-disorder ($\sigma = 0.25$) LDOS is almost zero and is spatially featureless. When disorder is increased, regions where the LDOS is nonvanishing form even at $E = 0$. At moderate levels of disorder ($\sigma = 0.50$) these regions tend to be isolated, surrounded by a sea of vanishing DOS. These are sufficient however to produce a finite DOS when averaged over the entire system. When disorder is tuned to be strong ($\sigma = 1.00$), the LDOS map at $E = 0$ displays considerable randomness: patches where the LDOS vanishes coexist with regions where the DOS is visibly nonzero, thereby resulting in a nonzero average DOS.
       
       As energies are increased the $\sigma = 0.25$ maps start exhibiting modulations in the LDOS that arise from quasiparticle interference in the presence of weak disorder. As disorder is increased, this structure becomes less and less visible: the $\sigma = 1.00$ maps at $E = 0.150$ and $E = 0.300$ show randomness that is not much different than the maps obtained at $E = 0$. The strong-disorder maps show at higher energies similar structures as the zero-energy case, with regions where the LDOS is heavily suppressed existing alongside areas with nonzero DOS. The presence of these patches where the LDOS is almost zero at large $\sigma$ is responsible for the overall suppression of the averaged DOS relative to less disordered cases.
       
       We repeat this analysis for the unitary-scatterer disorder model. For this form of disorder we show the quasiparticle DOS as a function of energy $E$ in Fig.~\ref{fig:mp}. When a small number of impurities are present (\emph{e.g.}, $p = 0.125\%$), the DOS is barely altered from the clean case: the DOS tends toward zero at $E = 0$, increases linearly for a broad energy range, and displays sharp coherence peaks at $E \approx 0.300$. The same behavior holds for higher concentration of levels such as $p = 0.25\%$ and $p = 0.50\%$. We can see that the coherence peaks become slightly lower for these cases. 
       
       A major feature of these plots for a broad range of $p$ is the rounding off of the DOS at an energy scale that appears to be dependent on the concentration. Near $E=0$, the DOS scales linearly. As $p$ is increased, the $d$-wave gap fills in a particular manner: more spectral weight accumulates at a characteristic energy scale, so that instead of a linear DOS as in the clean case, one sees the DOS encountering a ``hump'' that becomes more pronounced when $p$ is increased. With increasing $p$ the DOS surrounding $E = 0$ starts accumulating larger values of DOS, all while the coherence peaks become shorter and flatten, showing a transfer of spectral weight from the coherence peaks towards the region around the Fermi energy. It is interesting to note that the way the gap is filled is different for the case of unitary scatterers than for random on-site disorder: for small $p$, spectral weight is moved from the coherence peaks towards the neighborhood of the Fermi energy, with a width roughly set by the impurity concentration, whereas for random Gaussian disorder the spectral weight is transfered to a far broader range of energies, with strong deviations from the clean case occuring even at energies away from $E = 0$. For higher values of $p$, the DOS resembles the large-$\sigma$ random-disorder cases discussed earlier. One feature that is consistently present---even at high values of $p$, with coherence peaks completely flattened and the DOS near the Fermi energy finite---is a visible dip \emph{at} $E = 0$. 
       
       Real-space maps of the LDOS for a $d$-wave superconductor subject to a variety of unitary-impurity concentrations are shown in Fig.~\ref{fig:mpqpdos}. At $p = 1.0\%$, the $E = 0$ LDOS map is largely almost zero, save for small areas that show large, nonzero values of the LDOS. A closer examination shows that these arise from interference effects from the presence of a few impurities bunched up together within a small area, arranged together such that a resonance forms. These resonances are very rare---in the $80\times80$ map we take, only one particular group of closely-spaced impurities generates such nonzero LDOS values at $E = 0$, whereas groups of a few impurities near one another do appear quite frequently. Despite their relative rarity, the presence of such regions with large average LDOS is enough to produce a small but nonzero average DOS for the entire sample. When the concentration is increased, we see behavior in the $E = 0$ maps that is strongly reminiscent of that seen in the maps from the Gaussian random disorder case. At $p = 4.0\%$, regions where the LDOS is nonzero appear more frequently, but they are isolated and are largely surrounded by areas where the LDOS is suppressed. The $p = 16.0\%$ case shows a remarkably large number of lattice sites with large values of the LDOS. Clearly in this case the large impurity concentration means that there is a large probability that an impurity is placed in close proximity to another impurity, resulting in a nonzero LDOS.
       
       At higher energies the $p = 1.0\%$ and $p = 4.0\%$ cases show modulations that are due to quasiparticle scattering interference (QPI) from multiple impurities. In particular the $p = 1.0\%$ map at $E = 0.300$ shows strikingly prominent modulations in the LDOS due to the presence of disorder; the $p = 4.0\%$ map at the same energy also shows visible modulations, but the larger number of impurities results in an average DOS that is lower than the $p = 1.0\%$ case. The $p = 16.0\%$ case, on the other hand, shows almost no visible traces of patterns arising from QPI. Instead what one sees is a very inhomogeneous map featuring both sites with very strong suppression of the LDOS and sites at which the LDOS is large. For this particular concentration, the degree of inhomogeneity does not change markedly upon increasing $E$.
       
       The suppression of the DOS at $E = 0$ for both random-potential and unitary-scatterer disorder has been discussed at length by Senthil and Fisher with field-theoretic methods\cite{senthil1999quasiparticle} and by Yashenkin \emph{et al.} using diagrammatic $T$-matrix techniques.\cite{yashenkin2001nesting} This suppression---found to be logarithmic in both approaches---can understood as being due to the inclusion of diffusive modes that, in the absence of symmetries other than spin rotation invariance, lead to an overall suppression of the DOS. Yashenkin \emph{et al.} also find that the addition of artificial nesting symmetries (\emph{e.g.}, a particle-hole-symmetric normal-state band structure in the presence of unitary scatterers ) can lead rise to additional diffusive modes that \emph{enhance} the DOS at the Fermi energy. It is interesting to note that even in strong-disorder regimes where these approximations do not hold---diagrammatic and field-theoretical treatments both implicitly rely on a relatively narrow distribution of disorder for them to be sensible---this logarithmic suppression at the Fermi energy is still very much evident for both random-potential and unitary-scatterer disorder.
       
       We finally discuss the case of smooth disorder. We first focus on the case where the dopants have the same sign of the impurity strength---\emph{i.e.}, the full potential is given by Eq.~\ref{eq:vs}. Fig.~\ref{fig:ss} shows the quasiparticle DOS for a $d$-wave superconductor with such disorder, for various doping concentrations $p$. The behavior of the DOS near $E = 0$ has a number of interesting features when $p$ is increased. First, at low $p$, the DOS is close to zero. As $p$ is increased, the DOS gradually acquires a finite value, and at higher concentrations ($p = 20\%$ and $p = 40\%$) the DOS has a small bump at $E = 0$ relative to the value of the clean DOS. The neighborhood of the Fermi energy shows a gradual roundening of the DOS from a sharp V-shape in the clean and mildly disordered cases to a smooth U-shape for higher impurity concentrations. For all $p$, coherence peaks are present and quite prominent, but these shorten and move towards the Fermi energy as $p$ is increased. This can be attributed to the fact that for this particular form of disorder, the mean of the disorder potential is nonzero, and the chemical potential is shifted away---only slightly for lower $p$, and considerably more strongly for larger and larger $p$, as seen in Fig.~\ref{fig:SSh}. It is interesting to note that despite the fact that this form of potential seemingly represents a strong modification to the $d$-wave superconductor, the effect is mainly to transfer spectral weight from the coherence peaks to the Fermi energy, with a corresponding rounding of the DOS, without impacting the DOS that much in the intermediate-energy regimes. There is also no visible suppression at $E = 0$, as was the case in the pointlike disorder models we discussed earlier. It seems that the overall effect of this particular form of disorder, at least as the quasiparticle DOS is concerned, is qualitatively \emph{much} weaker than the random Gaussian on-site energy and the multiple unitary-scatterer models at roughly similar disorder widths or impurity concentrations.
       
      Real-space plots are shown in Fig.~\ref{fig:ssqpdos}. The plots at $E = 0$ show how a nonzero DOS is generated in the neighborhood of the Fermi energy. At $p = 10\%$, the effect is only mild, as the LDOS is almost spatially uniform. With increasing  concentration visible patterns start to show up in the LDOS maps. These patterns are interesting because they correspond to only a small portion of the entire system, but do generate, upon averaging over space, an overall nonzero DOS centered around $E = 0$. Unlike similar maps for the pointlike disorder cases, the patterns---which manifest themselves as streaks of nonzero DOS amid a featureless, almost-zero background---display a smoothness that is not present in the highly disordered pointlike cases. While displaying patchiness, it exhibits spatial variations that are much more ragged than in the smooth case. Meanwhile the maps taken at higher energies show crisscrossing patterns which arise naturally from quasiparticle intereference due to scattering off of a highly random smooth disorder potential. Unlike the maps showing pointlike disorder, the modulations here are much smoother, owing to the fact that these arise from small-momenta scattering processes. 
       
       We next turn to the case where there is an equal number of positive- and negative-strength dopants---\emph{i.e.}, the disorder potential shown in Eq.~\ref{eq:vz}. This will prove to be a much more interesting case than the smooth-disorder scenario we had just discussed. We show plots of the DOS for this disorder potential in Fig.~\ref{fig:ssn}. A number of remarkable features are present in these plots which we will now discuss in detail. We focus first on the region around $E = 0$. At low $p$, the DOS vanishes, but at $p = 10\%$ the DOS acquires a value that is appreciably larger than that of the clean or low-doping cases. At this doping the DOS at $E = 0$ has a slight upward hump, and the DOS surrounding the Fermi energy has a U-shape and is considerably rounded off compared to the shape of the clean DOS. At higher dopings, a very prominent spike in the DOS at $E = 0$ start to form: this spike is localized at $E = 0$, and falls off quickly towards the base of a ``valley.'' It can be seen that the area around the Fermi energy hosts a considerable amount of spectral weight relative to the clean case as $p$ is increased.
       
       These effects near the Fermi energy are far more pronounced because elsewhere there are no significant deviations from the clean DOS. Even for very large dopings (\emph{e.g.}, $p = 40\%$), the DOS at intermediate and high energies are almost unchanged from that of the clean case. The main significant change at these energy ranges happens at the coherence peaks ($E \approx 0.3$), which become shorter and more rounded with increasing disorder. However the rounding and shortening are nowhere near as pronounced or as strong as those in the random-potential or unitary-scatterer cases. Recall that in these other cases, the coherence peaks are destroyed at some level of disorder ($\sigma \approx 0.5$ for random potential disorder, and $p \approx 8\%$ for unitary scatterers). However, even at $p = 40\%$ doping, smooth disorder preserves coherence peaks. More emphatically, the global structure of the $d$-wave DOS is preserved even for very large dopings. 
       
       This is remarkable given how randomly distributed the disorder potential is. This can be seen in histograms of the disorder potential values for this particular form of smooth disorder, which we show in Fig.~\ref{fig:SSNh}. One can see that they are almost normally distributed, with widths not far off from the weaker incarnations of the random-potential case we discussed earlier. The difference of course lies in the presence of spatial correlations in the smooth disorder potential, which are completely absent for pointlike disorder. Evidently, unlike random-potential or unitary-scatterer disorder, which show dramatic spectral-weight transfers from the coherence peaks to a broad range of energies, for this particular form of smooth disorder only moderate spectral weight transfer occurs, with the bulk accumulating near the Fermi energy and almost none in intermediate-energy regimes.
       
       The $E = 0$ maps in Fig.~\ref{fig:ssnqpdos} show how a spike in the average DOS is generated. At low $p$, few if any streaks are visible, and these faint streaks occur against a background where the LDOS is heavily suppressed. As $p$ increases, more of these streaks are visible, and in the $p = 40\%$ case these streaks are strong enough that averaging over the LDOS yields a finite value. The $E = 0.150$ maps show, as in the other smooth-disorder case we studied, diagonal crisscrossing patterns that can be attributed to quasiparticle scattering interference. Note that the modulations in real space are slowly varying, which as before can be attributed to the fact that, in this disorder scenario, nearly all scattering is forward. The fact that mostly diagonal streaks can be seen is due to the fact that scattering occurs heavily within one node only, and the only $\mathbf{q}$-vector corresponding to such intranodal scattering is $\mathbf{q}_7$, which is diagonal and small. At the coherence-peak energies ($E = 0.300$), the diagonal streaks are now mainly replaced by moduations in the vertical and horizontal directions---a reflection of the fact that these LDOS maps are still heavily determined by quasiparticle scattering interference. At this energy regime the vertical/horizontal momentum $\mathbf{q}_1$ becomes most dominant, leading to the prominent modulations in the horizontal and vertical directions. The maps at higher energies show a remarkable degree of similarity with each other, despite vastly different amounts of doping, indicating that the transfer of spectral weight away from these energies is largely muted. This is very different from what we have seen for random-potential or unitary-scatterer disorder.
       
       The origin of the sharply enhanced DOS at $E = 0$ is unknown, but we will try to characterize this effect as fully as possible numerically. First of all, the resonances are sharply located at $E = 0$, and are very narrow. Fig.~\ref{fig:resonancezoom} shows a close-up view of the DOS within a small window of the Fermi energy. We find that the resonances, which are most visible at $p = 20\%$ and $p = 40\%$, have a width of $\Delta E \approx 0.006$ centered about $E = 0$, and that these subsequently plateau into a flat profile a short distance away from the Fermi energy. From our numerical results it appears that these zero-energy resonances are uncorrelated with the underlying smooth disorder potential. It is an intrinsically many-impurity effect, since results from single-impurity simulations do not show a sharp spike in the local DOS at zero energy. It also depends rather sensitively on the length scales associated with the smooth disorder potential. In Fig.~\ref{fig:rhovsl} we plot the DOS at $E = 0$ versus the screening length $L$ for three different impurity concentrations $p$, keeping the positions of the impurities at a given $p$ fixed. In these plots we change $V_0$ as $L$ is varied in Eq.~\ref{eq:vn} so that $V(\mathbf{r = 0})$ remains the same for all values of $L$ we consider. This choice ensures that the resulting smooth disorder potentials feature the same degree of spatial variations, even as $L$ is varied. As we have seen in the $L = 4$ case heavily discussed earlier, at fixed $L$ the $E = 0$ DOS depends on $p$, with the DOS increasing as $p$ is increased.  More remarkably, however, we can see that at fixed $p$, the zero-energy DOS increases monotonically as $L$ is increased. This is interesting because at face value the smooth-disorder potentials at various $L$ appear to be very similar to each other. This is seen in Fig.~\ref{fig:smoothpotentialvsl}, which shows the different smooth disorder potentials used at fixed $p = 10\%$. These are similar in appearance, but evidently lead to considerable differences in the values of $\rho(E = 0)$. This suggests that the range of the potential plays an important role in the emergence of these resonances at zero energy.
       
       We note that a mechanism for the \emph{enhancement} of the DOS at $E = 0$ was discussed by Yashenkin \emph{et al.}, who point out that diffusion modes due to additional symmetries could lead to an increase in the DOS at $E = 0$. \cite{yashenkin2001nesting} It is not clear at all if this mechanism has any relation with the real-space streaks which generate the spike at the Fermi energy in our numerics. It was argued that symmetries such as particle-hole symmetry in the normal state lead to this enhancement; however, the normal-state band structure we use does not have any special symmetries, so this cannot explain this phenomenon. It should be noted too that Yashenkin \emph{et al.}'s analysis relies on pointlike scatterers treated within a self-consistent $T$-matrix approximation, which does not describe the smooth disorder potentials which generate the enhanced DOS at $E = 0$. It is thus an interesting, if possibly very difficult, problem to apply the analysis of Yashenkin \emph{et al.} to smooth impurity potentials. Treating smooth disorder analytically is a formidable challenge, unlike random-potential and unitary-scatterer disorder, and tractability is generally possible only in the nodal approximation, at which the Born or $T$-matrix approximations can be used. We will thus leave an explanation of these strong zero-energy enhancements of the DOS due to smooth disorder as an open problem.
       
       \section{Correlation Between the Local Density of States and the Disorder Potential} 
       
         \begin{figure*}
         	\centering
         	\includegraphics[width=.3\textwidth]{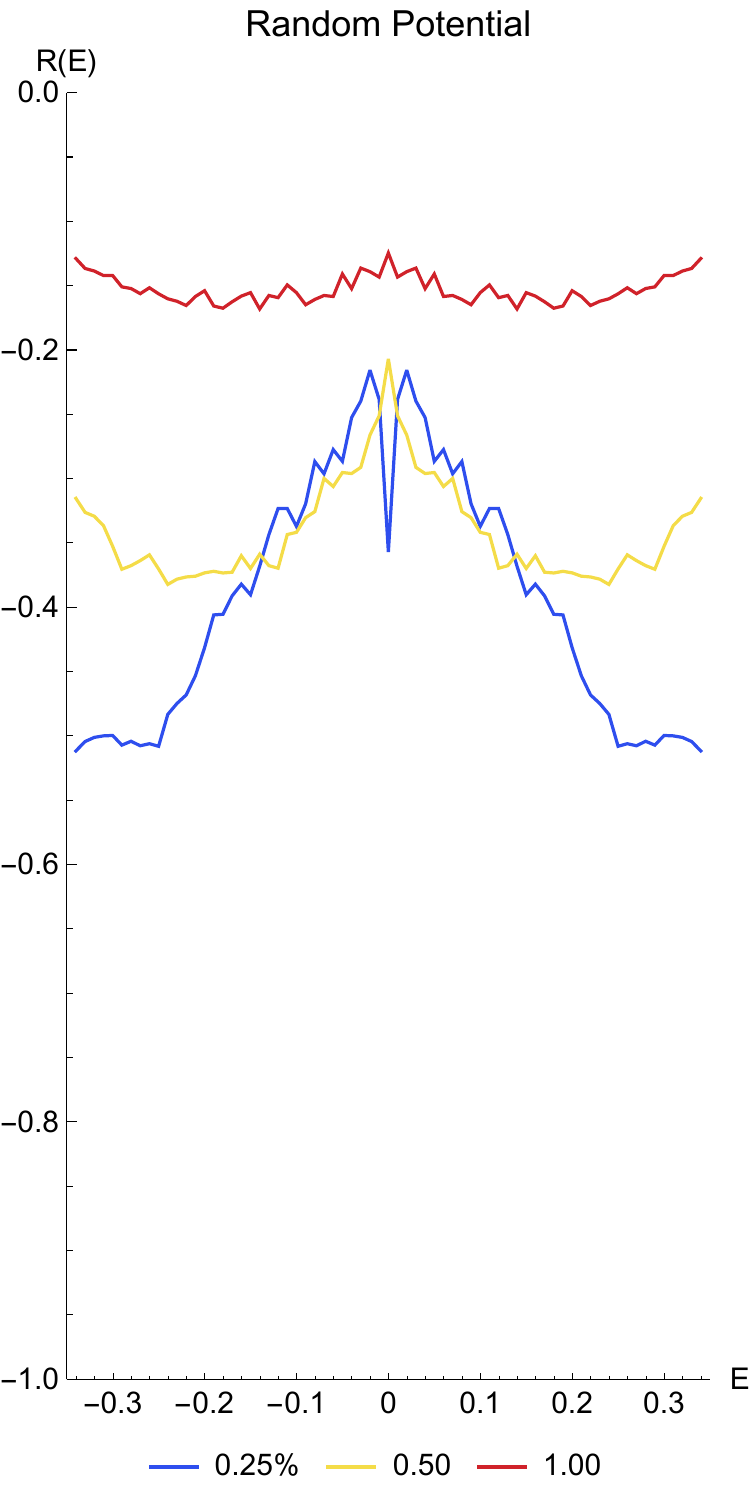}
         	\includegraphics[width=.3\textwidth]{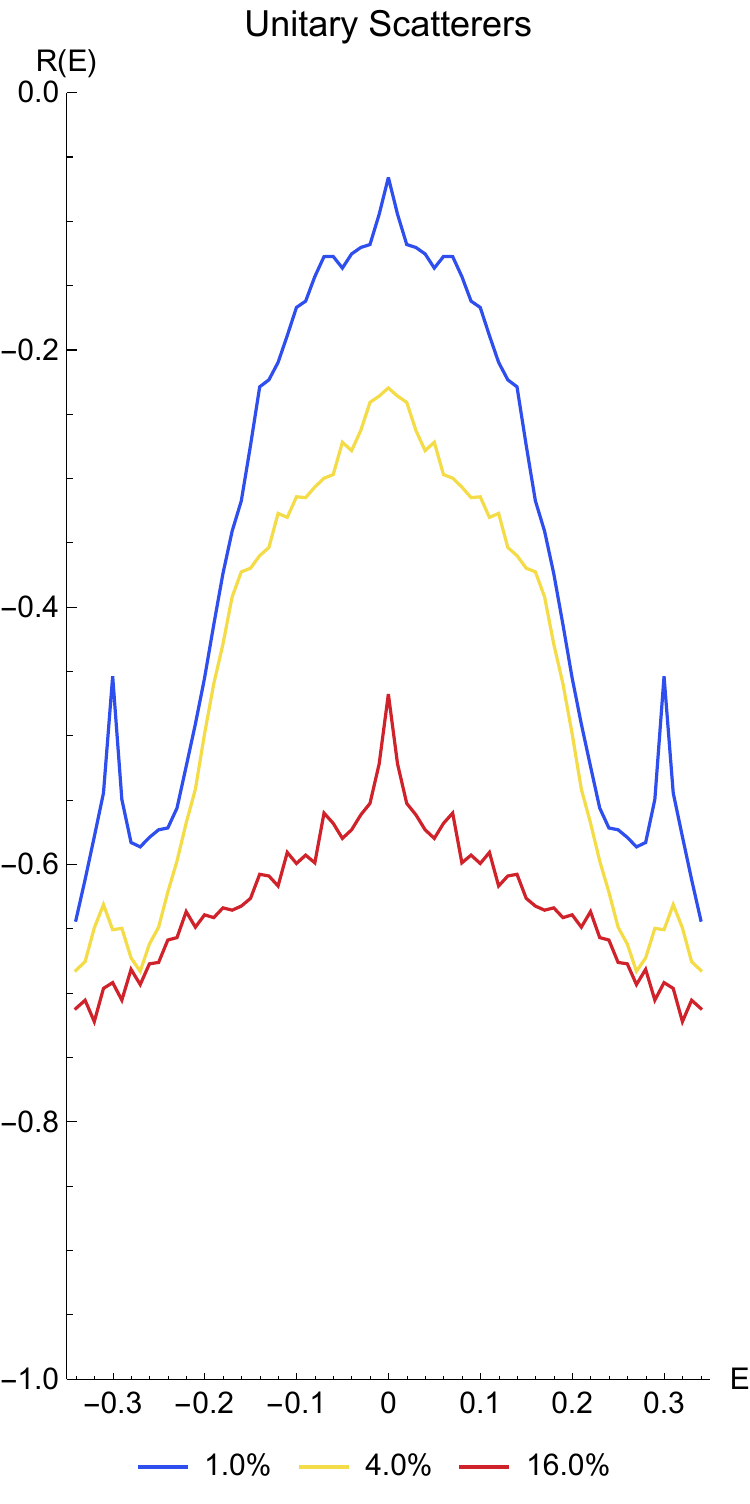}
         	\includegraphics[width=.3\textwidth]{"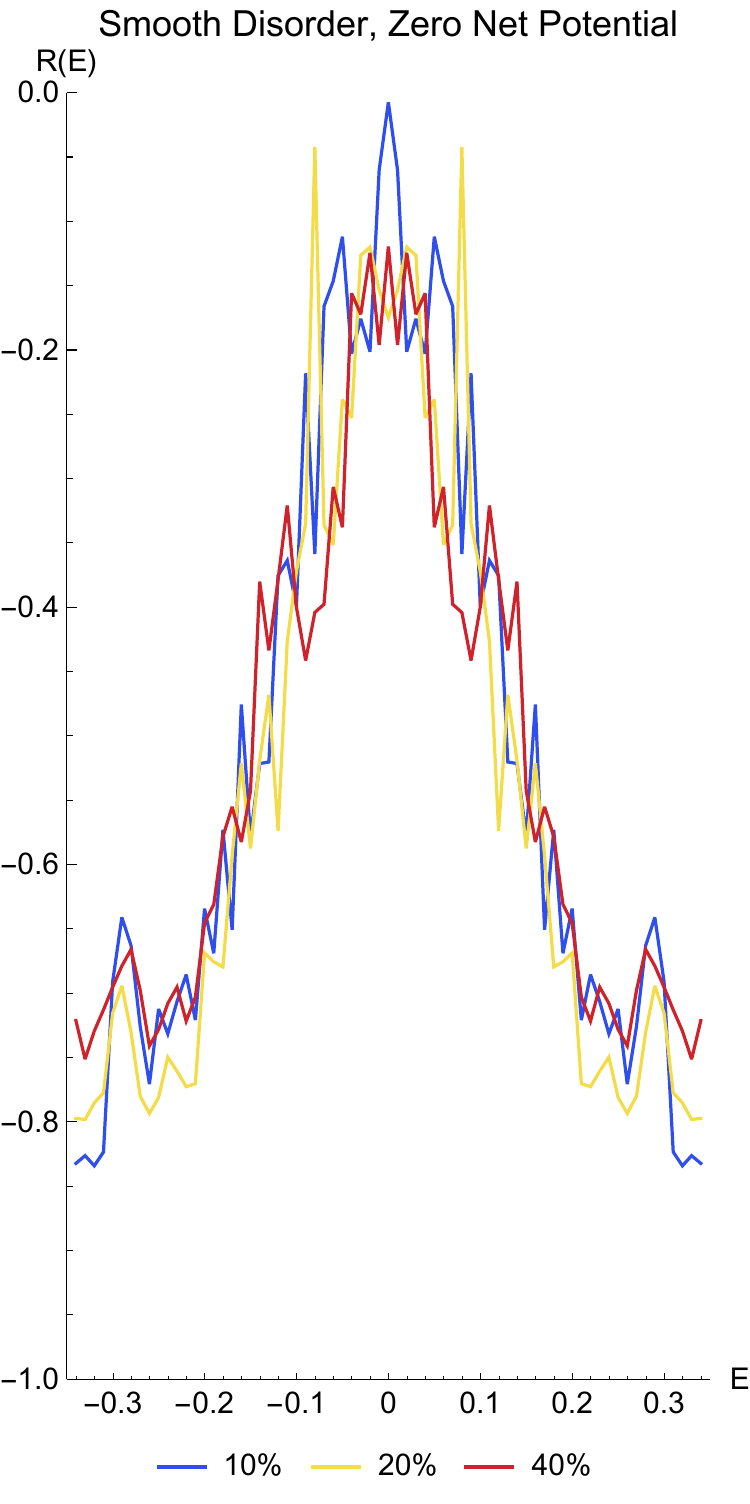}
         	\caption{Plot of the correlation coefficient $R$ between the local density of states in the middlemost $80\times80$ patch of the system and the disorder potential in that region for different types of disorder, for varying disorder strength, as a function of energy. For all three plots the correlation coefficient is negative---that is, there is an overall \emph{anticorrelation} between the LDOS and the disorder potential.}
         	\label{fig:corr}
         \end{figure*}
       
       As discussed earlier, the behavior of the real-space LDOS varies as the amount of disorder is increased, with low-disorder cases exhibiting more visible modulations in the LDOS that are due to QPI. At high energies these modulations follow closely the details of the disorder potential. As disorder is increased, these modulations become less prominent. We can get some insight into how ``strong'' the disorder in the system is by computing the coefficient of correlation $R(E)$ between the local density of states at energy $E$ and the disorder potential. $R(E)$ is defined in the following manner:
       \begin{equation}
       R(E) = \frac{\sum_{ij} (V(i, j) - \overline{V})(\rho(i, j, E) - \overline{\rho(E)})}{\sqrt{(\sum_{ij} (V(i, j) - \overline{V})^2)(\sum_{ij} (\rho(i, j,E) - \overline{\rho(E)})^2)}}.
       \label{eq:rdefinition}
       \end{equation}
		Here $V(i,j)$ is the disorder potential at site$(i,j)$, $\rho(i,j,E)$ is the quasiparticle DOS at site $(i,j)$ and energy $E$, and  $\overline{V}$ and $\overline{\rho(E)}$ are the average values of the disorder potential and the DOS, respectively, over the area where we perform the calculation. We compute $R$ between the middlemost $80 \times 80$ LDOS patch of the system at energy $E$  and the disorder potential in that same patch of the system. Plots of $R(E)$ are shown in Fig.~\ref{fig:corr}. This is motivated by a similar analysis performed by McElroy \emph{et al.} on experimentally-obtained LDOS data from BSCCO; they find that there is moderate \emph{anticorrelation} between the locations of the dopant defects and LDOS minima.\cite{mcelroy2005atomic} Our analysis differs from theirs in that we know the details of the disorder potential directly, and the cross-correlation is between the potential and the LDOS, not between the impurity location and the LDOS.
       
	   In the case of random-potential disorder, what we find is that the LDOS is only moderately anticorrelated with the disorder potential, even for weak disorder. When $\sigma = 0.25$, $R$ decreases from a small value ($R \approx -0.2$) until it saturates at $R \approx -0.5$ at $E \approx 0.25$, indicating that the high-energy LDOS displays more similarity with the underlying disorder potential than the low-energy LDOS.  As $\sigma$ is increased, the LDOS and the disorder potential become even less anticorrelated. $R(E)$ at $\sigma = 0.5$ shows only a moderate degree of dependence on energy, and at $\sigma = 1.00$ $R(E)$ is almost energy-independent and has a small value, indicating that the two variables are only weakly anticorrelated.
	   
	   For multiple unitary scatterers the situation becomes markedly different. The $R(E)$ obtained for the $p = 1.0\%$ case exhibits a very visible dependence on energy. At low energies the LDOS is very weakly anticorrelated with the disorder potential, but this anticorrelation increases sharply as energy is increased, a sign that higher-energy LDOS maps match the features of the disorder potential more than the lower-energy maps do; for instance, $R \approx -0.6$ at $E \approx 0.25$. This trend is even noticeable once $E$ is increased past the $d$-wave gap edge, where it can be seen that $R$ continues to be more and more anticorrelated with increasing $E$. This behavior can be seen to a good extent in the $p = 4.0\%$ case, for which $R$ shows a similar degree of energy-dependence in the intermediate- and high-energy ranges as in the $p = 1.0\%$ case. The $p = 16.0\%$ case is interesting, as in that case $R$ is much less energy-dependent than in the cases involving lower concentrations, similar to the strong-disorder ($\sigma = 1.00$) case of the random-potential model, but the overall coefficient indicates that \emph{stronger} anticorrelation is present between the two variables. This can be explained by the fact that unitary pointlike scatterers suppress the LDOS \emph{at} the impurity sites, which contributes to the overall anticorrelation between the LDOS and the disorder potential.
	   
	   The smooth-disorder cases feature behavior that is starkly different from the random-potential or unitary-scatterer models. For one, we obtain strongly energy-dependent $R(E)$ at all concentrations we consider ($10\%$, $20\%$, and $40\%$). In addition, the behavior of $R$ does not appear to vary as $p$ is altered. At low energies, there is almost no anticorrelation between the LDOS and the disorder potential, but the anticorrelation sharply increases as $E$ is increased. $R$ reaches very large values at high energies---for instance, $R \approx -0.7$ at $E \approx 0.3$---and in these regimes the LDOS maps bear a remarkable resemblance to plots of the smooth disorder potential, with regions where the LDOS is suppressed coinciding with patches at which the disorder potential is positive and vice versa. Another interesting aspect is that strong fluctuations in $R$ exist, independently of $p$. This is unlike random-potential or unitary-scatterer disorder, for which we saw that the fluctuations in $R$ are minimal. It is important to note from these plots that the resonances in the DOS at $E = 0$ are almost completely \emph{uncorrelated} with the disorder potential---the origin of these resonance streaks at zero energy appears not to originate from local features of the disorder potential.
       
       \section{Properties of the Density of States near $E = 0$}
       
       \begin{figure*}
       	\centering
       	\includegraphics[width=.45\textwidth]{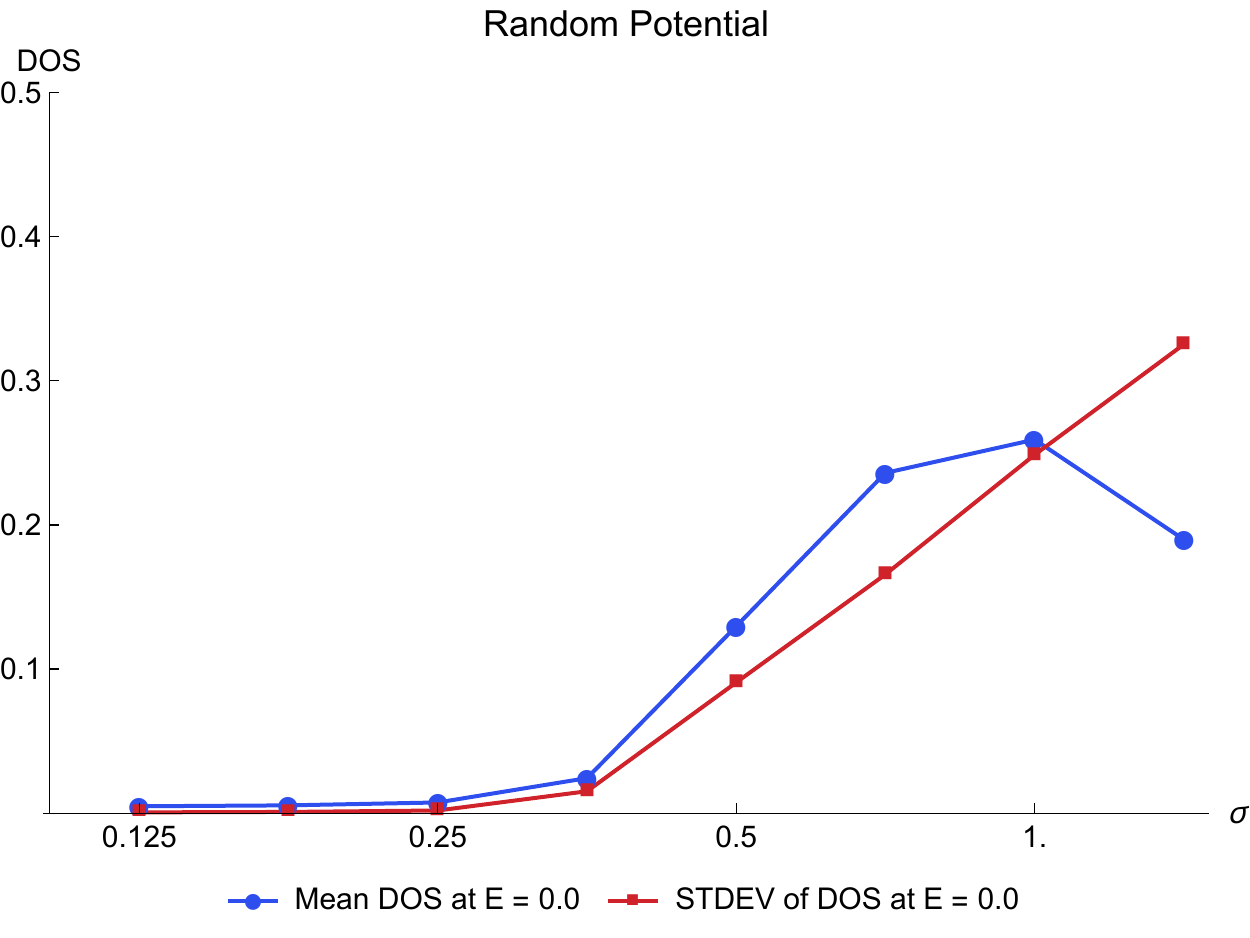}
       	\includegraphics[width=.45\textwidth]{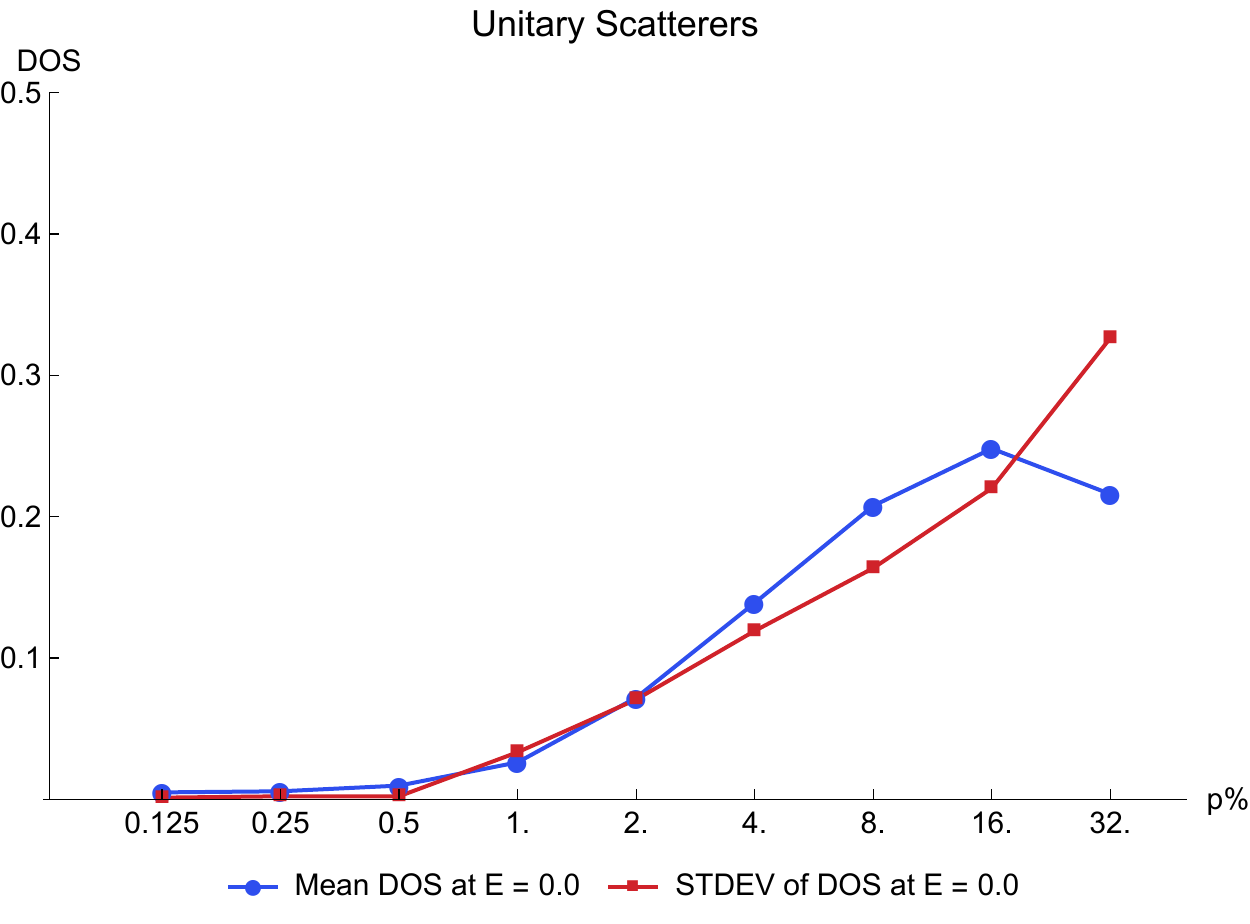}
       	\includegraphics[width=.45\textwidth]{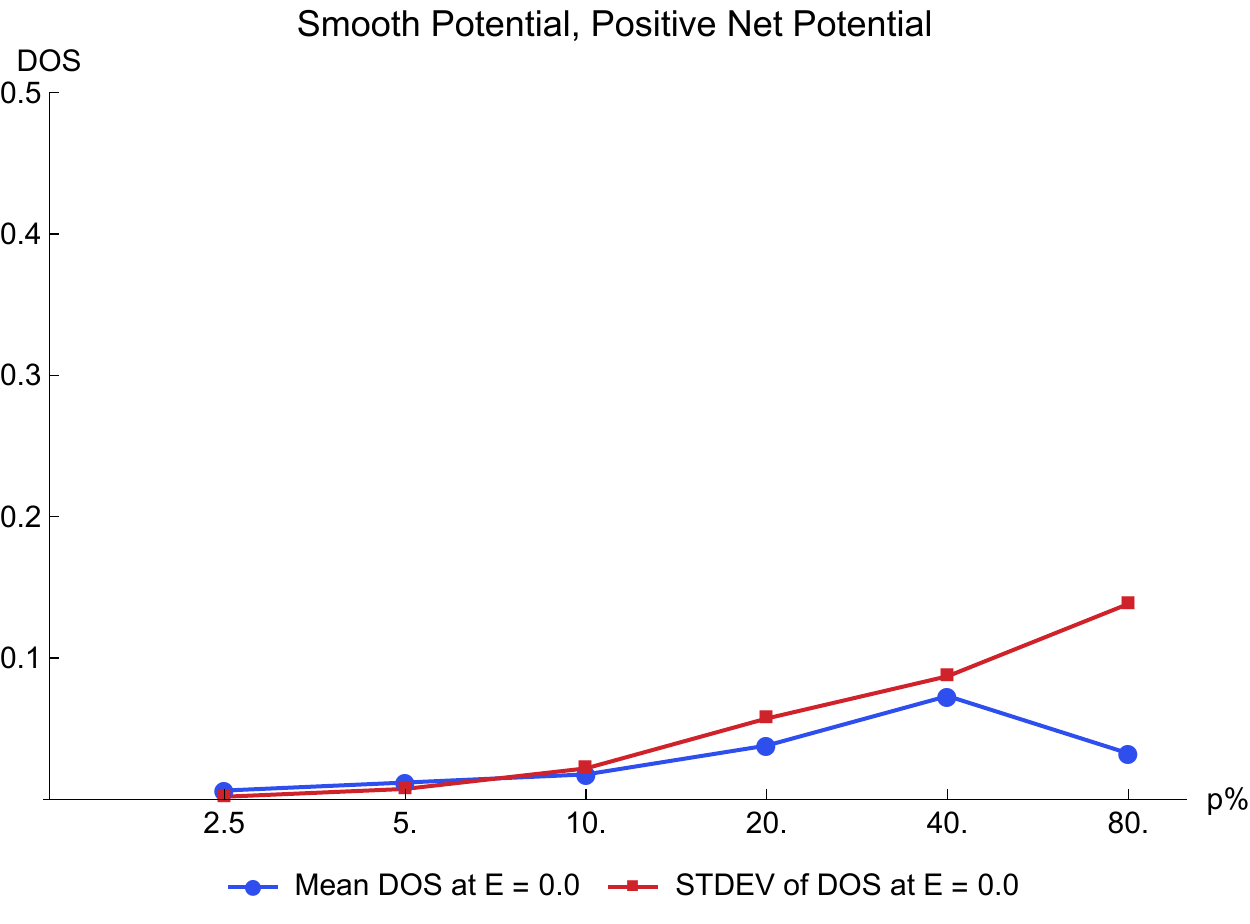}
       	\includegraphics[width=.45\textwidth]{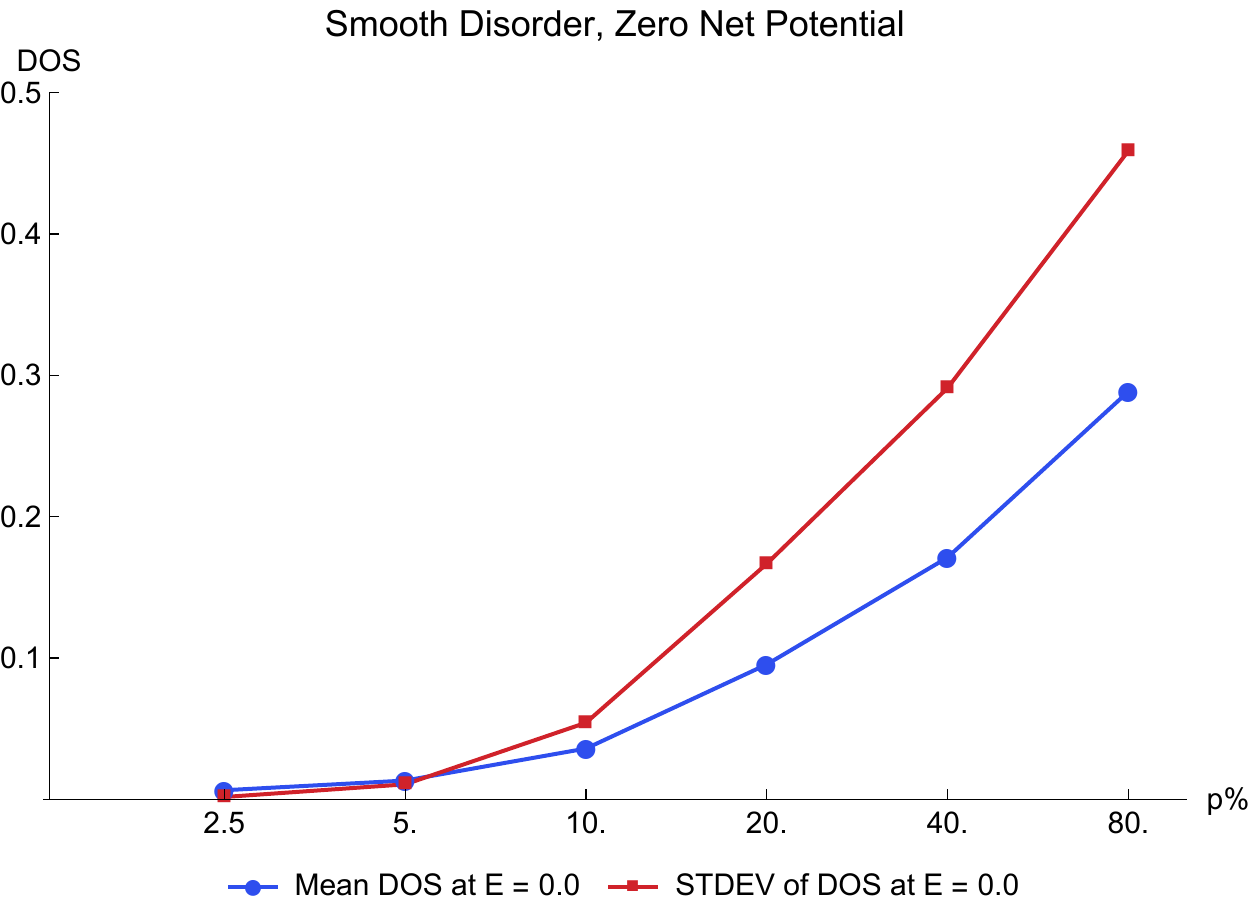}
       	\caption{Plots of the mean and standard deviation of the quasiparticle \emph{local} DOS at $E = 0$ for different types of disorder. Five realizations are utilized for each value of the disorder strength parameter for each type of disorder; an average over $5\times10^5$ values of the local DOS is taken.}
       	\label{fig:fecp}
       \end{figure*}
       
       \begin{figure*}
       	\centering
       	\includegraphics[width=.45\textwidth]{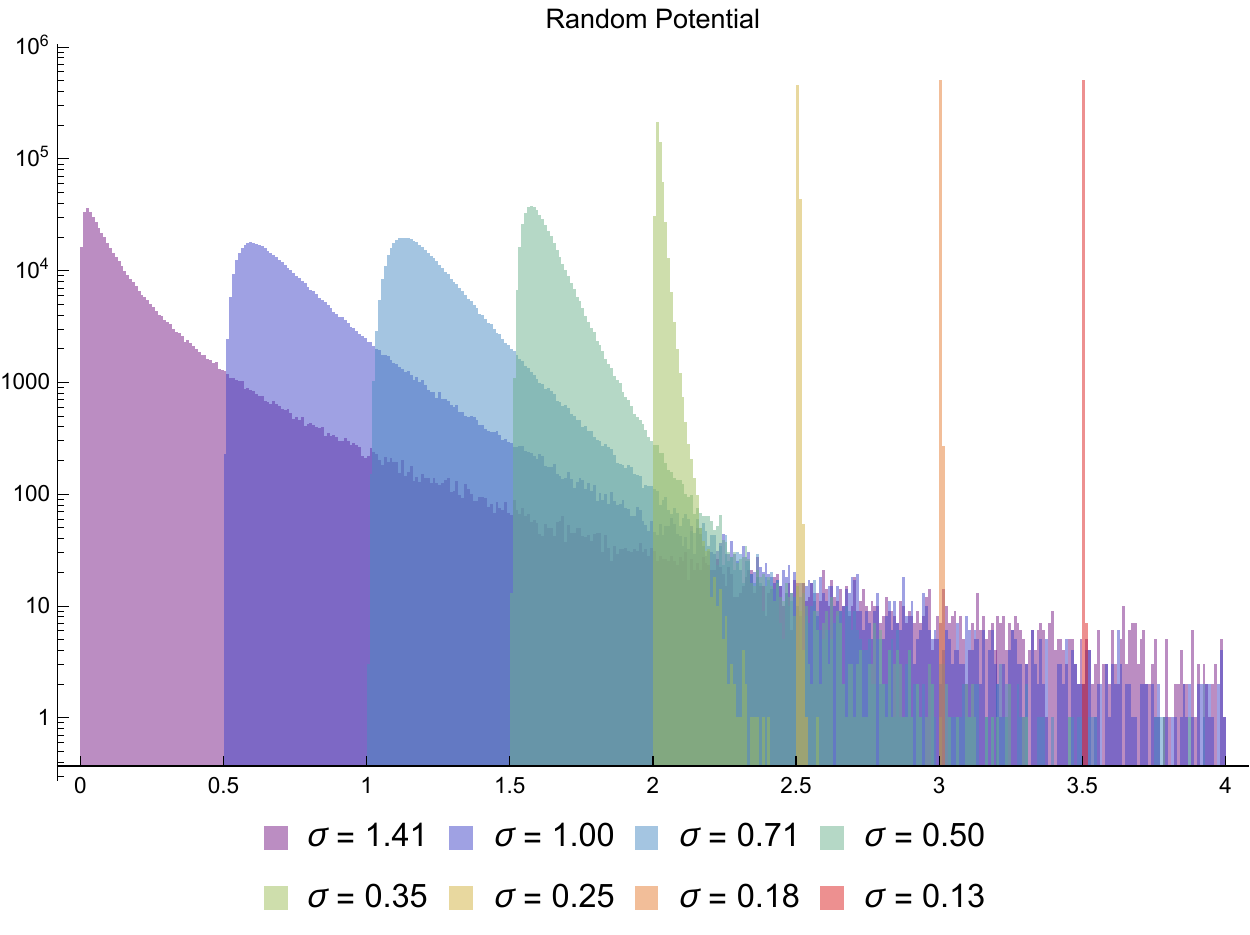}
       	\includegraphics[width=.45\textwidth]{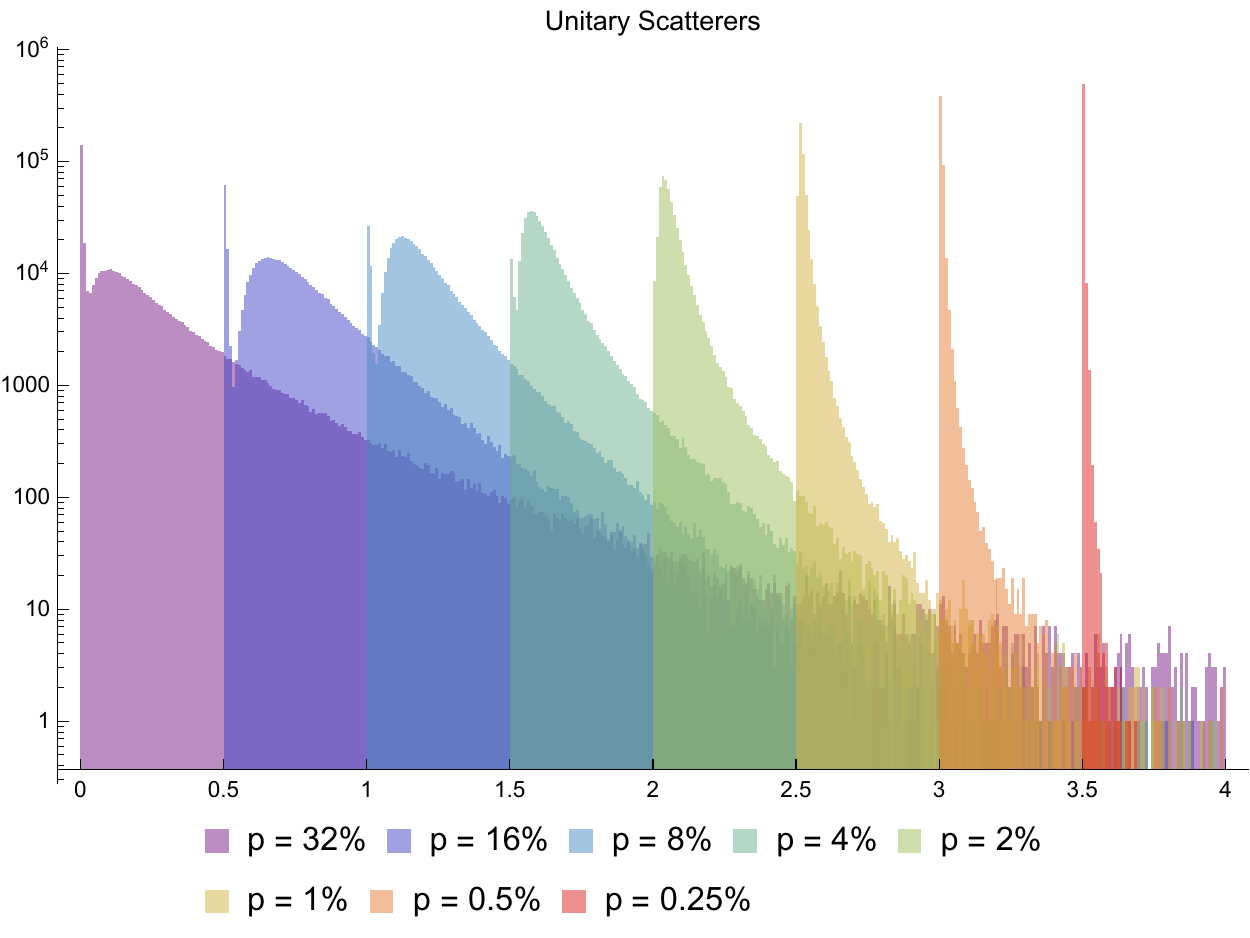}
       	\includegraphics[width=.45\textwidth]{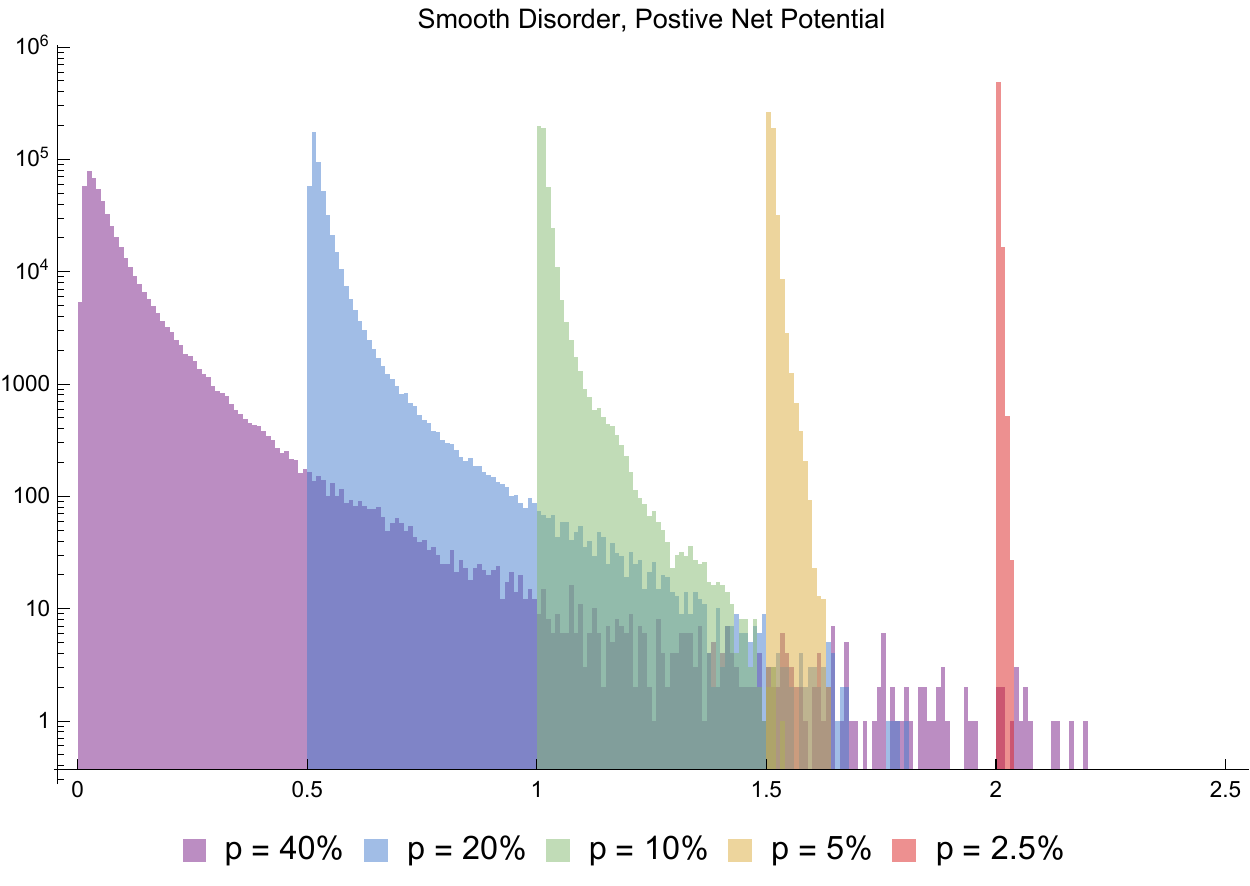}
       	\includegraphics[width=.45\textwidth]{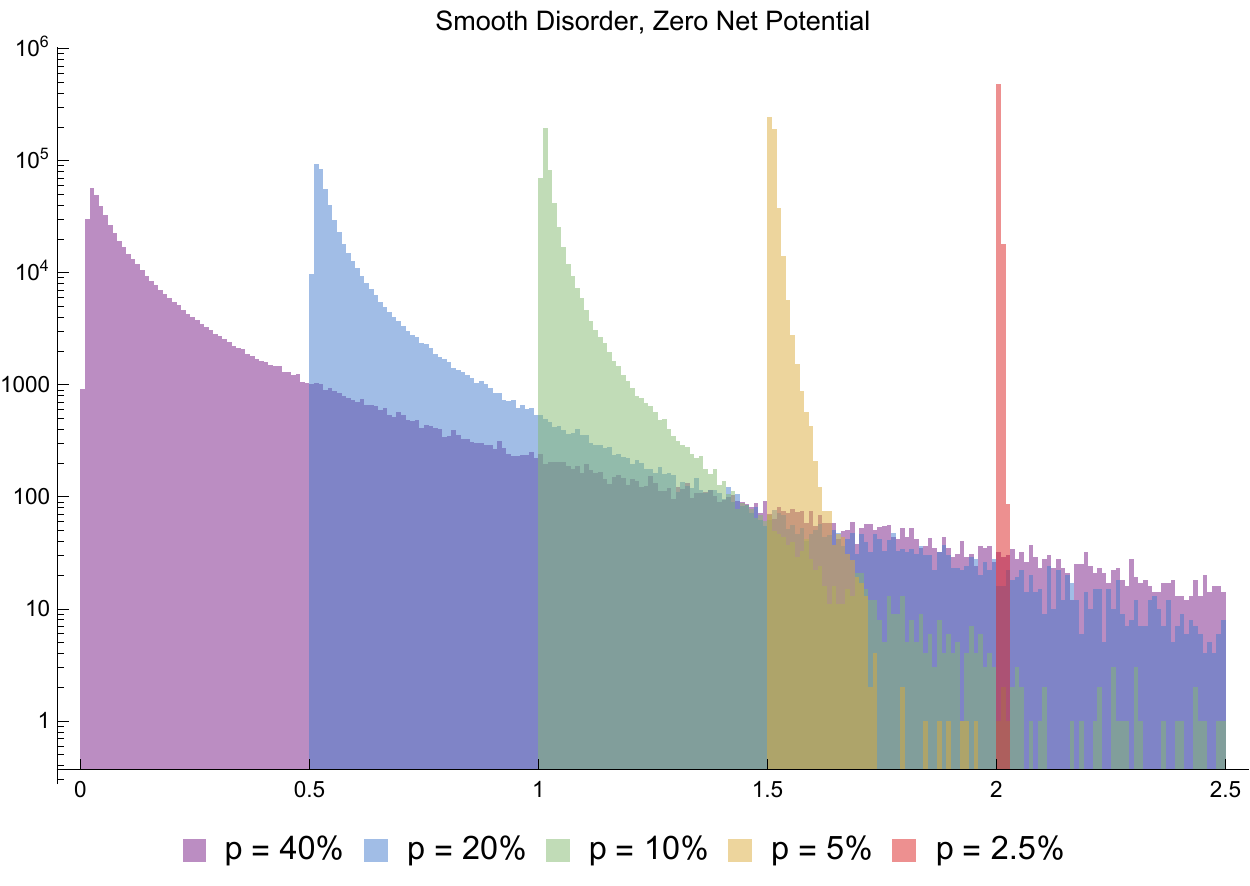}
       	\caption{Histogram of the \emph{local} DOS at $E = 0$ for different types of disorder. A logarithmic scale is used for the $y$-axis. Five realizations are used per value of disorder strength parameter for each type of disorder. $5\times10^5$ values of the LDOS for each value of disorder parameter are shown here. To show the variation in the behavior of the distributions for various disorder strengths, the histogram for a particular value of disorder strength is offset from the preceeding one. The bin width is 0.01.}
       	\label{fig:ldos}
       \end{figure*}
       
       As a considerable number of properties of the cuprate superconductors rely on the physics of the low-energy quasiparticles near the Fermi energy, we will examine more closely the behavior of the DOS near $E = 0$ as disorder is increased. We have seen that, in the random-potential and unitary-scatterer models of disorder, when the amount of disorder is increased, $\rho(E = 0)$ acquires a finite value, then drops once more towards zero after a certain disorder strength is reached. To see if this behavior is robust, we show in Fig.~\ref{fig:fecp} plots of the mean and standard deviation of $\rho(E = 0)$ as the amount of disorder is increased for each of the four models of disorder we use, with five realizations used per value of the disorder strength parameter. All in all, a total of 500,000 LDOS values for each value of the disorder strength parameter are used to generate this plot. A similar if considerably more detailed analysis of LDOS distributions on the Anderson model was performed by Schubert \emph{et al.} in order to obtain critera for Anderson localization using finite-size scaling.\cite{schubert2010distribution} We will not repeat their finite-size analysis here. It should be noted that, under certain conditions, information about the distribution of the LDOS at $E = 0$ can be extracted by obtaining the $^{17}$O Knight shift values from nuclear magnetic resonance experiments.\cite{ouazi2006impurity, zhou2017quasiparticle, zhou2017spin} In particular, Zhou \emph{et al.} find an asymmetric distribution of Knight shifts in YBCO with charge order, which suggests that the LDOS at the Fermi energy is distributed similarly, and argue that a likely explanation of this is quasiparticle scattering off of defects.\cite{zhou2017quasiparticle}
       
       Let us discuss first the random-potential model. In the weak-disorder regime, the mean and standard deviation of the DOS are both close to zero and exhibit almost no dependence on $\sigma$. Starting at approximately $\sigma = 0.35$ the mean DOS becomes finite, increasing as $\sigma$ is increased, and, more interestingly, the standard deviation of $\rho(E = 0)$ depends strongly on the value of $\sigma$. This trend continues until $\sigma = 1.00$: as disorder is increased past that point, the mean DOS starts to decrease, while the standard deviation continues to increase until $\sigma = 1.4$ is reached. In these strong-disorder regimes, the way that $\rho(E = 0) \to 0$ is of a fundamentally different nature than the way the weak-disorder DOS tends toward zero: the distribution of the strong-disorder DOS, while heavily weighted towards zero, exhibits very large spatial variations. The weak-disorder case on the other hand is almost fully concentrated at zero, with almost negligible variations in space.
       
       Surprisingly similar behavior can be seen in the unitary-scatterer model. One can see that in the low-concentration regime (\emph{i.e.}, up to $p \approx 0.5$\%), both the mean and the standard deviation of the LDOS are almost zero. Then at around $p = 1.0$\% both the mean and standard deviation display a strong dependence on $p$, with both increasing as the impurity concentrations are increased. This behavior stops at around $p = 16.0$\%, at which point the mean LDOS reaches the largest value (out of the values of $p$ we consider), and the mean starts to decrease once $p$ is increased. The standard deviation continues to increase past $p = 16.0$\% up until $p = 32.0$\%, signaling that despite the decrease in the mean LDOS, the spatial variations remain considerable. It is interesting to note that both the mean and standard deviation of the DOS at $E = 0$ in this case depend on $p$ very similarly to the way the same two quantities depend on $\sigma$ in the Gaussian random-potential case discussed before, despite the considerable differences present between the two disorder scenarios. 
       
      Despite the huge difference in the effects seen in the quasiparticle DOS and local DOS maps between smooth and pointlike disorder, the DOS at $E = 0$ for the smooth-disorder case does display a similar dependence on the disorder strength as for pointlike disorder. For the positive-net-potential case, low doping concentrations show a mean LDOS close to zero, with a corresponding small standard deviation indicating small spatial variations in the LDOS. Both the mean and standard deviation exhibit a dependence on $p$ up to the (quite unphysical) doping $p = 40\%$. At that point the mean LDOS becomes a maximum, but the standard deviation continues to increase past that point. The zero-net-potential case meanwhile shows much more spatial variation than the positive-net-potential case. Low dopings show small mean and standard deviations, and as $p$ is increased these two quantities depend strongly on $p$. Interestingly, at $p = 10\%$ the standard deviation starts to depend more strongly on $p$; consequently, at intermediate and high dopings the LDOS at $E = 0$ has a considerable amount of spatial variation. The mean LDOS also has a strong dependence on $p$.
       
       The extent to which the LDOS at the Fermi energy varies over space can be visualized neatly by taking histograms of these LDOS values for various values of the disorder strength parameter. These histograms are shown in Fig.~\ref{fig:ldos}. To facilitate comparisons between LDOS distributions corresponding to different disorder strengths, we use the same bin width for each histogram. For random-potential disorder, it can be seen that the weak-disorder cases feature very narrowly distributed LDOS values. When disorder is increased until $\sigma = 1.00$ the distributions start to broaden, and as a consequence the peaks of the distributions shift rightwards, becoming lower, with the mean moving away from zero. For values of $\sigma > 1.00$ the distribution starts to narrow, with much of the distribution being concentrated near zero, but there remains a large amount of spatial variation. Because of the large weight at and near zero, $\rho(E = 0)$ is suppressed in these cases, but the distribution is much more variable than in the weak-disorder case. We note in passing that throughout the range of disorder strengths we consider, $\rho(E = 0)$ is consistently distributed log-normally, which is remarkable given how dramatically different the overall statistics of these distributions are as disorder is varied.
       
       Moving on to unitary-scatterer disorder, at small $p$ the distribution is centered mainly around $\rho(E=0) \approx 0$, but with a small number of LDOS values with larger values arising from the random interference effects discussed earlier. These effects become more and more numerous as $p$ is increased, leading to broader and flatter distributions at intermediate impurity concentrations. The behavior of the LDOS distributions in the multiple unitary-scatterer case parallels very closely that of the Gaussian random-potential disorder, with distributions for both cases widening and then subsequently narrowing once more as $p$ or $\sigma$ is increased. The main difference here is that the distribution of the LDOS for unitary scatterers is \emph{bimodal} for moderate and large values of $p$: as the LDOS is suppressed almost completely \emph{at} impurity sites, these represent a considerable number of LDOS values that are zero, and these peaks in the distributions are present independently of the variations arising from the very presence of these impurities. When one takes these impurity-site LDOS values out of consideration, the LDOS distribution is log-normal, similar to the case of random Gaussian disorder (which, unlike the unitary-scatterer model, does not exhibit a special subset of lattice sites at which the LDOS is maximally suppressed).
       
       In the case of smooth disorder with positive net potential, one can see that the generation of a finite DOS is achieved by an increase in the spatial variation, resulting in the broadening of the distribution. Similarly, when we consider smooth disorder with zero net potential, as $p$ is increased, the LDOS distributions at $E = 0$ become very broad. While this effect is also seen in the other pointlike forms of disorder we looked at earlier, here the broadening is more pronounced, and much more so compared to the positive-net-potential case. We also do not hit the strong-disorder regime where these LDOS distributions start to narrow while still exhibiting strong spatial variations, which we encountered in the random-potential and unitary-scatterer disorder models.
       
       We end this section by noting that our results for weak disorder match closely with what field-theoretic treatments of disorder find, which is that the DOS at $E = 0$ vanishes.\cite{nersesyan1994disorder, senthil1999quasiparticle, altland2002theories} A crucial assumption made in the construction of these field theories is that the distribution of the disorder is narrow.\cite{altland2002theories} Indeed, we find that weak disorder of whatever form leads to a very small DOS at the Fermi energy. What our numerical results suggest however is that the DOS is not vanishing only up to some threshold value of disorder which invalidates the construction of these field-theoretic models. Instead what we find is that the DOS at $E = 0$ varies smoothly as the amount of disorder is increased, suggesting that crossovers, rather than sharp transitions, occur as one moves from weak to intermediate disorder and from intermediate to strong disorder.
       
    \section{Low-Temperature Specific Heat}

     \begin{figure*}
     	\centering
     	\includegraphics[height=.5\textwidth]{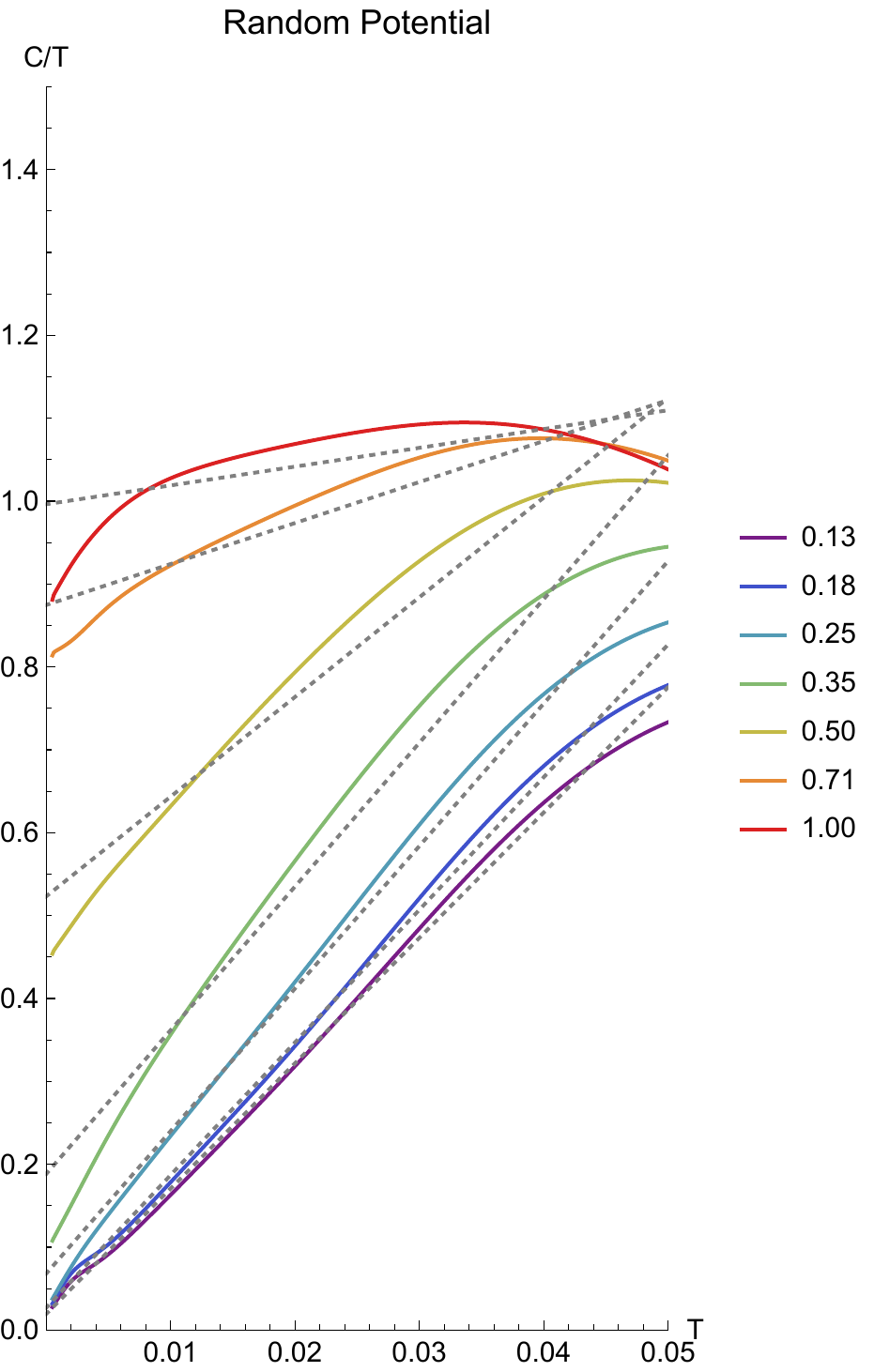}
     	\includegraphics[height=.5\textwidth]{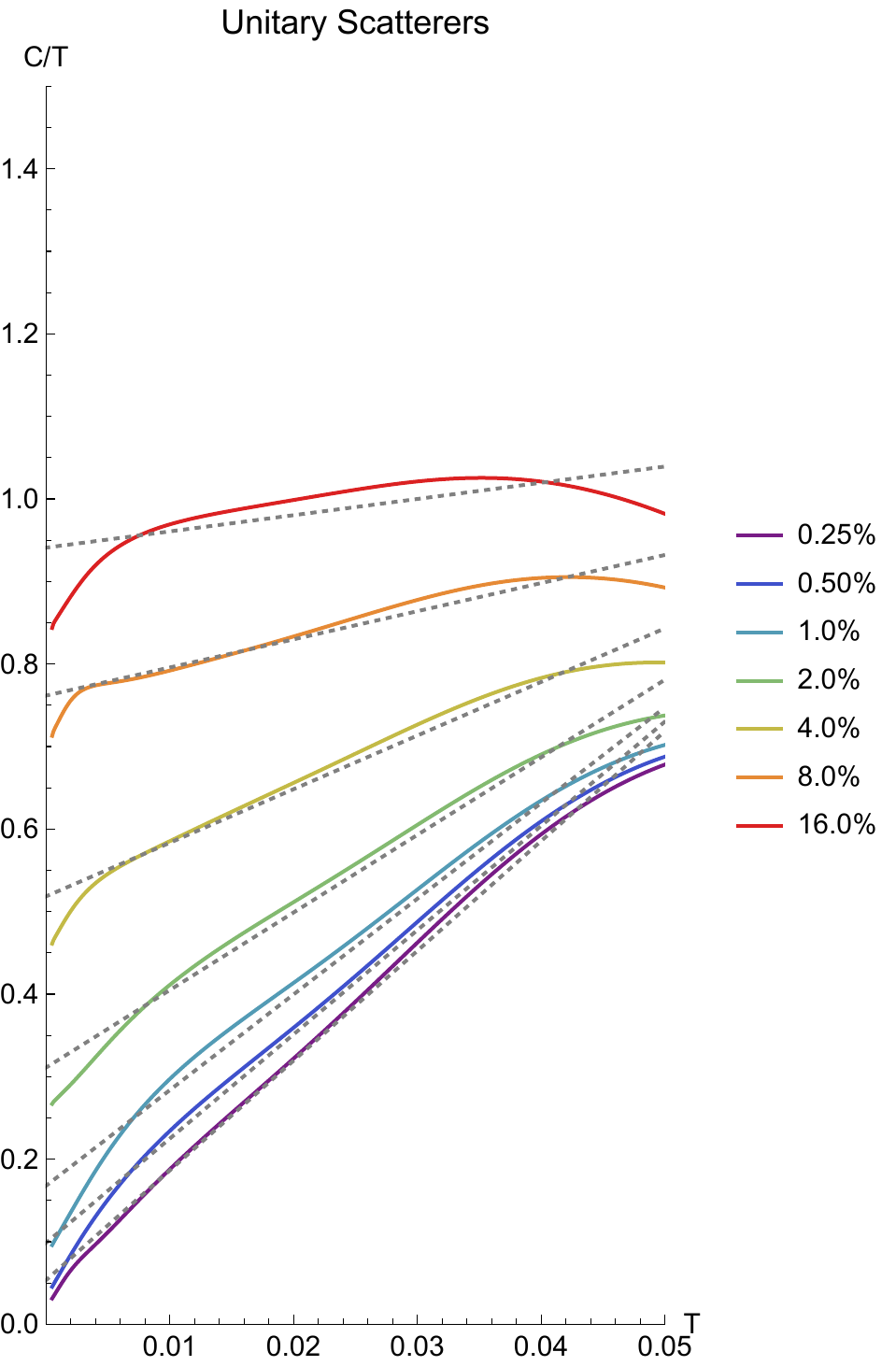}
     	\includegraphics[height=.5\textwidth]{"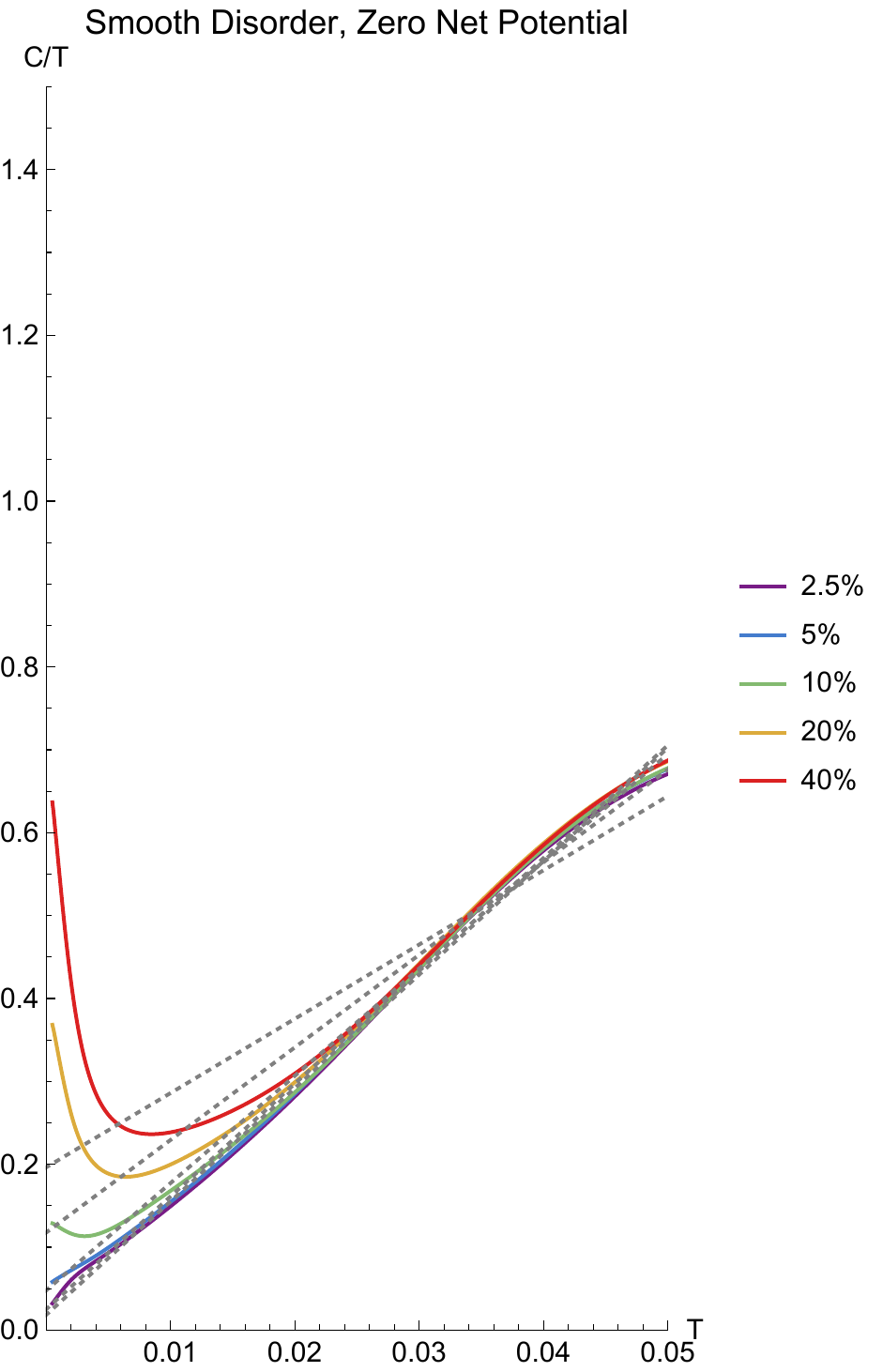}
     	\caption{Plots of $C/T$ as a function of temperature $T$ for different types of disorder. Gray dotted lines indicate fits of the $C/T$ curves to the form $C/T = \gamma_0 + \alpha T$, the scaling expected from $d$-wave quasiparticles with a nonzero DOS at $E = 0$. The numerically-obtained $C/T$ exhibits visible deviations from this scaling.}
     	\label{fig:sh}
     \end{figure*}   
     
     The next quantity we will consider is the low-temperature specific heat. We will examine the contributions of the $d$-wave quasiparticles to the specific heat, neglecting the effect of phonons which arise at higher temperatures. As mentioned earlier, a clean $d$-wave superconductor has a DOS which vanishes at $E = 0$ linearly, and this gives rise to a $T^2$-dependent term in the specific heat $C$. Interestingly, in specific heat experiments, it is found that this $T^2$-dependent term is difficult to disentangle from the signal.\cite{riggs2011heat} Instead the most prominent contributions to the specific heat are the phonon contribution (scaling as $T^3$) and the contribution due to a finite density of states at zero energy, which scales as $T$, similar to a normal metal. We thus begin our discussion of specific heat with the necessary warning that it is difficult to match the dependence on temperature of $C$ from our numerical calculations with that found in specific heat experiments. What can be unambiguously compared between simulation and experiment, however, is the magnitude of $\gamma_0$, the coefficient of the linear-in-$T$ term in $C$ which is proportional to the DOS at $E = 0$.
     
     Shown in Fig.~\ref{fig:sh} are plots of $C/T$ versus $T$ for various types of disorder. We first discuss random-potential disorder. It can be seen that when $\sigma$ is small, the specific heat scales as $C \propto \gamma_0 T + \alpha T^2$, with $\gamma_0$ very small, reflecting the fact that the DOS at the Fermi energy at weak random-potential disorder is suppressed. The behavior of $\gamma_0$ closely follows that of the DOS at $E = 0$, as a large jump in $\gamma_0$ is found at $\sigma \approx 0.35$. Even at moderately strong disorder, the specific heat is still found to scale as $C \propto \gamma_0 T + \alpha T^2$, at least up to $T \approx 0.03$ (approximately $50$ K). When disorder is strong enough, the scaling finally starts to deviate considerably from that found in the weak-disorder cases. For instance, when $\sigma \approx 1.00$, $C/T$ becomes concave downward. The large value of $C/T$ as $T \to 0$ seen in that case is a reflection of the very large DOS at $E = 0$.
     
     For the case of multiple unitary scatterers, the specific heat results are by and large similar to the random-potential case. Low concentrations of unitary scatterers show a very small value of $\gamma_0$, and with large values of $\gamma_0$ reached only until $p \approx 2.0 \%$ is reached. It bears noting that at low temperatures the specific heat roughly scales as $C \propto \gamma_0 T + \alpha T^2$ at low and moderate concentrations of unitary scatterers. The unitary-scatterer cases however feature mild kinks in the $C/T$-versus-$T$ plots at low temperatures which are not present in the random-potential cases. These kinks arise from the particular form of the DOS profiles in the unitary-scatterer cases, which show both a rounding of the DOS at energy scales set by the scattering rate, and ultimately its suppression at $E = 0$. The kink in the $C/T$ profile becomes more prominent with increasing $p$, and in the strong-disorder regime the plot becomes, as in the random-potential case, concave downward.
     
     Finally, smooth-potential disorder gives rise to specific heat behavior that is rather demonstrably different from that arising from random-potential or unitary-scatterer disorder. Low concentrations of smooth scatterers (\emph{e.g.}, $p \approx 2.5\%$ or $p \approx 5.0\%$)  show $C \propto \gamma_0 T + \alpha T^2$ scaling of the specific heat, with correspondingly small values of $\gamma_0$, reflecting the relatively small DOS at the Fermi energy due to these levels of smooth disorder. However, the unusual behavor of the DOS at $E = 0$ at higher concentrations $p$ manifests itself in a strange kink in the plot of $C/T$ versus $T$, showing strong deviations from the scaling one would expect from both $d$-wave dispersion and a finite DOS at the Fermi energy. The large value of $C/T$ as $T \to 0$ results naturally from the enhancement of the DOS at $E = 0$, and as $T$ is increased $C/T$ dips, then rises linearly once more past a certain temperature. It is worth noting that the deviations from the expected scaling are fairly localized within a small region near $T = 0$, with the specific heat returning to quadratic scaling $C \propto \alpha T^2$ once temperature is raised past some threshold value.
     
     Given the aforementioned difficulty of measuring precisely $\alpha$ from experiment, we cannot say much about how consistent with experiment our numerically-obtained scaling for $C$ is. However, what we obtain for $\gamma_0$ can be compared with that found from experiment with definiteness. We will return to a comparison with results from specific heat experiments at the conclusion of this paper.
     
     \section{Quasiparticle Localization}
     
          \begin{figure*}
          	\centering
          	\includegraphics[width=.3\textwidth]{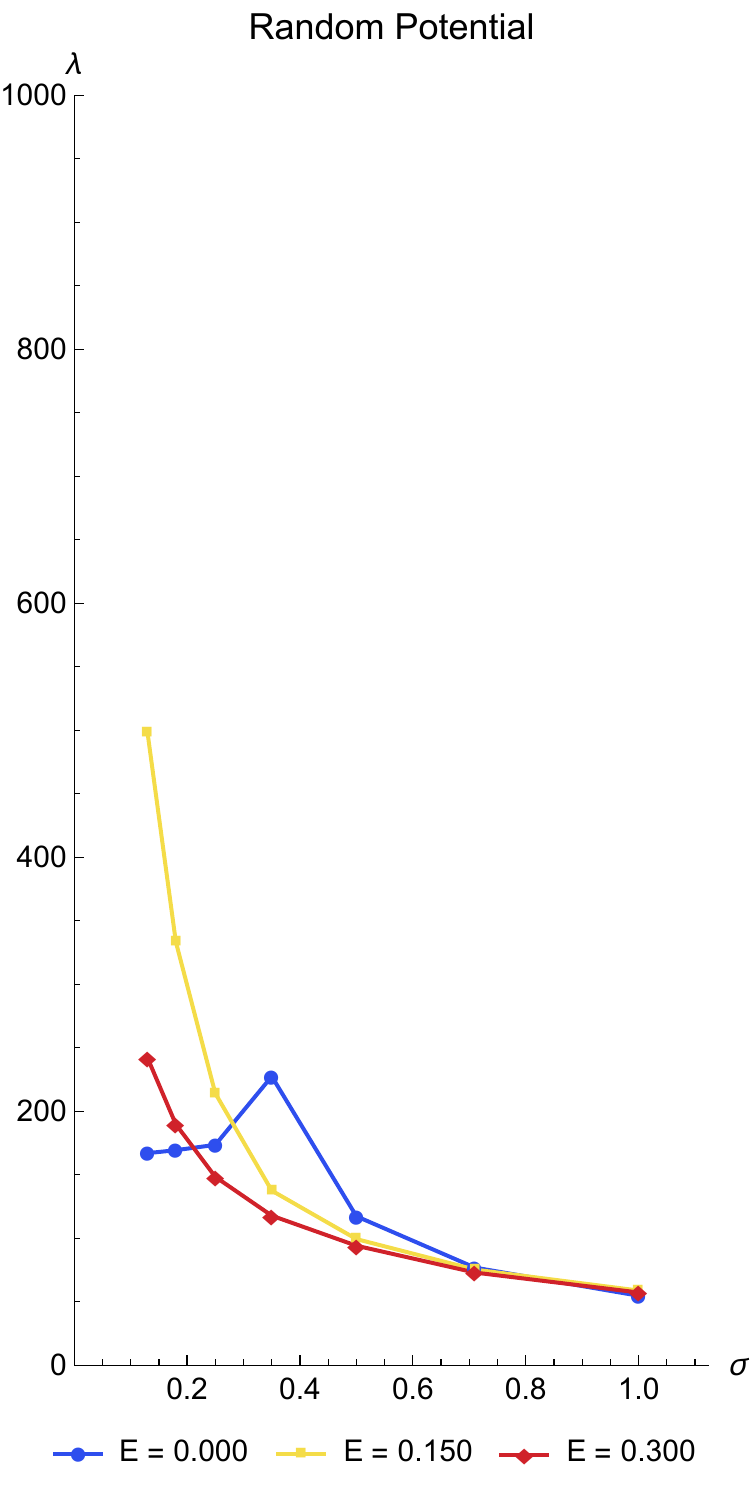}
          	\includegraphics[width=.3\textwidth]{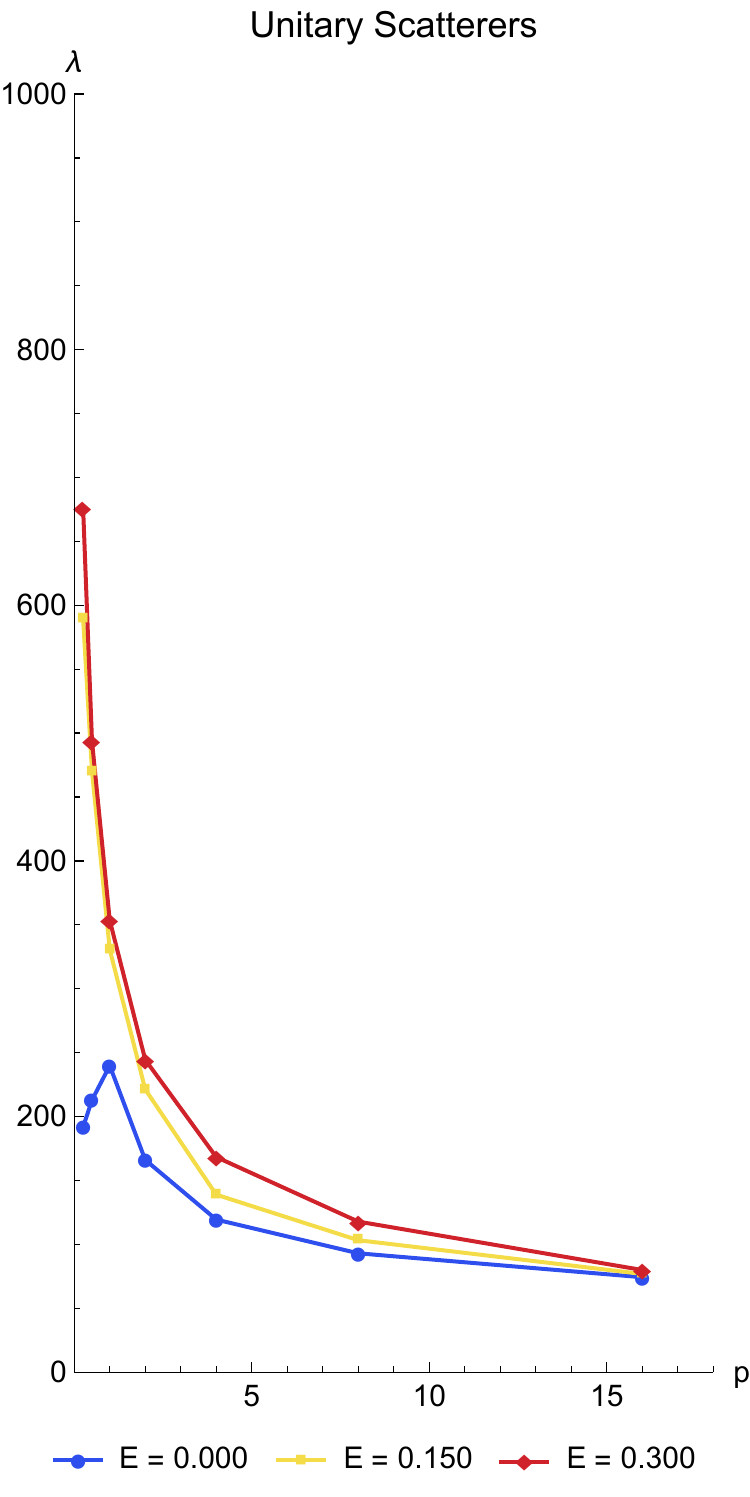}
          	\includegraphics[width=.3\textwidth]{"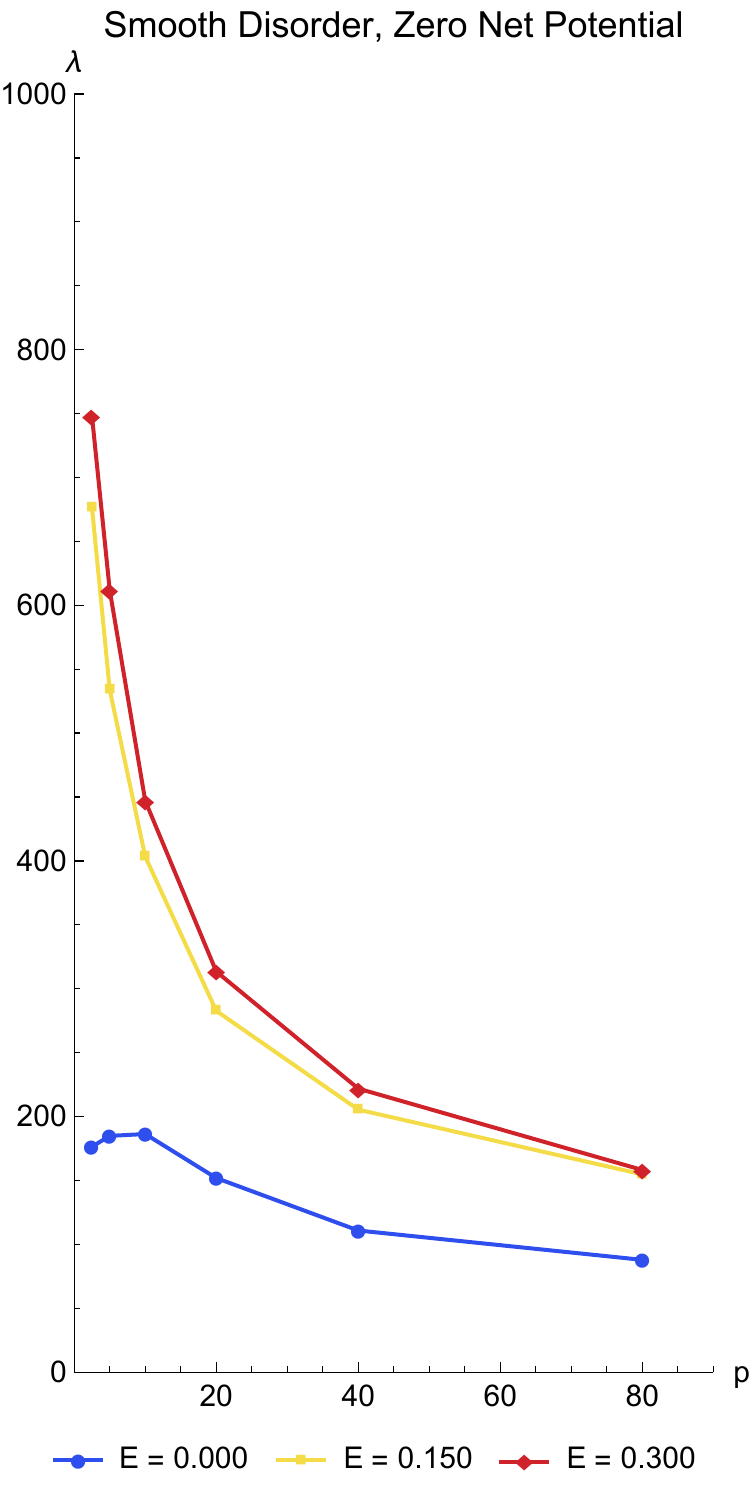}
          	\caption{Plots of the localization length $\lambda$ (in units where the lattice constant $a = 1$)  for different types of disorder, taken at three different energies $E$. The $x$-axis shows the disorder strength parameter, given by $\sigma$ for random-potential disorder (leftmost plot) and the impurity concentration $p$ for unitary-scatterer and smooth disorder (middle and rightmost plots).}
          	\label{fig:ll}
          \end{figure*}   
    
     \begin{figure*}
     	\centering
     	\includegraphics[width=.3\textwidth]{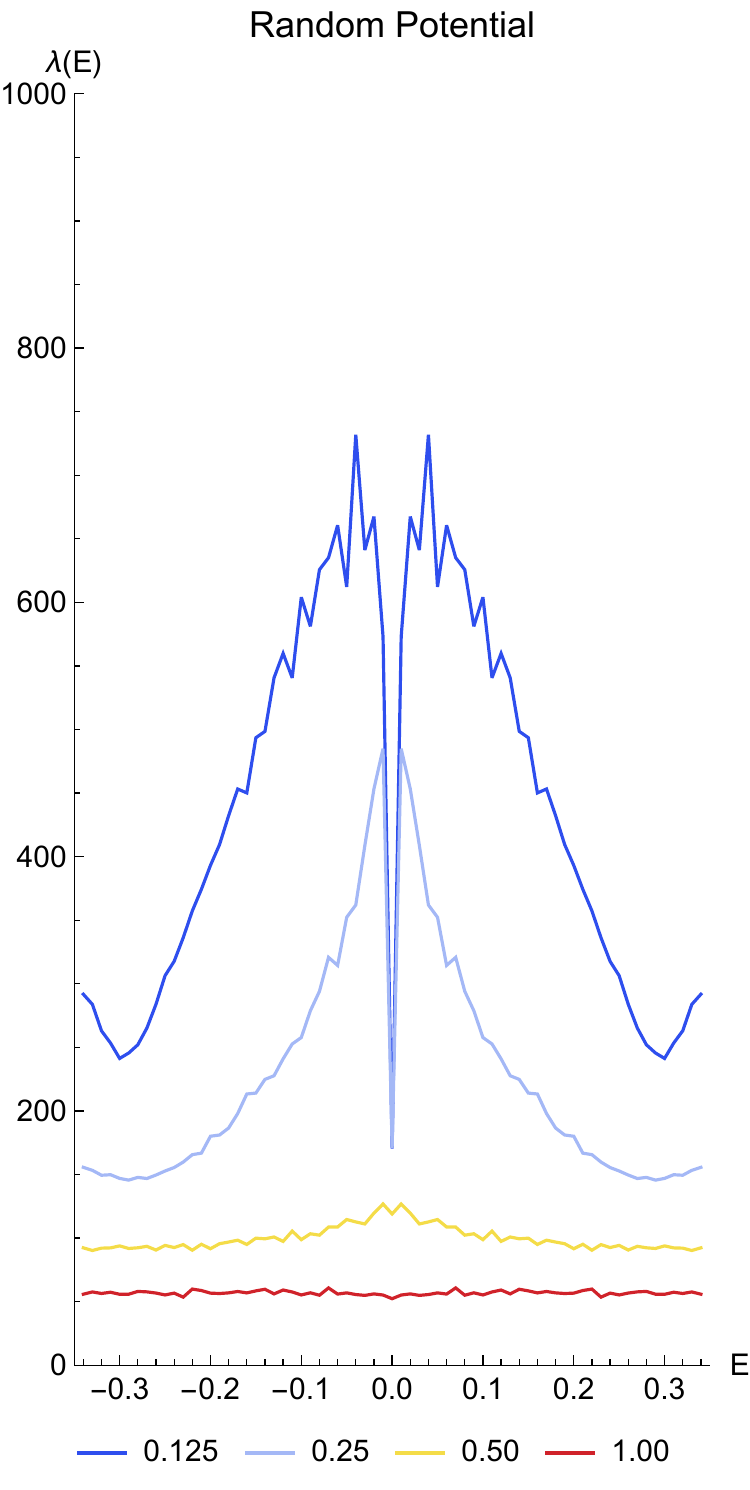}
     	\includegraphics[width=.3\textwidth]{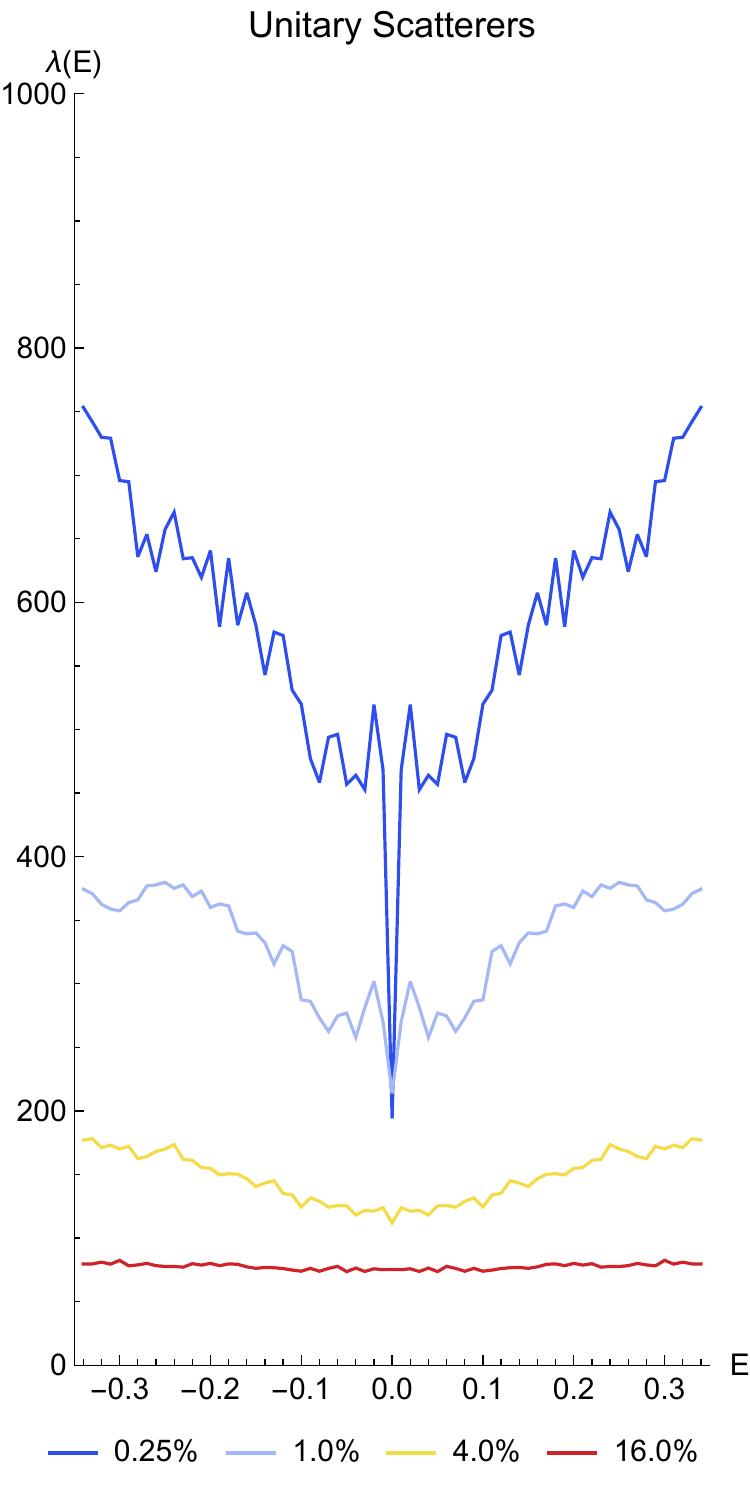}
     	\includegraphics[width=.3\textwidth]{"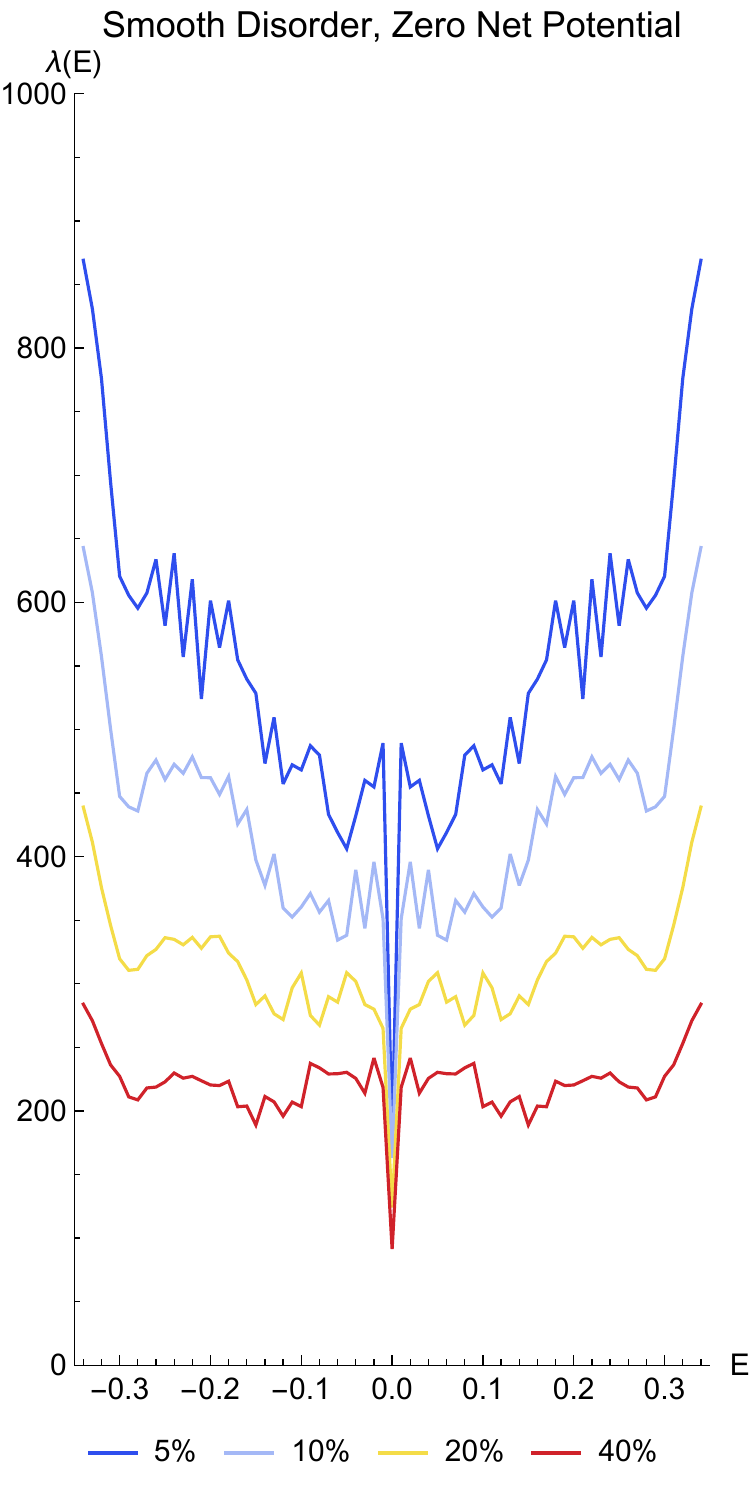}
     	\caption{Plot of the localization length $\lambda$ as a function of energy $E$ for different types of disorder.}
     	\label{fig:ll_energy}
     \end{figure*}   
     
     The final quantity of interest to us is the localization length $\lambda$. Unlike the DOS and the specific heat, the localization length is not an experimental observable; no experiment exists which measures the quantity described by Eq.~\ref{eq:loclength}. However it is a very important quantity in that it gives information as to how localized the states at a particular energy are. It is a rather difficult quantity to measure in finite-size simulations of lattice systems because more often than not $\lambda$ is much bigger than the system size. The numerical method we use however circumvents this difficulty by allowing one dimension of the system to be much longer than the other. Thus we can use one definition of the localization length which involves the transmission probability between two ends of an elongated two-dimensional system.\cite{mackinnon1981one, buŀka1985mobility, kramer1993localization, xiang1995effect} This enables us to \emph{directly} and exactly calculate the localization length for the full disordered system. As a first exercise we calculate the localization length $\lambda$ using Eq.~\ref{eq:loclength} on the same set of disorder configurations as used in Figs.~\ref{fig:fecp} and~\ref{fig:ldos}. In Fig.~\ref{fig:ll} we show $\lambda$ for three different values of $E$: $E = 0$ (corresponding to states at and near the Fermi energy), $E = 0.15$ (for states far from either the Fermi energy or the coherence peaks, but still within the $d$-wave gap), and $E = 0.3$ (states at and near the coherence peaks). We also plot in Fig.~\ref{fig:ll_energy} the localization length as a function of energy for various forms of disorder using the same configurations used in Figs.~\ref{fig:rp},~\ref{fig:mp},~\ref{fig:ss}, and~\ref{fig:ssn}.
     
     Let us discuss localization in the random-potential model first. We begin with the states at and near the Fermi energy. The dependence of $\lambda$ at $E = 0$ on $\sigma$ appears to be unusual: it is approximately constant from $\sigma = 0.13$ to $\sigma = 0.25$, then hits a peak at around $\sigma = 0.35$ before decreasing with increasing disorder. The localization lengths for these states are small at weak disorder ($\lambda \approx 170$), and the strong-disorder $\lambda$ is even smaller---$\lambda \approx 50$ at $\sigma = 1.00$, smaller in fact than the transverse dimension of the system.
     
     The localization lengths at intermediate and high energies show more consistent behavior than the low-energy case. These decrease monotonically as disorder is increased. It is worth noting that while these states are quasi-extended at low disorder, with a larger localization length than for the $E = 0$ states, there is a range of $\sigma$ where these higher-energy states have a smaller $\lambda$ than states near the Fermi energy, which coincides at the range where $\lambda (E = 0)$ peaks. We will later show that the contrast in behavior seen here between the $E = 0$ case and that for higher energies is also seen in other forms of disorder. However a remarkable fact about random-potential disorder is that, of the various types of disorder we consider, this has the most dramatic impact on the behavior of the localization length. For one, it can be seen from the results that $\lambda(E = 0.15) > \lambda(E = 0.3)$ for all values of $\sigma$ we use, implying that the intermediate-energy states are less localized than the higher-energy ones---a feature that is not seen in other forms of disorder we consider. Also, the closeness of the values of $\lambda$ at different $E$ for all $\sigma$ is much less pronounced in the unitary- or smooth-scatterer cases. These cases exhibit a more visible and rigid separation of $\lambda$ as a function of energy for a wide range of disorder strengths---\emph{i.e.}, $\lambda(E = 0) < \lambda(E = 0.15) < \lambda(E = 0.3)$ for these cases, which the random-potential case clearly does not show. There is a disorder strength---$\sigma \approx 1.00$---at which the localization lengths for the three different energies are approximately the same number; this corresponds to the onset of the strong-disorder regime.
     
     We can see these effects more clearly when the localization length is plotted versus energy. Notice that for all disorder strengths we consider, the states near the Fermi energy are strongly localized, and their localization lengths at $E = 0$ are close in value to one another even as the amount of disorder is varied. For weak disorder ($\sigma = 0.125$ and $\sigma = 0.25$) the localization length rises from a small value at $E = 0$ into a prominent peak at some small energy ($E \approx 0.02$ for $\sigma = 0.125$ and $E \approx 0.01$ for $\sigma = 0.25$ ), after which it decreases as energy is increased. It bears noting that the localization lengths at intermediate and high energies at these disorder levels are still quite large, at around 200-600 lattice constants. At $E \approx 0.3$ (the coherence-peak energy), the localization length for the $\sigma = 0.125$ case starts to increase; this effect is not visible when disorder is stronger.  When disorder is increased, the localization length stops exhibiting these energy-dependent features: when $\sigma = 0.50$, $\lambda$ is almost energy-independent, and this is even more the case for $\sigma = 1.00$, indicating that the states are strongly localized at all energies.
     
     We next discuss unitary-scatterer disorder. Focusing first on the $E = 0$ case, we see that it exhibits the same unusual dependence on $p$ as the random-potential case at the same energy does on $\sigma$. At low impurity concentrations $\lambda(E = 0)$ increases slightly with increasing $p$, reaching a peak at $p = 1.0\%$ before decreasing monotonically as a function of $p$. This is in stark contrast with the behavior of $\lambda$ at higher energies, which monotonically decrease with increasing $p$ for \emph{all} $p$ we consider. It is worth comparing these plots to the ones derived for the random-potential case. Here we can see that, in the unitary-scatterer model, the low-disorder cases at intermediate and high energies have a far larger localization length than in the random-potential model. The impact of unitary scatterers is less pronouced than Gaussian random-potential disorder at low disorder, but with stronger disorder the behavior of the localization length for this case starts to become similar to that of the random-potential case. At higher impurity concentrations, the values of $\lambda$ for different $E$ approach each other as $p$ is increased, with $\lambda(E = 0) \approx \lambda(E = 0.15) \approx \lambda(E = 0.3)$ at $p = 16.0\%$, which corresponds to the strong-disorder regime of this particular form of disorder.
     
     The localization length for the unitary-scatterer model exhibits a very different dependence on energy from the Gaussian random-potential case, at least for small amounts of disorder. Near the Fermi energy, the states are strongly localized, and as with the previous disorder model we discussed the localization lengths at $E = 0$ are close in value to each other. At low concentrations, the localization length increases from $E = 0$ up to some energy, then after that point it increases once more with increasing energy, but at a decreased rate. This is seen in the $p = 0.25\%$ and $p = 1.0\%$ cases. Evidently, past a certain threshold energy the states become far less localized, with very large localization lengths at intermediate and high energies (around 300-700 lattice constants), and states at higher energies are less localized than those at intermediate energies---in stark contrast to the Gaussian random-potential case. When $p$ is increased, however, these energy-dependent features become far less noticeable, as can be seen when $p = 4.0\%$, indicating that when disorder is large enough, the effects of localization become visible at all energies, and not just at small energies. At these large-$p$ regimes the behavior of the localization length with increasing energy becomes very similar to that seen in the strong-disorder random-potential case in that little, if any, dependence on energy can be discerned, and in that all states are localized, even at high energies.
  
     We finally consider the localization due to smooth disorder. Here we will consider smooth disorder potentials whose spatial average is zero---\emph{i.e.}, disorder potentials described by Eq.~\ref{eq:vz}. Here the smoothness of the disorder potential makes itself particularly manifest. First, for the states near the Fermi energy, it can be seen that $\lambda(E = 0)$ does not exhibit a sharp peak at some disorder strength, unlike what is seen for random-potential or unitary-scatterer disorder. Instead its profile is flat at low $p$, and it then smoothly decreases as $p$ increases. It is interesting to note that $\lambda(E = 0)$ manages to be fairly large even at high impurity concentrations. Notably, when one has unphysically high $p$ (\emph{e.g.}, $p = 40\%$ or $p = 80\%$), the localization length at the Fermi energy is still $\lambda(E = 0) \approx 80$-$100$. For comparison's sake, that point is reached with random-potential disorder at $\sigma \approx 0.50$ and with unitary-scatterer disorder at $p \approx 8.0\%$---levels of disorder which are strong enough to destroy coherence peaks. From just the consideration of states near the chemical potential, the impact of smooth disorder on $\lambda$ is much less pronounced than either of these other cases.
     
     The absence of any strong impact on the localization lengths is even visible at higher energies. Here it can be seen that the localization lengths for $E = 0.15$ and $E = 0.3$ are very large---$\lambda \approx 500$ for low $p$. Even at $p = 20.0\%$ we find that $\lambda \approx 300$. Such large values of $\lambda$ are seen only at low levels of disorder for the random-potential model ($\sigma \approx 0.18$) and the unitary-scatterer model ($p \approx 1.0\%$). Even at very high smooth-impurity concentrations such as $p = 40\%$ and $p = 80\%$, we find that $\lambda \approx 150$-$200$; these localization lengths correspond to low disorder levels in the random-potential and unitary-scatterer models of disorder. Another notable observation is the fact that $\lambda(E = 0.15)$ and $\lambda(E = 0.3)$ are quite close to each other for all $p$. We \emph{never} reach the onset of the strong-disorder regime that we observe in the other pointlike models of disorder---that is, the disorder strength at which $\lambda(E = 0) \approx \lambda(E = 0.15) \approx \lambda(E = 0.3)$. We find that at the absurdly unphysical $p = 80\%$ concentration $\lambda(E = 0.15) \approx \lambda(E = 0.3)$, but $\lambda(E = 0)$ remains parametrically much smaller than either. This is a clear sign that, even with very large off-plane impurity concentrations, the impact of this form of disorder on the localization of states at all energies is much more muted than in random-potential or unitary-scatterer disorder---\emph{especially} at intermediate or high energies.
     
     The plots of $\lambda$ versus $E$ for the smooth-disorder case exhibit a number of differences from the other two forms of disorder we have considered. First, the states near the Fermi energy are strongly localized, but as the energy is increased the localization length increases sharply for all $p$ we consider until some value of $E$ is encountered, at which point the localization length exhibits a far less pronounced dependence on $E$. At low concentrations (\emph{e.g.}, $p = 5\%$), the localization length by and large increases as energy is increased, but with considerable random fluctuations. When the concentration is increased, the localization length grows more slowly with $E$. It is interesting to note that the localization length trends upward past $E \approx 0.3$, the energy where coherence peaks are found, indicating that states at energies higher than the coherence-peak positions are quite extended in space. These behaviors are different from those seen in unitary-scatterer or random-potential disorder, although there are similarities---at low energies $\lambda$ for smooth disorder behaves similarly as in unitary-scatterer disorder, while at higher energies there is a noticeable increase in $\lambda$ starting at $E \approx 0.3$, similar to what is seen in weak random-potential disorder. Even at very large values of $p$ the behavior of the localization length is still similar to that at low concentrations; at $p = 40\%$ $\lambda$ is still visibly energy-dependent, suggesting once more that even in these regimes disorder of this form has a far weaker effect than the other two types of disorder we have considered. It is instructive to compare smooth disorder at $p = 20\%$ to, say, unitary-scatterer disorder at $p = 1.0\%$ or random-potential disorder at $\sigma = 0.25$---the localization lengths for these three cases occupy a similar range to each other.
     
     Our numerical results for the localization length are in good qualitative agreement with the analytical results obtained by Lee, who performed self-consistent calculations for weak Gaussian random-potential and dilute unitary-scatterer disorder in the $d$-wave superconducting state.\cite{lee1993localized} Some caveats need to be mentioned, however, as our numerics exhibit more detail and structure about the localization properties of these disorder models.  Lee argued that the states near the Fermi energy are localized, although the extent to which these states are localized away from $E = 0$ (instead of being quasi-extended) was found to depend on whether the scattering is in the Born limit or the unitary limit. In the Born-scattering limit of Gaussian random disorder it was found that localization is negligible away from $E = 0$, whereas for unitary scatterers localization can be observed at energies $E < \Gamma_0$, where $\Gamma_0$ is the scattering rate in the superconducting state as $E \to 0$. In our numerical results we find that the states within the vicinity of $E = 0$ are special in being much more localized than their neighbors in energy space for \emph{all} weak-disorder models we consider. We find that the dip in the localization length at $E = 0$ for the unitary-scatter case is narrower than Lee's calculations suggest---that is, the energy range over which the quasiparticles are sharply localized is considerably narrower than Lee's estimate of the scattering rate $\Gamma_0$. Away from $E = 0$ the behavior of the localization length is in much more quantitative agreement with Lee's predictions: $\lambda(\omega) \approx v_F/\Gamma(\omega) \approx 1/\omega$ for random-potential disorder in the Born limit and  $\lambda(\omega) \approx \omega$ for unitary scatterers, which are behaviors similar to what we can observe in the weak-disorder cases we discussed earlier. Our numerical results are also in good agreement with earlier numerical work on random-potential and unitary-scatterer models of disorder.\cite{xiang1995effect, franz1996impurity,zhu2000quasiparticle}
     
     The behavior of the localization length as a function of $E$ at weak disorder resembles that predicted by Senthil and Fisher from field theory.\cite{senthil1999quasiparticle} Their inclusion of diffusive modes---as elucidated in the complementary diagrammatic approach by Yashenkin \emph{et al.}\cite{yashenkin2001nesting}---implies that additional behavior due to these modes, not captured by self-consistent diagrammatic theory, should account for the differences between these approaches.\cite{altland2002theories} Senthil and Fisher argue that, at least in the case of unitary scatterers in the dilute limit, there are three regimes: the ballistic regime, the diffusive regime (at $E \approx \Gamma_0$), and finally the localized regime near $E = 0$. The distinction between the ballistic and diffusive regimes cannot be clearly delineated from our numerics, but the crossover from the ballistic/diffisive regimes to the sharply localized regime can be seen very clearly in the weak-disorder cases we consider. Also, we find, in agreement with Senthil and Fisher's results, that the localization length as $E \to 0$ in fact approaches a finite constant---in striking contrast to the predictions by Nersesyan \emph{et al.}, who find a diverging localization length as $E \to 0$. \cite{nersesyan1994disorder} Our calculations find that this constant localization length at the Fermi energy is independent of the disorder strength in the weak-disorder regime, and stands in contrast to the behavior of the localization length at higher energies, which is found to be dependent on the disorder strength.

\section{Discussion and Conclusion}

We have revisited the effects of disorder in high-temperature superconductors using exact real-space methods which allow large system sizes to be studied, and have ensured that the parameters we have used in our models hew closely to what is known about the cuprates from experiment. We have focused primarily on the density of states and the localization length, two quantities that are of central importance in the study of disordered systems, and made use of various models of site-energy disorder---random Gaussian potentials, multiple unitary scatterers, and off-plane dopants---which are found to result in vastly different behavior depending on which particular model is used.

Our main motivation for looking at the density of states of disordered $d$-wave superconductors once more is the observation---seen consistently in experiments as disparate as specific heat measurements, ARPES, and STS---that there appears to be a nonzero density of states in the cuprate superconductors, even those for which the samples can be made very clean, such as YBCO.  The persistent appearance of such a signal has prompted a number of explanations that do not invoke disorder, and at the very least suggests the possiblity that physics beyond the usual paradigm of $d$-wave superconductivity has to explain this. We reconsider the possibility that disorder is responsible for this nonzero density of states, and find that disorder of a form rarely considered in the older literature on the subject can in fact be a plausible explanation for this phenomenon.

The idea that the cuprates host different variants of disorder is not strange or even new, as STS experiments can directly visualize the disorder present in these materials and find that throughout the phase diagram of BSCCO, the signatures of disorder are present---whether in the form of quasiparticle interference in the superconducting state, or the real-space inhomogeneities in the DOS and pairing gaps in the pseudogap regime. However, the very chemistry of the cuprates naturally precludes the possiblity that disorder is present within the CuO$_2$ planes. The most natural form of disorder, at least from a chemical standpoint, appears to be dopants located some distance from the conducting planes. Doped cuprates host a nonzero number of oxygens at off-plane sites, and they exert an effect on the physics within the CuO$_2$ planes by means of a screened Coulomb potential that modifies the chemical potential at sites located within the conducting planes. The longer-ranged nature of these potentials makes them trickier to model than unitary scatterers or random-potential disorder, but the numerical methods presented here allow the effects of these forms of disorder to be simulated with great efficiency. We have also been able to obtain the localization length, a quantity that, thanks to its large size, is unable to be extracted from exact diagonalization studies of small systems, and closely examine its behavior as a function of disorder strength and energy for different models of disorder used.

Examining first random-potential disorder, we find that its effect on the DOS is to flatten the coherence peaks at the edge of the $d$-wave gap, and that the dominant spectral-weight transfer processes appear to be from the coherence peaks to intermediate energies, with not much spectral weight transferred to the region near the Fermi energy. A large finite DOS at $E = 0$ is not generated until fairly strong levels of disorder are reached. We consistently see that the DOS at the Fermi energy is suppressed relative to that at nearby energies; that the DOS profile at that region is V-shaped, in stark contrast to what is seen in STS experiments; and that coherence peaks are considerably flattened, even when disorder is weak. For this form of disorder the localization length exhibits an interesting dependence on energy and disorder strength: states near $E = 0$ are localized, but the localization length sharply increases moving away from the Fermi energy, until it starts decreasing monotonically as energy is increased.

Multiple unitary scatterers are found to exhibit spectral-weight transfers from the coherence peaks to a particular energy scale, resulting in the presence of a hump-like feature in the DOS at small energies, with otherwise small deviations from the clean case at small impurity concentrations. The DOS consistently exhibits suppression at $E = 0$, and manages to acquire a large finite value only when fairly large concentrations are reached. As the concentration is increased the $d$-wave gap gets filled and the coherence peaks become more and more flattened. The behavior of the localization length for this form of disorder is drastically different from the random-potential case, especially at low levels of disorder. The localization length is small for states near the Fermi energy, then increases sharply until some energy is reached, and subsequently increases once more, but at a far slower rate.

Off-plane scatterers are the most interesting case, insofar as even a large concentration of such dopants turns out not to destroy the $d$-wave profile of the DOS---spectral weight transfers are minimal at best---while generating a finite DOS at $E = 0$ at levels of disorder that are not far off from what is seen in experiment. For the parameters we have used in our disorder potential, concentrations around $10$-$20\%$ result a small but visibly finite DOS at the Fermi energy and a U-shaped DOS profile for small energies, which are consistent with experiment. At higher concentrations, an unusual resonance forms at $E = 0$; this appears to be an intrinsically many-impurity effect, as there is no obvious correlation between the disorder potential and the resulting resonant DOS. The localization length is found to be much bigger than that seen in the previous two disorder models used. While the states near $E = 0$ have a short localization length, away from that region the localization lengths are very large, even when the concentrations are sizable---for comparison's sake we have found that a concentration of $20\%$ off-plane scatters has roughly the same effect on the localization length for a broad energy range as an ensemble of unitary scatterers with concentration $1.0\%$, or random-potential disorder with $\sigma = 0.25$. These results all point to the fact that smooth scatterers have far less of an effect on the DOS and the localization properties of a $d$-wave superconductor than the other two disorder models, even when the amount of smooth disorder is large.

It is worth asking whether we can make any definitive conclusions regarding the nature of disorder in the cuprates from our results. Disorder makes itself felt in a panoply of effects seen in various experiments, but isolating its effect with any definiteness is difficult given the vast array of strongly correlated phenomena present in the cuprates. We have focused mostly on single-particle properties in the form of the DOS, and and it bears noting that many of the effects due to disorder we have described could be due to other effects as well. Disorder broadens the DOS, but so do interactions (in the form of self-energies) and finite temperatures. We work in the $T \to 0$ limit, so the latter alternative is ruled out, but even then we cannot rule out the possibility that nontrivial physics beyond the mean-field model of a $d$-wave superconductor we work with can explain the bulk of what is seen in experiment. The best we could do in the meantime is to look at the extent to which disorder---and disorder alone---reproduces experiment.

How does one square the presence of a finite DOS at $E = 0$ with the amount of disorder present in the cuprates, assuming that disorder alone is responsible for the broadening? Zero-field specific heat measurements on YBCO find a residual $T$-linear term in the specific heat whose coefficient is $\gamma \approx 2$ mJ$\cdot$mol$^{-1}\cdot$K$^{-2}$.\cite{moler1994magnetic, moler1997specific, riggs2011heat} Using Eq.~\ref{eq:sommerfeld}, we find that $\rho(E = 0) \approx 0.1$. Interestingly, angle-resolved photoemission spectroscopy provides a similar value for the residual DOS at $E = 0$.  The widths of energy distribution curves taken from ARPES experiments on clean BSCCO suggest that the scattering rate in the superconducting state is around $\Gamma \approx 15$ meV near zero binding energy.\cite{kaminski2000quasiparticles, valla2006fine, koralek2006laser} Using the formula
\begin{equation}
\rho(E = 0) = \sum_{\mathbf{k}} \frac{\Gamma}{\epsilon^2_{\mathbf{k}} + \Delta^2_{\mathbf{k}} + \Gamma^2},
\end{equation}
this too leads to $\rho(E = 0) \approx 0.1$. These provide constraints in the amount of disorder in the cuprates, assuming that this finite value of the DOS is due purely to disorder.

Unitary scatterers can be safely ruled out. STS experiments show few, if any, signals of unitary scatterers in real-space conductance maps of clean BSCCO. They do not show the resonances one sees in zinc-doped BSCCO. The presence of vacancies however could be one source of unitary-scatterer disorder in the cuprates. How numerous would they have to be to produce a finite density of states consistent with experiment? From our numerics it appears that $p = 2.0\%$ and $p = 4.0\%$ are the closest matches to this, but these concentrations of unitary scatterers appear to be too high to describe clean BSCCO. In fact, these are too large to describe even zinc-doped BSCCO---the STS experiments on these doped materials use a zinc-dopant concentration of $p = 0.6\%$,\cite{pan2000imaging} and conductance maps from these studies show very prominent resonances that are not present in clean BSCCO.

Weak random-potential disorder can also be ruled out as a primary source of the finite DOS ultimately for two reasons. First, by the argument we used above for unitary-scatterer disorder, the level of Gaussian disorder needed to reproduce $\rho(E = 0) \approx 0.1$ is around $\sigma = 0.50$. At this level of disorder, the coherence peaks are completely flattened and smeared out. This is in contrast to what is seen in STS experiments, which consistently find a spatially-averaged LDOS with prominent coherence peaks in the superconducting state of BSCCO. Second, at this level disorder is strong enough that the usual telltale signatures of QPI are no longer present. As discussed before, this form of disorder is consistent with QPI when $\sigma$ is very small.\cite{sulangi2017revisiting} When disorder of this sort is weak, peaks in the power spectrum of the LDOS corresponding to what the octet model predicts are visible and prominent, and the real-space maps show crisscrossing patterns consistent with experiment. However this is destroyed when disorder is increased, and STS studies of BSCCO show that disorder is never strong enough to prevent the formation of modulations governed by QPI---disorder has to be weak enough that QPI is preserved. The strong levels of disorder that would produce a finite DOS at $E = 0$ consistent with the large self-energies found in ARPES would on the other hand not result in QPI. This suggests that QPI due to weak random-potential disorder occurs \emph{on top} of other effects that are primarily responsible for the finite DOS at $E = 0$.

This leaves us with smooth disorder due to off-plane dopants. Many aspects of these dopants remain mysterious, and important properties---the screening length, the strength of the potential, and even the exact placement of these dopants---are not known with any degree of accuracy. Nevertheless, in our treatment of these dopants we have attempted to be consistent with a number of crucial facts. First, the dopant concentration is generally large, and second, the dopants are located some distance away from the CuO$_2$ planes, which leads to small-angle scattering. We find that the effects of smooth disorder on the DOS are much more muted than in the other two disorder cases, with minimal impact on the heights of the coherence peaks and only small spectral-weight transfers to the region near the Fermi energy. This is seen too in our calculations of the localization length in the presence of this form of disorder, which is found to remain quite large for a wide energy range even for large impurity concentrations $p$. We find that $\rho(E = 0)$ acquires a value within the range $[0.05, 0.10]$ for a rather wide range of $p$---this would correspond to $p \approx 10$-$20\%$, depending on which disorder scenario one has. The more realistic scenario, in which the impurity strengths of the scatterers all have the same sign, features considerably more suppression of the DOS at the Fermi energy than the case where the spatially-averaged disorder potential is zero. The zero-average scenario has a number of very interesting features at large concentrations ($p \approx 20\%$, for instance), such as resonances \emph{at} $E = 0$ whose origins appear to be unrelated to the exact details of the disorder potential. While these prominent resonances are not seen in experiment, lower impurity concentrations show much more muted LDOS patterns at $E = 0$, which, while yielding a nonzero DOS at the Fermi energy when averaged, are far less observable than at higher concentrations, and the value of the DOS appears to be fairly consistent with experiment.

Having said this, studies of quasiparticle scattering interference in BSCCO do consistently demonstrate that small- \emph{and} large-momenta scattering processes occur in BSCCO, which is something that \emph{purely} smooth disorder cannot take into account on its own. Purely smooth disorder such as what we discussed in this section has been shown to give rise to Fourier-transformed maps where large-momenta peaks are missing.\cite{nunner2006fourier, sulangi2017revisiting} Because so much of the chemistry of the cuprates is consistent with off-plane disorder, and because strong, pointlike potentials are rarely encountered in BSCCO, it is a bit of a mystery why the observed QPI exhibits large-momenta peaks. It is of course entirely possible that these effects occur in tandem with each other---smooth potentials cause the finite DOS, while relatively weak pointlike disorder causes QPI---but a full resolution still awaits, and possibly requires a much more microscopic modeling of the tunneling process.\cite{kreisel2015interpretation}

We additionally caution the reader that our work has focused on strictly two-dimensional $d$-wave superconductors, and as such we have neglected the effects of coupling to the third dimension. The suppression of the DOS in the presence of in-plane pointlike disorder has been shown in field-theoretical work to occur strictly in 2D, and the logarithmic divergences responsible for this effect are cut off when interlayer coupling is included.\cite{nersesyan1994disorder} The observed dips we see in the in-plane disorder cases would be lost the more three-dimensional the system becomes, and this leaves open the possibility that, in the presence of interlayer coupling, this finite DOS could be due in part to the presence of pointlike forms of disorder. We thus stress that our results do not by any means suggest that smooth disorder is the be-all and end-all cause of the finite DOS at the Fermi energy. However, as noted earlier, YBCO is noted to have clean CuO$_2$ planes, so any influence of in-plane disorder on the DOS is likely to be very weak, regardless of the presence of interlayer coupling.

The possibility that the finite DOS at the Fermi energy in the superconducting state of the cuprates is due to disorder---smooth disorder, in particular---does not leave other explanations wanting, however, and one should not rule these out completely. It is possible that disorder is present alongside other, more exotic effects involving strong interactions (quantum criticality, for instance). In such a scenario there would be even more broadening involved. When the self-energies incorporating both disorder and interactions contain a nontrivial dependence on frequence or temperature, numerous interesting effects could conceivably occur. It would be interesting to see if alternative explanations invoking, say, quantum criticality or coexisting order result in the preservation of crucial aspects of the $d$-wave state, as the smooth-disorder scenario does.

On a completely different note, our results suggest a number of avenues for future work. First, the incorporation of full self-consistency is one possibility, albeit a very technically challenging one, at least from the point of view of our methods. While self-consistency \emph{may} not be completely necessary---it might very well be that the superconductivity in the cuprates is decidedly non-BCS-like---it would be very interesting to see how smooth disorder affects the superconducting order parameter. The non-self-consistent results in this paper suggest that smooth disorder has a far more muted effect on the single-particle properties of the $d$-wave superconductor than unitary-scatterer or random-potential disorder, so it is reasonable to guess that a fully self-consistent treatment would result in the preservation of $d$-wave superconductivity up to very high off-plane impurity concentrations, and consequently a large $T_c$ even when the superconductor is disordered. A second possibility is to revisit the exact calculation of the superfluid stiffness, $T_c$, and optical conductivity in the superconducting state\cite{scalapino1992superfluid, scalapino1993insulator} in the presence of off-plane disorder, and to examine if superconductivity is ever destroyed by smooth disorder. Our results suggest that even something as relatively anodyne as disorder---especially a relatively overlooked form of disorder like off-plane dopants---can produce surprisingly rich physics that accounts for many observed experimental properties of the cuprate high-temperature superconductors.

\begin{acknowledgments}
We thank M. P. Allan, M.-H. Julien, and R.-J. Slager for very helpful discussions. This work was supported by the Netherlands Organisation for Scientific Research (NWO/OCW) as part of the Frontiers of Nanoscience (NanoFront) program.
\end{acknowledgments}

\nocite{*}

\bibliography{paper_QDOS}

\providecommand{\noopsort}[1]{}\providecommand{\singleletter}[1]{#1}%
\begin{thebibliography}{64}%
\makeatletter
\providecommand \@ifxundefined [1]{%
 \@ifx{#1\undefined}
}%
\providecommand \@ifnum [1]{%
 \ifnum #1\expandafter \@firstoftwo
 \else \expandafter \@secondoftwo
 \fi
}%
\providecommand \@ifx [1]{%
 \ifx #1\expandafter \@firstoftwo
 \else \expandafter \@secondoftwo
 \fi
}%
\providecommand \natexlab [1]{#1}%
\providecommand \enquote  [1]{``#1''}%
\providecommand \bibnamefont  [1]{#1}%
\providecommand \bibfnamefont [1]{#1}%
\providecommand \citenamefont [1]{#1}%
\providecommand \href@noop [0]{\@secondoftwo}%
\providecommand \href [0]{\begingroup \@sanitize@url \@href}%
\providecommand \@href[1]{\@@startlink{#1}\@@href}%
\providecommand \@@href[1]{\endgroup#1\@@endlink}%
\providecommand \@sanitize@url [0]{\catcode `\\12\catcode `\$12\catcode
  `\&12\catcode `\#12\catcode `\^12\catcode `\_12\catcode `\%12\relax}%
\providecommand \@@startlink[1]{}%
\providecommand \@@endlink[0]{}%
\providecommand \url  [0]{\begingroup\@sanitize@url \@url }%
\providecommand \@url [1]{\endgroup\@href {#1}{\urlprefix }}%
\providecommand \urlprefix  [0]{URL }%
\providecommand \Eprint [0]{\href }%
\providecommand \doibase [0]{http://dx.doi.org/}%
\providecommand \selectlanguage [0]{\@gobble}%
\providecommand \bibinfo  [0]{\@secondoftwo}%
\providecommand \bibfield  [0]{\@secondoftwo}%
\providecommand \translation [1]{[#1]}%
\providecommand \BibitemOpen [0]{}%
\providecommand \bibitemStop [0]{}%
\providecommand \bibitemNoStop [0]{.\EOS\space}%
\providecommand \EOS [0]{\spacefactor3000\relax}%
\providecommand \BibitemShut  [1]{\csname bibitem#1\endcsname}%
\let\auto@bib@innerbib\@empty
\bibitem [{\citenamefont {Hoffman}\ \emph {et~al.}(2002)\citenamefont
  {Hoffman}, \citenamefont {McElroy}, \citenamefont {Lee}, \citenamefont
  {Lang}, \citenamefont {Eisaki}, \citenamefont {Uchida},\ and\ \citenamefont
  {Davis}}]{hoffman2002imaging}%
  \BibitemOpen
  \bibfield  {author} {\bibinfo {author} {\bibfnamefont {J.~E.}\ \bibnamefont
  {Hoffman}}, \bibinfo {author} {\bibfnamefont {K.}~\bibnamefont {McElroy}},
  \bibinfo {author} {\bibfnamefont {D.-H.}\ \bibnamefont {Lee}}, \bibinfo
  {author} {\bibfnamefont {K.~M.}\ \bibnamefont {Lang}}, \bibinfo {author}
  {\bibfnamefont {H.}~\bibnamefont {Eisaki}}, \bibinfo {author} {\bibfnamefont
  {S.}~\bibnamefont {Uchida}}, \ and\ \bibinfo {author} {\bibfnamefont {J.~C.}\
  \bibnamefont {Davis}},\ }\href@noop {} {\bibfield  {journal} {\bibinfo
  {journal} {Science}\ }\textbf {\bibinfo {volume} {297}},\ \bibinfo {pages}
  {1148} (\bibinfo {year} {2002})}\BibitemShut {NoStop}%
\bibitem [{\citenamefont {McElroy}\ \emph {et~al.}(2003)\citenamefont
  {McElroy}, \citenamefont {Simmonds}, \citenamefont {Hoffman}, \citenamefont
  {Lee}, \citenamefont {Orenstein}, \citenamefont {Eisaki}, \citenamefont
  {Uchida},\ and\ \citenamefont {Davis}}]{mcelroy2003relating}%
  \BibitemOpen
  \bibfield  {author} {\bibinfo {author} {\bibfnamefont {K.}~\bibnamefont
  {McElroy}}, \bibinfo {author} {\bibfnamefont {R.~W.}\ \bibnamefont
  {Simmonds}}, \bibinfo {author} {\bibfnamefont {J.~E.}\ \bibnamefont
  {Hoffman}}, \bibinfo {author} {\bibfnamefont {D.-H.}\ \bibnamefont {Lee}},
  \bibinfo {author} {\bibfnamefont {J.}~\bibnamefont {Orenstein}}, \bibinfo
  {author} {\bibfnamefont {H.}~\bibnamefont {Eisaki}}, \bibinfo {author}
  {\bibfnamefont {S.}~\bibnamefont {Uchida}}, \ and\ \bibinfo {author}
  {\bibfnamefont {J.~C.}\ \bibnamefont {Davis}},\ }\href@noop {} {\bibfield
  {journal} {\bibinfo  {journal} {Nature}\ }\textbf {\bibinfo {volume} {422}},\
  \bibinfo {pages} {592} (\bibinfo {year} {2003})}\BibitemShut {NoStop}%
\bibitem [{\citenamefont {Kohsaka}\ \emph {et~al.}(2008)\citenamefont
  {Kohsaka}, \citenamefont {Taylor}, \citenamefont {Wahl}, \citenamefont
  {Schmidt}, \citenamefont {Lee}, \citenamefont {Fujita}, \citenamefont
  {Alldredge}, \citenamefont {McElroy}, \citenamefont {Lee}, \citenamefont
  {Eisaki}, \citenamefont {Uchida}, \citenamefont {Lee},\ and\ \citenamefont
  {Davis}}]{kohsaka2008cooper}%
  \BibitemOpen
  \bibfield  {author} {\bibinfo {author} {\bibfnamefont {Y.}~\bibnamefont
  {Kohsaka}}, \bibinfo {author} {\bibfnamefont {C.}~\bibnamefont {Taylor}},
  \bibinfo {author} {\bibfnamefont {P.}~\bibnamefont {Wahl}}, \bibinfo {author}
  {\bibfnamefont {A.}~\bibnamefont {Schmidt}}, \bibinfo {author} {\bibfnamefont
  {J.}~\bibnamefont {Lee}}, \bibinfo {author} {\bibfnamefont {K.}~\bibnamefont
  {Fujita}}, \bibinfo {author} {\bibfnamefont {J.~W.}\ \bibnamefont
  {Alldredge}}, \bibinfo {author} {\bibfnamefont {K.}~\bibnamefont {McElroy}},
  \bibinfo {author} {\bibfnamefont {J.}~\bibnamefont {Lee}}, \bibinfo {author}
  {\bibfnamefont {H.}~\bibnamefont {Eisaki}}, \bibinfo {author} {\bibfnamefont
  {S.}~\bibnamefont {Uchida}}, \bibinfo {author} {\bibfnamefont {D.-H.}\
  \bibnamefont {Lee}}, \ and\ \bibinfo {author} {\bibfnamefont {J.~C.}\
  \bibnamefont {Davis}},\ }\href@noop {} {\bibfield  {journal} {\bibinfo
  {journal} {Nature}\ }\textbf {\bibinfo {volume} {454}},\ \bibinfo {pages}
  {1072} (\bibinfo {year} {2008})}\BibitemShut {NoStop}%
\bibitem [{\citenamefont {Lee}\ \emph {et~al.}(2009)\citenamefont {Lee},
  \citenamefont {Fujita}, \citenamefont {Schmidt}, \citenamefont {Kim},
  \citenamefont {Eisaki}, \citenamefont {Uchida},\ and\ \citenamefont
  {Davis}}]{lee2009spectroscopic}%
  \BibitemOpen
  \bibfield  {author} {\bibinfo {author} {\bibfnamefont {J.}~\bibnamefont
  {Lee}}, \bibinfo {author} {\bibfnamefont {K.}~\bibnamefont {Fujita}},
  \bibinfo {author} {\bibfnamefont {A.~R.}\ \bibnamefont {Schmidt}}, \bibinfo
  {author} {\bibfnamefont {C.~K.}\ \bibnamefont {Kim}}, \bibinfo {author}
  {\bibfnamefont {H.}~\bibnamefont {Eisaki}}, \bibinfo {author} {\bibfnamefont
  {S.}~\bibnamefont {Uchida}}, \ and\ \bibinfo {author} {\bibfnamefont {J.~C.}\
  \bibnamefont {Davis}},\ }\href@noop {} {\bibfield  {journal} {\bibinfo
  {journal} {Science}\ }\textbf {\bibinfo {volume} {325}},\ \bibinfo {pages}
  {1099} (\bibinfo {year} {2009})}\BibitemShut {NoStop}%
\bibitem [{\citenamefont {Wang}\ and\ \citenamefont
  {Lee}(2003)}]{wang2003quasiparticle}%
  \BibitemOpen
  \bibfield  {author} {\bibinfo {author} {\bibfnamefont {Q.-H.}\ \bibnamefont
  {Wang}}\ and\ \bibinfo {author} {\bibfnamefont {D.-H.}\ \bibnamefont {Lee}},\
  }\href@noop {} {\bibfield  {journal} {\bibinfo  {journal} {Phys. Rev. B}\
  }\textbf {\bibinfo {volume} {67}},\ \bibinfo {pages} {020511} (\bibinfo
  {year} {2003})}\BibitemShut {NoStop}%
\bibitem [{\citenamefont {Capriotti}\ \emph {et~al.}(2003)\citenamefont
  {Capriotti}, \citenamefont {Scalapino},\ and\ \citenamefont
  {Sedgewick}}]{capriotti2003wave}%
  \BibitemOpen
  \bibfield  {author} {\bibinfo {author} {\bibfnamefont {L.}~\bibnamefont
  {Capriotti}}, \bibinfo {author} {\bibfnamefont {D.~J.}\ \bibnamefont
  {Scalapino}}, \ and\ \bibinfo {author} {\bibfnamefont {R.~D.}\ \bibnamefont
  {Sedgewick}},\ }\href@noop {} {\bibfield  {journal} {\bibinfo  {journal}
  {Phys. Rev. B}\ }\textbf {\bibinfo {volume} {68}},\ \bibinfo {pages} {014508}
  (\bibinfo {year} {2003})}\BibitemShut {NoStop}%
\bibitem [{\citenamefont {Zhu}\ \emph {et~al.}(2004)\citenamefont {Zhu},
  \citenamefont {Atkinson},\ and\ \citenamefont {Hirschfeld}}]{zhu2004power}%
  \BibitemOpen
  \bibfield  {author} {\bibinfo {author} {\bibfnamefont {L.}~\bibnamefont
  {Zhu}}, \bibinfo {author} {\bibfnamefont {W.~A.}\ \bibnamefont {Atkinson}}, \
  and\ \bibinfo {author} {\bibfnamefont {P.~J.}\ \bibnamefont {Hirschfeld}},\
  }\href@noop {} {\bibfield  {journal} {\bibinfo  {journal} {Phys. Rev. B}\
  }\textbf {\bibinfo {volume} {69}},\ \bibinfo {pages} {060503} (\bibinfo
  {year} {2004})}\BibitemShut {NoStop}%
\bibitem [{\citenamefont {Nunner}\ \emph {et~al.}(2006)\citenamefont {Nunner},
  \citenamefont {Chen}, \citenamefont {Andersen}, \citenamefont {Melikyan},\
  and\ \citenamefont {Hirschfeld}}]{nunner2006fourier}%
  \BibitemOpen
  \bibfield  {author} {\bibinfo {author} {\bibfnamefont {T.~S.}\ \bibnamefont
  {Nunner}}, \bibinfo {author} {\bibfnamefont {W.}~\bibnamefont {Chen}},
  \bibinfo {author} {\bibfnamefont {B.~M.}\ \bibnamefont {Andersen}}, \bibinfo
  {author} {\bibfnamefont {A.}~\bibnamefont {Melikyan}}, \ and\ \bibinfo
  {author} {\bibfnamefont {P.~J.}\ \bibnamefont {Hirschfeld}},\ }\href@noop {}
  {\bibfield  {journal} {\bibinfo  {journal} {Phys. Rev. B}\ }\textbf {\bibinfo
  {volume} {73}},\ \bibinfo {pages} {104511} (\bibinfo {year}
  {2006})}\BibitemShut {NoStop}%
\bibitem [{\citenamefont {Vishik}\ \emph {et~al.}(2009)\citenamefont {Vishik},
  \citenamefont {Nowadnick}, \citenamefont {Lee}, \citenamefont {Shen},
  \citenamefont {Moritz}, \citenamefont {Devereaux}, \citenamefont {Tanaka},
  \citenamefont {Sasagawa},\ and\ \citenamefont {Fujii}}]{vishik2009momentum}%
  \BibitemOpen
  \bibfield  {author} {\bibinfo {author} {\bibfnamefont {I.~M.}\ \bibnamefont
  {Vishik}}, \bibinfo {author} {\bibfnamefont {E.~A.}\ \bibnamefont
  {Nowadnick}}, \bibinfo {author} {\bibfnamefont {W.~S.}\ \bibnamefont {Lee}},
  \bibinfo {author} {\bibfnamefont {Z.~X.}\ \bibnamefont {Shen}}, \bibinfo
  {author} {\bibfnamefont {B.}~\bibnamefont {Moritz}}, \bibinfo {author}
  {\bibfnamefont {T.~P.}\ \bibnamefont {Devereaux}}, \bibinfo {author}
  {\bibfnamefont {K.}~\bibnamefont {Tanaka}}, \bibinfo {author} {\bibfnamefont
  {T.}~\bibnamefont {Sasagawa}}, \ and\ \bibinfo {author} {\bibfnamefont
  {T.}~\bibnamefont {Fujii}},\ }\href@noop {} {\bibfield  {journal} {\bibinfo
  {journal} {Nature Physics}\ }\textbf {\bibinfo {volume} {5}},\ \bibinfo
  {pages} {718} (\bibinfo {year} {2009})}\BibitemShut {NoStop}%
\bibitem [{\citenamefont {Kreisel}\ \emph {et~al.}(2015)\citenamefont
  {Kreisel}, \citenamefont {Choubey}, \citenamefont {Berlijn}, \citenamefont
  {Ku}, \citenamefont {Andersen},\ and\ \citenamefont
  {Hirschfeld}}]{kreisel2015interpretation}%
  \BibitemOpen
  \bibfield  {author} {\bibinfo {author} {\bibfnamefont {A.}~\bibnamefont
  {Kreisel}}, \bibinfo {author} {\bibfnamefont {P.}~\bibnamefont {Choubey}},
  \bibinfo {author} {\bibfnamefont {T.}~\bibnamefont {Berlijn}}, \bibinfo
  {author} {\bibfnamefont {W.}~\bibnamefont {Ku}}, \bibinfo {author}
  {\bibfnamefont {B.~M.}\ \bibnamefont {Andersen}}, \ and\ \bibinfo {author}
  {\bibfnamefont {P.~J.}\ \bibnamefont {Hirschfeld}},\ }\href@noop {}
  {\bibfield  {journal} {\bibinfo  {journal} {Phys. Rev. Lett.}\ }\textbf
  {\bibinfo {volume} {114}},\ \bibinfo {pages} {217002} (\bibinfo {year}
  {2015})}\BibitemShut {NoStop}%
\bibitem [{\citenamefont {Sulangi}\ \emph {et~al.}(2017)\citenamefont
  {Sulangi}, \citenamefont {Allan},\ and\ \citenamefont
  {Zaanen}}]{sulangi2017revisiting}%
  \BibitemOpen
  \bibfield  {author} {\bibinfo {author} {\bibfnamefont {M.~A.}\ \bibnamefont
  {Sulangi}}, \bibinfo {author} {\bibfnamefont {M.~P.}\ \bibnamefont {Allan}},
  \ and\ \bibinfo {author} {\bibfnamefont {J.}~\bibnamefont {Zaanen}},\
  }\href@noop {} {\bibfield  {journal} {\bibinfo  {journal} {Phys. Rev. B}\
  }\textbf {\bibinfo {volume} {96}},\ \bibinfo {pages} {134507} (\bibinfo
  {year} {2017})}\BibitemShut {NoStop}%
\bibitem [{\citenamefont {Gorkov}\ and\ \citenamefont
  {Kalugin}(1985)}]{gorkov1985defects}%
  \BibitemOpen
  \bibfield  {author} {\bibinfo {author} {\bibfnamefont {L.~P.}\ \bibnamefont
  {Gorkov}}\ and\ \bibinfo {author} {\bibfnamefont {P.~A.}\ \bibnamefont
  {Kalugin}},\ }\href@noop {} {\bibfield  {journal} {\bibinfo  {journal} {JETP
  Letters}\ }\textbf {\bibinfo {volume} {41}},\ \bibinfo {pages} {253}
  (\bibinfo {year} {1985})}\BibitemShut {NoStop}%
\bibitem [{\citenamefont {Hirschfeld}\ \emph {et~al.}(1988)\citenamefont
  {Hirschfeld}, \citenamefont {W{\"o}lfle},\ and\ \citenamefont
  {Einzel}}]{hirschfeld1988consequences}%
  \BibitemOpen
  \bibfield  {author} {\bibinfo {author} {\bibfnamefont {P.~J.}\ \bibnamefont
  {Hirschfeld}}, \bibinfo {author} {\bibfnamefont {P.}~\bibnamefont
  {W{\"o}lfle}}, \ and\ \bibinfo {author} {\bibfnamefont {D.}~\bibnamefont
  {Einzel}},\ }\href@noop {} {\bibfield  {journal} {\bibinfo  {journal} {Phys.
  Rev. B}\ }\textbf {\bibinfo {volume} {37}},\ \bibinfo {pages} {83} (\bibinfo
  {year} {1988})}\BibitemShut {NoStop}%
\bibitem [{\citenamefont {Lee}(1993)}]{lee1993localized}%
  \BibitemOpen
  \bibfield  {author} {\bibinfo {author} {\bibfnamefont {P.~A.}\ \bibnamefont
  {Lee}},\ }\href@noop {} {\bibfield  {journal} {\bibinfo  {journal} {Phys.
  Rev. Lett.}\ }\textbf {\bibinfo {volume} {71}},\ \bibinfo {pages} {1887}
  (\bibinfo {year} {1993})}\BibitemShut {NoStop}%
\bibitem [{\citenamefont {Durst}\ and\ \citenamefont
  {Lee}(2000)}]{durst2000impurity}%
  \BibitemOpen
  \bibfield  {author} {\bibinfo {author} {\bibfnamefont {A.~C.}\ \bibnamefont
  {Durst}}\ and\ \bibinfo {author} {\bibfnamefont {P.~A.}\ \bibnamefont
  {Lee}},\ }\href@noop {} {\bibfield  {journal} {\bibinfo  {journal} {Phys.
  Rev. B}\ }\textbf {\bibinfo {volume} {62}},\ \bibinfo {pages} {1270}
  (\bibinfo {year} {2000})}\BibitemShut {NoStop}%
\bibitem [{\citenamefont {Yashenkin}\ \emph {et~al.}(2001)\citenamefont
  {Yashenkin}, \citenamefont {Atkinson}, \citenamefont {Gornyi}, \citenamefont
  {Hirschfeld},\ and\ \citenamefont {Khveshchenko}}]{yashenkin2001nesting}%
  \BibitemOpen
  \bibfield  {author} {\bibinfo {author} {\bibfnamefont {A.~G.}\ \bibnamefont
  {Yashenkin}}, \bibinfo {author} {\bibfnamefont {W.~A.}\ \bibnamefont
  {Atkinson}}, \bibinfo {author} {\bibfnamefont {I.~V.}\ \bibnamefont
  {Gornyi}}, \bibinfo {author} {\bibfnamefont {P.~J.}\ \bibnamefont
  {Hirschfeld}}, \ and\ \bibinfo {author} {\bibfnamefont {D.~V.}\ \bibnamefont
  {Khveshchenko}},\ }\href@noop {} {\bibfield  {journal} {\bibinfo  {journal}
  {Phys. Rev. Lett.}\ }\textbf {\bibinfo {volume} {86}},\ \bibinfo {pages}
  {5982} (\bibinfo {year} {2001})}\BibitemShut {NoStop}%
\bibitem [{\citenamefont {Nersesyan}\ \emph {et~al.}(1994)\citenamefont
  {Nersesyan}, \citenamefont {Tsvelik},\ and\ \citenamefont
  {Wenger}}]{nersesyan1994disorder}%
  \BibitemOpen
  \bibfield  {author} {\bibinfo {author} {\bibfnamefont {A.~A.}\ \bibnamefont
  {Nersesyan}}, \bibinfo {author} {\bibfnamefont {A.~M.}\ \bibnamefont
  {Tsvelik}}, \ and\ \bibinfo {author} {\bibfnamefont {F.}~\bibnamefont
  {Wenger}},\ }\href@noop {} {\bibfield  {journal} {\bibinfo  {journal} {Phys.
  Rev. Lett.}\ }\textbf {\bibinfo {volume} {72}},\ \bibinfo {pages} {2628}
  (\bibinfo {year} {1994})}\BibitemShut {NoStop}%
\bibitem [{\citenamefont {Senthil}\ \emph {et~al.}(1998)\citenamefont
  {Senthil}, \citenamefont {Fisher}, \citenamefont {Balents},\ and\
  \citenamefont {Nayak}}]{senthil1998quasiparticle}%
  \BibitemOpen
  \bibfield  {author} {\bibinfo {author} {\bibfnamefont {T.}~\bibnamefont
  {Senthil}}, \bibinfo {author} {\bibfnamefont {M.~P.~A.}\ \bibnamefont
  {Fisher}}, \bibinfo {author} {\bibfnamefont {L.}~\bibnamefont {Balents}}, \
  and\ \bibinfo {author} {\bibfnamefont {C.}~\bibnamefont {Nayak}},\
  }\href@noop {} {\bibfield  {journal} {\bibinfo  {journal} {Phys. Rev. Lett.}\
  }\textbf {\bibinfo {volume} {81}},\ \bibinfo {pages} {4704} (\bibinfo {year}
  {1998})}\BibitemShut {NoStop}%
\bibitem [{\citenamefont {Senthil}\ and\ \citenamefont
  {Fisher}(1999)}]{senthil1999quasiparticle}%
  \BibitemOpen
  \bibfield  {author} {\bibinfo {author} {\bibfnamefont {T.}~\bibnamefont
  {Senthil}}\ and\ \bibinfo {author} {\bibfnamefont {M.~P.~A.}\ \bibnamefont
  {Fisher}},\ }\href@noop {} {\bibfield  {journal} {\bibinfo  {journal} {Phys.
  Rev. B}\ }\textbf {\bibinfo {volume} {60}},\ \bibinfo {pages} {6893}
  (\bibinfo {year} {1999})}\BibitemShut {NoStop}%
\bibitem [{\citenamefont {Altland}\ \emph {et~al.}(2002)\citenamefont
  {Altland}, \citenamefont {Simons},\ and\ \citenamefont
  {Zirnbauer}}]{altland2002theories}%
  \BibitemOpen
  \bibfield  {author} {\bibinfo {author} {\bibfnamefont {A.}~\bibnamefont
  {Altland}}, \bibinfo {author} {\bibfnamefont {B.~D.}\ \bibnamefont {Simons}},
  \ and\ \bibinfo {author} {\bibfnamefont {M.~R.}\ \bibnamefont {Zirnbauer}},\
  }\href@noop {} {\bibfield  {journal} {\bibinfo  {journal} {Physics Reports}\
  }\textbf {\bibinfo {volume} {359}},\ \bibinfo {pages} {283} (\bibinfo {year}
  {2002})}\BibitemShut {NoStop}%
\bibitem [{\citenamefont {Moler}\ \emph {et~al.}(1994)\citenamefont {Moler},
  \citenamefont {Baar}, \citenamefont {Urbach}, \citenamefont {Liang},
  \citenamefont {Hardy},\ and\ \citenamefont {Kapitulnik}}]{moler1994magnetic}%
  \BibitemOpen
  \bibfield  {author} {\bibinfo {author} {\bibfnamefont {K.~A.}\ \bibnamefont
  {Moler}}, \bibinfo {author} {\bibfnamefont {D.~J.}\ \bibnamefont {Baar}},
  \bibinfo {author} {\bibfnamefont {J.~S.}\ \bibnamefont {Urbach}}, \bibinfo
  {author} {\bibfnamefont {R.}~\bibnamefont {Liang}}, \bibinfo {author}
  {\bibfnamefont {W.~N.}\ \bibnamefont {Hardy}}, \ and\ \bibinfo {author}
  {\bibfnamefont {A.}~\bibnamefont {Kapitulnik}},\ }\href@noop {} {\bibfield
  {journal} {\bibinfo  {journal} {Phys. Rev. Lett.}\ }\textbf {\bibinfo
  {volume} {73}},\ \bibinfo {pages} {2744} (\bibinfo {year}
  {1994})}\BibitemShut {NoStop}%
\bibitem [{\citenamefont {Moler}\ \emph {et~al.}(1997)\citenamefont {Moler},
  \citenamefont {Sisson}, \citenamefont {Urbach}, \citenamefont {Beasley},
  \citenamefont {Kapitulnik}, \citenamefont {Baar}, \citenamefont {Liang},\
  and\ \citenamefont {Hardy}}]{moler1997specific}%
  \BibitemOpen
  \bibfield  {author} {\bibinfo {author} {\bibfnamefont {K.~A.}\ \bibnamefont
  {Moler}}, \bibinfo {author} {\bibfnamefont {D.~L.}\ \bibnamefont {Sisson}},
  \bibinfo {author} {\bibfnamefont {J.~S.}\ \bibnamefont {Urbach}}, \bibinfo
  {author} {\bibfnamefont {M.~R.}\ \bibnamefont {Beasley}}, \bibinfo {author}
  {\bibfnamefont {A.}~\bibnamefont {Kapitulnik}}, \bibinfo {author}
  {\bibfnamefont {D.~J.}\ \bibnamefont {Baar}}, \bibinfo {author}
  {\bibfnamefont {R.}~\bibnamefont {Liang}}, \ and\ \bibinfo {author}
  {\bibfnamefont {W.~N.}\ \bibnamefont {Hardy}},\ }\href@noop {} {\bibfield
  {journal} {\bibinfo  {journal} {Phys. Rev. B}\ }\textbf {\bibinfo {volume}
  {55}},\ \bibinfo {pages} {3954} (\bibinfo {year} {1997})}\BibitemShut
  {NoStop}%
\bibitem [{\citenamefont {Riggs}\ \emph {et~al.}(2011)\citenamefont {Riggs},
  \citenamefont {Vafek}, \citenamefont {Kemper}, \citenamefont {Betts},
  \citenamefont {Migliori}, \citenamefont {Balakirev}, \citenamefont {Hardy},
  \citenamefont {Liang}, \citenamefont {Bonn},\ and\ \citenamefont
  {Boebinger}}]{riggs2011heat}%
  \BibitemOpen
  \bibfield  {author} {\bibinfo {author} {\bibfnamefont {S.~C.}\ \bibnamefont
  {Riggs}}, \bibinfo {author} {\bibfnamefont {O.}~\bibnamefont {Vafek}},
  \bibinfo {author} {\bibfnamefont {J.~B.}\ \bibnamefont {Kemper}}, \bibinfo
  {author} {\bibfnamefont {J.~B.}\ \bibnamefont {Betts}}, \bibinfo {author}
  {\bibfnamefont {A.}~\bibnamefont {Migliori}}, \bibinfo {author}
  {\bibfnamefont {F.~F.}\ \bibnamefont {Balakirev}}, \bibinfo {author}
  {\bibfnamefont {W.~N.}\ \bibnamefont {Hardy}}, \bibinfo {author}
  {\bibfnamefont {R.}~\bibnamefont {Liang}}, \bibinfo {author} {\bibfnamefont
  {D.~A.}\ \bibnamefont {Bonn}}, \ and\ \bibinfo {author} {\bibfnamefont
  {G.~S.}\ \bibnamefont {Boebinger}},\ }\href@noop {} {\bibfield  {journal}
  {\bibinfo  {journal} {Nature Physics}\ }\textbf {\bibinfo {volume} {7}},\
  \bibinfo {pages} {332} (\bibinfo {year} {2011})}\BibitemShut {NoStop}%
\bibitem [{\citenamefont {Berg}\ \emph {et~al.}(2008)\citenamefont {Berg},
  \citenamefont {Chen},\ and\ \citenamefont {Kivelson}}]{berg2008stability}%
  \BibitemOpen
  \bibfield  {author} {\bibinfo {author} {\bibfnamefont {E.}~\bibnamefont
  {Berg}}, \bibinfo {author} {\bibfnamefont {C.-C.}\ \bibnamefont {Chen}}, \
  and\ \bibinfo {author} {\bibfnamefont {S.~A.}\ \bibnamefont {Kivelson}},\
  }\href@noop {} {\bibfield  {journal} {\bibinfo  {journal} {Phys. Rev. Lett.}\
  }\textbf {\bibinfo {volume} {100}},\ \bibinfo {pages} {027003} (\bibinfo
  {year} {2008})}\BibitemShut {NoStop}%
\bibitem [{\citenamefont {Allais}\ and\ \citenamefont
  {Senthil}(2012)}]{allais2012loop}%
  \BibitemOpen
  \bibfield  {author} {\bibinfo {author} {\bibfnamefont {A.}~\bibnamefont
  {Allais}}\ and\ \bibinfo {author} {\bibfnamefont {T.}~\bibnamefont
  {Senthil}},\ }\href@noop {} {\bibfield  {journal} {\bibinfo  {journal} {Phys.
  Rev. B}\ }\textbf {\bibinfo {volume} {86}},\ \bibinfo {pages} {045118}
  (\bibinfo {year} {2012})}\BibitemShut {NoStop}%
\bibitem [{\citenamefont {Kivelson}\ and\ \citenamefont
  {Varma}(2012)}]{kivelson2012fermi}%
  \BibitemOpen
  \bibfield  {author} {\bibinfo {author} {\bibfnamefont {S.~A.}\ \bibnamefont
  {Kivelson}}\ and\ \bibinfo {author} {\bibfnamefont {C.~M.}\ \bibnamefont
  {Varma}},\ }\href@noop {} {\bibfield  {journal} {\bibinfo  {journal} {arXiv
  preprint arXiv:1208.6498}\ } (\bibinfo {year} {2012})}\BibitemShut {NoStop}%
\bibitem [{\citenamefont {Wang}\ and\ \citenamefont
  {Vafek}(2013)}]{wang2013quantum}%
  \BibitemOpen
  \bibfield  {author} {\bibinfo {author} {\bibfnamefont {L.}~\bibnamefont
  {Wang}}\ and\ \bibinfo {author} {\bibfnamefont {O.}~\bibnamefont {Vafek}},\
  }\href@noop {} {\bibfield  {journal} {\bibinfo  {journal} {Phys. Rev. B}\
  }\textbf {\bibinfo {volume} {88}},\ \bibinfo {pages} {024506} (\bibinfo
  {year} {2013})}\BibitemShut {NoStop}%
\bibitem [{\citenamefont {Collocott}\ and\ \citenamefont
  {Driver}(1990)}]{collocott1990specific}%
  \BibitemOpen
  \bibfield  {author} {\bibinfo {author} {\bibfnamefont {S.~J.}\ \bibnamefont
  {Collocott}}\ and\ \bibinfo {author} {\bibfnamefont {R.}~\bibnamefont
  {Driver}},\ }\href@noop {} {\bibfield  {journal} {\bibinfo  {journal}
  {Physica C: Superconductivity}\ }\textbf {\bibinfo {volume} {167}},\ \bibinfo
  {pages} {598} (\bibinfo {year} {1990})}\BibitemShut {NoStop}%
\bibitem [{\citenamefont {Junod}\ \emph {et~al.}(1994)\citenamefont {Junod},
  \citenamefont {Wang}, \citenamefont {Tsukamoto}, \citenamefont {Triscone},
  \citenamefont {Revaz}, \citenamefont {Walker},\ and\ \citenamefont
  {Muller}}]{junod1994specific}%
  \BibitemOpen
  \bibfield  {author} {\bibinfo {author} {\bibfnamefont {A.}~\bibnamefont
  {Junod}}, \bibinfo {author} {\bibfnamefont {K.-Q.}\ \bibnamefont {Wang}},
  \bibinfo {author} {\bibfnamefont {T.}~\bibnamefont {Tsukamoto}}, \bibinfo
  {author} {\bibfnamefont {G.}~\bibnamefont {Triscone}}, \bibinfo {author}
  {\bibfnamefont {B.}~\bibnamefont {Revaz}}, \bibinfo {author} {\bibfnamefont
  {E.}~\bibnamefont {Walker}}, \ and\ \bibinfo {author} {\bibfnamefont
  {J.}~\bibnamefont {Muller}},\ }\href@noop {} {\bibfield  {journal} {\bibinfo
  {journal} {Physica C: Superconductivity}\ }\textbf {\bibinfo {volume}
  {229}},\ \bibinfo {pages} {209} (\bibinfo {year} {1994})}\BibitemShut
  {NoStop}%
\bibitem [{\citenamefont {Urbach}\ \emph {et~al.}(1989)\citenamefont {Urbach},
  \citenamefont {Mitzi}, \citenamefont {Kapitulnik}, \citenamefont {Wei},\ and\
  \citenamefont {Morris}}]{urbach1989low}%
  \BibitemOpen
  \bibfield  {author} {\bibinfo {author} {\bibfnamefont {J.~S.}\ \bibnamefont
  {Urbach}}, \bibinfo {author} {\bibfnamefont {D.~B.}\ \bibnamefont {Mitzi}},
  \bibinfo {author} {\bibfnamefont {A.}~\bibnamefont {Kapitulnik}}, \bibinfo
  {author} {\bibfnamefont {J.~Y.~T.}\ \bibnamefont {Wei}}, \ and\ \bibinfo
  {author} {\bibfnamefont {D.~E.}\ \bibnamefont {Morris}},\ }\href@noop {}
  {\bibfield  {journal} {\bibinfo  {journal} {Phys. Rev. B}\ }\textbf {\bibinfo
  {volume} {39}},\ \bibinfo {pages} {12391} (\bibinfo {year}
  {1989})}\BibitemShut {NoStop}%
\bibitem [{\citenamefont {Atkinson}\ \emph
  {et~al.}(2000{\natexlab{a}})\citenamefont {Atkinson}, \citenamefont
  {Hirschfeld}, \citenamefont {MacDonald},\ and\ \citenamefont
  {Ziegler}}]{atkinson2000details}%
  \BibitemOpen
  \bibfield  {author} {\bibinfo {author} {\bibfnamefont {W.~A.}\ \bibnamefont
  {Atkinson}}, \bibinfo {author} {\bibfnamefont {P.~J.}\ \bibnamefont
  {Hirschfeld}}, \bibinfo {author} {\bibfnamefont {A.~H.}\ \bibnamefont
  {MacDonald}}, \ and\ \bibinfo {author} {\bibfnamefont {K.}~\bibnamefont
  {Ziegler}},\ }\href@noop {} {\bibfield  {journal} {\bibinfo  {journal} {Phys.
  Rev. Lett.}\ }\textbf {\bibinfo {volume} {85}},\ \bibinfo {pages} {3926}
  (\bibinfo {year} {2000}{\natexlab{a}})}\BibitemShut {NoStop}%
\bibitem [{\citenamefont {Pan}\ \emph {et~al.}(2000)\citenamefont {Pan},
  \citenamefont {Hudson}, \citenamefont {Lang}, \citenamefont {Eisaki},
  \citenamefont {Uchida},\ and\ \citenamefont {Davis}}]{pan2000imaging}%
  \BibitemOpen
  \bibfield  {author} {\bibinfo {author} {\bibfnamefont {S.~H.}\ \bibnamefont
  {Pan}}, \bibinfo {author} {\bibfnamefont {E.~W.}\ \bibnamefont {Hudson}},
  \bibinfo {author} {\bibfnamefont {K.~M.}\ \bibnamefont {Lang}}, \bibinfo
  {author} {\bibfnamefont {H.}~\bibnamefont {Eisaki}}, \bibinfo {author}
  {\bibfnamefont {S.}~\bibnamefont {Uchida}}, \ and\ \bibinfo {author}
  {\bibfnamefont {J.~C.}\ \bibnamefont {Davis}},\ }\href@noop {} {\bibfield
  {journal} {\bibinfo  {journal} {Nature}\ }\textbf {\bibinfo {volume} {403}},\
  \bibinfo {pages} {746} (\bibinfo {year} {2000})}\BibitemShut {NoStop}%
\bibitem [{\citenamefont {McElroy}\ \emph {et~al.}(2005)\citenamefont
  {McElroy}, \citenamefont {Lee}, \citenamefont {Slezak}, \citenamefont {Lee},
  \citenamefont {Eisaki}, \citenamefont {Uchida},\ and\ \citenamefont
  {Davis}}]{mcelroy2005atomic}%
  \BibitemOpen
  \bibfield  {author} {\bibinfo {author} {\bibfnamefont {K.}~\bibnamefont
  {McElroy}}, \bibinfo {author} {\bibfnamefont {J.}~\bibnamefont {Lee}},
  \bibinfo {author} {\bibfnamefont {J.~A.}\ \bibnamefont {Slezak}}, \bibinfo
  {author} {\bibfnamefont {D.-H.}\ \bibnamefont {Lee}}, \bibinfo {author}
  {\bibfnamefont {H.}~\bibnamefont {Eisaki}}, \bibinfo {author} {\bibfnamefont
  {S.}~\bibnamefont {Uchida}}, \ and\ \bibinfo {author} {\bibfnamefont {J.~C.}\
  \bibnamefont {Davis}},\ }\href@noop {} {\bibfield  {journal} {\bibinfo
  {journal} {Science}\ }\textbf {\bibinfo {volume} {309}},\ \bibinfo {pages}
  {1048} (\bibinfo {year} {2005})}\BibitemShut {NoStop}%
\bibitem [{\citenamefont {Schmidt}\ \emph {et~al.}(2011)\citenamefont
  {Schmidt}, \citenamefont {Fujita}, \citenamefont {Kim}, \citenamefont
  {Lawler}, \citenamefont {Eisaki}, \citenamefont {Uchida}, \citenamefont
  {Lee},\ and\ \citenamefont {Davis}}]{schmidt2011electronic}%
  \BibitemOpen
  \bibfield  {author} {\bibinfo {author} {\bibfnamefont {A.~R.}\ \bibnamefont
  {Schmidt}}, \bibinfo {author} {\bibfnamefont {K.}~\bibnamefont {Fujita}},
  \bibinfo {author} {\bibfnamefont {E.-A.}\ \bibnamefont {Kim}}, \bibinfo
  {author} {\bibfnamefont {M.~J.}\ \bibnamefont {Lawler}}, \bibinfo {author}
  {\bibfnamefont {H.}~\bibnamefont {Eisaki}}, \bibinfo {author} {\bibfnamefont
  {S.}~\bibnamefont {Uchida}}, \bibinfo {author} {\bibfnamefont {D.-H.}\
  \bibnamefont {Lee}}, \ and\ \bibinfo {author} {\bibfnamefont {J.~C.}\
  \bibnamefont {Davis}},\ }\href@noop {} {\bibfield  {journal} {\bibinfo
  {journal} {New Journal of Physics}\ }\textbf {\bibinfo {volume} {13}},\
  \bibinfo {pages} {065014} (\bibinfo {year} {2011})}\BibitemShut {NoStop}%
\bibitem [{\citenamefont {Eisaki}\ \emph {et~al.}(2004)\citenamefont {Eisaki},
  \citenamefont {Kaneko}, \citenamefont {Feng}, \citenamefont {Damascelli},
  \citenamefont {Mang}, \citenamefont {Shen}, \citenamefont {Shen},\ and\
  \citenamefont {Greven}}]{eisaki2004effect}%
  \BibitemOpen
  \bibfield  {author} {\bibinfo {author} {\bibfnamefont {H.}~\bibnamefont
  {Eisaki}}, \bibinfo {author} {\bibfnamefont {N.}~\bibnamefont {Kaneko}},
  \bibinfo {author} {\bibfnamefont {D.~L.}\ \bibnamefont {Feng}}, \bibinfo
  {author} {\bibfnamefont {A.}~\bibnamefont {Damascelli}}, \bibinfo {author}
  {\bibfnamefont {P.~K.}\ \bibnamefont {Mang}}, \bibinfo {author}
  {\bibfnamefont {K.~M.}\ \bibnamefont {Shen}}, \bibinfo {author}
  {\bibfnamefont {Z.-X.}\ \bibnamefont {Shen}}, \ and\ \bibinfo {author}
  {\bibfnamefont {M.}~\bibnamefont {Greven}},\ }\href@noop {} {\bibfield
  {journal} {\bibinfo  {journal} {Phys. Rev. B}\ }\textbf {\bibinfo {volume}
  {69}},\ \bibinfo {pages} {064512} (\bibinfo {year} {2004})}\BibitemShut
  {NoStop}%
\bibitem [{\citenamefont {Nunner}\ \emph {et~al.}(2005)\citenamefont {Nunner},
  \citenamefont {Andersen}, \citenamefont {Melikyan},\ and\ \citenamefont
  {Hirschfeld}}]{nunner2005dopant}%
  \BibitemOpen
  \bibfield  {author} {\bibinfo {author} {\bibfnamefont {T.~S.}\ \bibnamefont
  {Nunner}}, \bibinfo {author} {\bibfnamefont {B.~M.}\ \bibnamefont
  {Andersen}}, \bibinfo {author} {\bibfnamefont {A.}~\bibnamefont {Melikyan}},
  \ and\ \bibinfo {author} {\bibfnamefont {P.~J.}\ \bibnamefont {Hirschfeld}},\
  }\href@noop {} {\bibfield  {journal} {\bibinfo  {journal} {Phys. Rev. Lett.}\
  }\textbf {\bibinfo {volume} {95}},\ \bibinfo {pages} {177003} (\bibinfo
  {year} {2005})}\BibitemShut {NoStop}%
\bibitem [{\citenamefont {Nunner}\ and\ \citenamefont
  {Hirschfeld}(2005)}]{nunner2005microwave}%
  \BibitemOpen
  \bibfield  {author} {\bibinfo {author} {\bibfnamefont {T.~S.}\ \bibnamefont
  {Nunner}}\ and\ \bibinfo {author} {\bibfnamefont {P.~J.}\ \bibnamefont
  {Hirschfeld}},\ }\href@noop {} {\bibfield  {journal} {\bibinfo  {journal}
  {Phys. Rev. B}\ }\textbf {\bibinfo {volume} {72}},\ \bibinfo {pages} {014514}
  (\bibinfo {year} {2005})}\BibitemShut {NoStop}%
\bibitem [{\citenamefont {Franz}\ \emph {et~al.}(1997)\citenamefont {Franz},
  \citenamefont {Kallin}, \citenamefont {Berlinsky},\ and\ \citenamefont
  {Salkola}}]{franz1997critical}%
  \BibitemOpen
  \bibfield  {author} {\bibinfo {author} {\bibfnamefont {M.}~\bibnamefont
  {Franz}}, \bibinfo {author} {\bibfnamefont {C.}~\bibnamefont {Kallin}},
  \bibinfo {author} {\bibfnamefont {A.~J.}\ \bibnamefont {Berlinsky}}, \ and\
  \bibinfo {author} {\bibfnamefont {M.~I.}\ \bibnamefont {Salkola}},\
  }\href@noop {} {\bibfield  {journal} {\bibinfo  {journal} {Phys. Rev. B}\
  }\textbf {\bibinfo {volume} {56}},\ \bibinfo {pages} {7882} (\bibinfo {year}
  {1997})}\BibitemShut {NoStop}%
\bibitem [{\citenamefont {Atkinson}\ \emph
  {et~al.}(2000{\natexlab{b}})\citenamefont {Atkinson}, \citenamefont
  {Hirschfeld},\ and\ \citenamefont {MacDonald}}]{atkinson2000gap}%
  \BibitemOpen
  \bibfield  {author} {\bibinfo {author} {\bibfnamefont {W.~A.}\ \bibnamefont
  {Atkinson}}, \bibinfo {author} {\bibfnamefont {P.~J.}\ \bibnamefont
  {Hirschfeld}}, \ and\ \bibinfo {author} {\bibfnamefont {A.~H.}\ \bibnamefont
  {MacDonald}},\ }\href@noop {} {\bibfield  {journal} {\bibinfo  {journal}
  {Phys. Rev. Lett.}\ }\textbf {\bibinfo {volume} {85}},\ \bibinfo {pages}
  {3922} (\bibinfo {year} {2000}{\natexlab{b}})}\BibitemShut {NoStop}%
\bibitem [{\citenamefont {Zhu}\ \emph {et~al.}(2000)\citenamefont {Zhu},
  \citenamefont {Sheng},\ and\ \citenamefont {Ting}}]{zhu2000quasiparticle}%
  \BibitemOpen
  \bibfield  {author} {\bibinfo {author} {\bibfnamefont {J.-X.}\ \bibnamefont
  {Zhu}}, \bibinfo {author} {\bibfnamefont {D.~N.}\ \bibnamefont {Sheng}}, \
  and\ \bibinfo {author} {\bibfnamefont {C.~S.}\ \bibnamefont {Ting}},\
  }\href@noop {} {\bibfield  {journal} {\bibinfo  {journal} {Phys. Rev. Lett.}\
  }\textbf {\bibinfo {volume} {85}},\ \bibinfo {pages} {4944} (\bibinfo {year}
  {2000})}\BibitemShut {NoStop}%
\bibitem [{\citenamefont {Godfrin}(1991)}]{godfrin1991method}%
  \BibitemOpen
  \bibfield  {author} {\bibinfo {author} {\bibfnamefont {E.~M.}\ \bibnamefont
  {Godfrin}},\ }\href@noop {} {\bibfield  {journal} {\bibinfo  {journal}
  {Journal of Physics: Condensed Matter}\ }\textbf {\bibinfo {volume} {3}},\
  \bibinfo {pages} {7843} (\bibinfo {year} {1991})}\BibitemShut {NoStop}%
\bibitem [{\citenamefont {Hod}\ \emph {et~al.}(2006)\citenamefont {Hod},
  \citenamefont {Peralta},\ and\ \citenamefont {Scuseria}}]{hod2006first}%
  \BibitemOpen
  \bibfield  {author} {\bibinfo {author} {\bibfnamefont {O.}~\bibnamefont
  {Hod}}, \bibinfo {author} {\bibfnamefont {J.~E.}\ \bibnamefont {Peralta}}, \
  and\ \bibinfo {author} {\bibfnamefont {G.~E.}\ \bibnamefont {Scuseria}},\
  }\href@noop {} {\bibfield  {journal} {\bibinfo  {journal} {The Journal of
  Chemical Physics}\ }\textbf {\bibinfo {volume} {125}},\ \bibinfo {pages}
  {114704} (\bibinfo {year} {2006})}\BibitemShut {NoStop}%
\bibitem [{\citenamefont {Reuter}\ and\ \citenamefont
  {Hill}(2012)}]{reuter2012efficient}%
  \BibitemOpen
  \bibfield  {author} {\bibinfo {author} {\bibfnamefont {M.~G.}\ \bibnamefont
  {Reuter}}\ and\ \bibinfo {author} {\bibfnamefont {J.~C.}\ \bibnamefont
  {Hill}},\ }\href@noop {} {\bibfield  {journal} {\bibinfo  {journal}
  {Computational Science \& Discovery}\ }\textbf {\bibinfo {volume} {5}},\
  \bibinfo {pages} {014009} (\bibinfo {year} {2012})}\BibitemShut {NoStop}%
\bibitem [{\citenamefont {Vishik}\ \emph {et~al.}(2010)\citenamefont {Vishik},
  \citenamefont {Lee}, \citenamefont {Schmitt}, \citenamefont {Moritz},
  \citenamefont {Sasagawa}, \citenamefont {Uchida}, \citenamefont {Fujita},
  \citenamefont {Ishida}, \citenamefont {Zhang}, \citenamefont {Devereaux},\
  and\ \citenamefont {Shen}}]{vishik2010doping}%
  \BibitemOpen
  \bibfield  {author} {\bibinfo {author} {\bibfnamefont {I.~M.}\ \bibnamefont
  {Vishik}}, \bibinfo {author} {\bibfnamefont {W.~S.}\ \bibnamefont {Lee}},
  \bibinfo {author} {\bibfnamefont {F.}~\bibnamefont {Schmitt}}, \bibinfo
  {author} {\bibfnamefont {B.}~\bibnamefont {Moritz}}, \bibinfo {author}
  {\bibfnamefont {T.}~\bibnamefont {Sasagawa}}, \bibinfo {author}
  {\bibfnamefont {S.}~\bibnamefont {Uchida}}, \bibinfo {author} {\bibfnamefont
  {K.}~\bibnamefont {Fujita}}, \bibinfo {author} {\bibfnamefont
  {S.}~\bibnamefont {Ishida}}, \bibinfo {author} {\bibfnamefont
  {C.}~\bibnamefont {Zhang}}, \bibinfo {author} {\bibfnamefont {T.~P.}\
  \bibnamefont {Devereaux}}, \ and\ \bibinfo {author} {\bibfnamefont {Z.-X.}\
  \bibnamefont {Shen}},\ }\href@noop {} {\bibfield  {journal} {\bibinfo
  {journal} {Phys. Rev. Lett.}\ }\textbf {\bibinfo {volume} {104}},\ \bibinfo
  {pages} {207002} (\bibinfo {year} {2010})}\BibitemShut {NoStop}%
\bibitem [{\citenamefont {MacKinnon}\ and\ \citenamefont
  {Kramer}(1981)}]{mackinnon1981one}%
  \BibitemOpen
  \bibfield  {author} {\bibinfo {author} {\bibfnamefont {A.}~\bibnamefont
  {MacKinnon}}\ and\ \bibinfo {author} {\bibfnamefont {B.}~\bibnamefont
  {Kramer}},\ }\href@noop {} {\bibfield  {journal} {\bibinfo  {journal} {Phys.
  Rev. Lett.}\ }\textbf {\bibinfo {volume} {47}},\ \bibinfo {pages} {1546}
  (\bibinfo {year} {1981})}\BibitemShut {NoStop}%
\bibitem [{\citenamefont {Bulka}\ \emph {et~al.}(1985)\citenamefont {Bulka},
  \citenamefont {Kramer},\ and\ \citenamefont
  {MacKinnon}}]{buŀka1985mobility}%
  \BibitemOpen
  \bibfield  {author} {\bibinfo {author} {\bibfnamefont {B.~R.}\ \bibnamefont
  {Bulka}}, \bibinfo {author} {\bibfnamefont {B.}~\bibnamefont {Kramer}}, \
  and\ \bibinfo {author} {\bibfnamefont {A.}~\bibnamefont {MacKinnon}},\
  }\href@noop {} {\bibfield  {journal} {\bibinfo  {journal} {Zeitschrift
  f{\"u}r Physik B Condensed Matter}\ }\textbf {\bibinfo {volume} {60}},\
  \bibinfo {pages} {13} (\bibinfo {year} {1985})}\BibitemShut {NoStop}%
\bibitem [{\citenamefont {Kramer}\ and\ \citenamefont
  {MacKinnon}(1993)}]{kramer1993localization}%
  \BibitemOpen
  \bibfield  {author} {\bibinfo {author} {\bibfnamefont {B.}~\bibnamefont
  {Kramer}}\ and\ \bibinfo {author} {\bibfnamefont {A.}~\bibnamefont
  {MacKinnon}},\ }\href@noop {} {\bibfield  {journal} {\bibinfo  {journal}
  {Reports on Progress in Physics}\ }\textbf {\bibinfo {volume} {56}},\
  \bibinfo {pages} {1469} (\bibinfo {year} {1993})}\BibitemShut {NoStop}%
\bibitem [{\citenamefont {Xiang}(1995)}]{xiang1995effect}%
  \BibitemOpen
  \bibfield  {author} {\bibinfo {author} {\bibfnamefont {T.}~\bibnamefont
  {Xiang}},\ }\href@noop {} {\bibfield  {journal} {\bibinfo  {journal} {Phys.
  Rev. B}\ }\textbf {\bibinfo {volume} {52}},\ \bibinfo {pages} {6204}
  (\bibinfo {year} {1995})}\BibitemShut {NoStop}%
\bibitem [{\citenamefont {Martin}\ \emph {et~al.}(2002)\citenamefont {Martin},
  \citenamefont {Balatsky},\ and\ \citenamefont {Zaanen}}]{martin2002impurity}%
  \BibitemOpen
  \bibfield  {author} {\bibinfo {author} {\bibfnamefont {I.}~\bibnamefont
  {Martin}}, \bibinfo {author} {\bibfnamefont {A.~V.}\ \bibnamefont
  {Balatsky}}, \ and\ \bibinfo {author} {\bibfnamefont {J.}~\bibnamefont
  {Zaanen}},\ }\href@noop {} {\bibfield  {journal} {\bibinfo  {journal} {Phys.
  Rev. Lett.}\ }\textbf {\bibinfo {volume} {88}},\ \bibinfo {pages} {097003}
  (\bibinfo {year} {2002})}\BibitemShut {NoStop}%
\bibitem [{\citenamefont {Zhu}\ \emph {et~al.}(2003)\citenamefont {Zhu},
  \citenamefont {Atkinson},\ and\ \citenamefont {Hirschfeld}}]{zhu2003two}%
  \BibitemOpen
  \bibfield  {author} {\bibinfo {author} {\bibfnamefont {L.}~\bibnamefont
  {Zhu}}, \bibinfo {author} {\bibfnamefont {W.~A.}\ \bibnamefont {Atkinson}}, \
  and\ \bibinfo {author} {\bibfnamefont {P.~J.}\ \bibnamefont {Hirschfeld}},\
  }\href@noop {} {\bibfield  {journal} {\bibinfo  {journal} {Phys. Rev. B}\
  }\textbf {\bibinfo {volume} {67}},\ \bibinfo {pages} {094508} (\bibinfo
  {year} {2003})}\BibitemShut {NoStop}%
\bibitem [{\citenamefont {Hill}\ \emph {et~al.}(2004)\citenamefont {Hill},
  \citenamefont {Lupien}, \citenamefont {Sutherland}, \citenamefont {Boaknin},
  \citenamefont {Hawthorn}, \citenamefont {Proust}, \citenamefont {Ronning},
  \citenamefont {Taillefer}, \citenamefont {Liang}, \citenamefont {Bonn},\ and\
  \citenamefont {Hardy}}]{hill2004transport}%
  \BibitemOpen
  \bibfield  {author} {\bibinfo {author} {\bibfnamefont {R.~W.}\ \bibnamefont
  {Hill}}, \bibinfo {author} {\bibfnamefont {C.}~\bibnamefont {Lupien}},
  \bibinfo {author} {\bibfnamefont {M.}~\bibnamefont {Sutherland}}, \bibinfo
  {author} {\bibfnamefont {E.}~\bibnamefont {Boaknin}}, \bibinfo {author}
  {\bibfnamefont {D.~G.}\ \bibnamefont {Hawthorn}}, \bibinfo {author}
  {\bibfnamefont {C.}~\bibnamefont {Proust}}, \bibinfo {author} {\bibfnamefont
  {F.}~\bibnamefont {Ronning}}, \bibinfo {author} {\bibfnamefont
  {L.}~\bibnamefont {Taillefer}}, \bibinfo {author} {\bibfnamefont
  {R.}~\bibnamefont {Liang}}, \bibinfo {author} {\bibfnamefont {D.~A.}\
  \bibnamefont {Bonn}}, \ and\ \bibinfo {author} {\bibfnamefont {W.~N.}\
  \bibnamefont {Hardy}},\ }\href@noop {} {\bibfield  {journal} {\bibinfo
  {journal} {Phys. Rev. Lett.}\ }\textbf {\bibinfo {volume} {92}},\ \bibinfo
  {pages} {027001} (\bibinfo {year} {2004})}\BibitemShut {NoStop}%
\bibitem [{\citenamefont {Pan}\ \emph {et~al.}(2001)\citenamefont {Pan},
  \citenamefont {O'Neal}, \citenamefont {Badzey}, \citenamefont {Chamon},
  \citenamefont {Ding}, \citenamefont {Engelbrecht}, \citenamefont {Wang},
  \citenamefont {Eisaki}, \citenamefont {Uchida}, \citenamefont {Gupta},
  \citenamefont {Ng}, \citenamefont {Hudson}, \citenamefont {Lang},\ and\
  \citenamefont {Davis}}]{pan2001microscopic}%
  \BibitemOpen
  \bibfield  {author} {\bibinfo {author} {\bibfnamefont {S.~H.}\ \bibnamefont
  {Pan}}, \bibinfo {author} {\bibfnamefont {J.~P.}\ \bibnamefont {O'Neal}},
  \bibinfo {author} {\bibfnamefont {R.~L.}\ \bibnamefont {Badzey}}, \bibinfo
  {author} {\bibfnamefont {C.}~\bibnamefont {Chamon}}, \bibinfo {author}
  {\bibfnamefont {H.}~\bibnamefont {Ding}}, \bibinfo {author} {\bibfnamefont
  {J.~R.}\ \bibnamefont {Engelbrecht}}, \bibinfo {author} {\bibfnamefont
  {Z.}~\bibnamefont {Wang}}, \bibinfo {author} {\bibfnamefont {H.}~\bibnamefont
  {Eisaki}}, \bibinfo {author} {\bibfnamefont {S.}~\bibnamefont {Uchida}},
  \bibinfo {author} {\bibfnamefont {A.~K.}\ \bibnamefont {Gupta}}, \bibinfo
  {author} {\bibfnamefont {K.-W.}\ \bibnamefont {Ng}}, \bibinfo {author}
  {\bibfnamefont {E.~W.}\ \bibnamefont {Hudson}}, \bibinfo {author}
  {\bibfnamefont {K.~M.}\ \bibnamefont {Lang}}, \ and\ \bibinfo {author}
  {\bibfnamefont {J.~C.}\ \bibnamefont {Davis}},\ }\href@noop {} {\bibfield
  {journal} {\bibinfo  {journal} {Nature}\ }\textbf {\bibinfo {volume} {413}},\
  \bibinfo {pages} {282} (\bibinfo {year} {2001})}\BibitemShut {NoStop}%
\bibitem [{\citenamefont {Bobowski}\ \emph {et~al.}(2010)\citenamefont
  {Bobowski}, \citenamefont {Baglo}, \citenamefont {Day}, \citenamefont
  {Semple}, \citenamefont {Dosanjh}, \citenamefont {Turner}, \citenamefont
  {Harris}, \citenamefont {Liang}, \citenamefont {Bonn},\ and\ \citenamefont
  {Hardy}}]{bobowski2010oxygen}%
  \BibitemOpen
  \bibfield  {author} {\bibinfo {author} {\bibfnamefont {J.~S.}\ \bibnamefont
  {Bobowski}}, \bibinfo {author} {\bibfnamefont {J.~C.}\ \bibnamefont {Baglo}},
  \bibinfo {author} {\bibfnamefont {J.}~\bibnamefont {Day}}, \bibinfo {author}
  {\bibfnamefont {L.}~\bibnamefont {Semple}}, \bibinfo {author} {\bibfnamefont
  {P.}~\bibnamefont {Dosanjh}}, \bibinfo {author} {\bibfnamefont {P.~J.}\
  \bibnamefont {Turner}}, \bibinfo {author} {\bibfnamefont {R.}~\bibnamefont
  {Harris}}, \bibinfo {author} {\bibfnamefont {R.}~\bibnamefont {Liang}},
  \bibinfo {author} {\bibfnamefont {D.~A.}\ \bibnamefont {Bonn}}, \ and\
  \bibinfo {author} {\bibfnamefont {W.~N.}\ \bibnamefont {Hardy}},\ }\href@noop
  {} {\bibfield  {journal} {\bibinfo  {journal} {Phys. Rev. B}\ }\textbf
  {\bibinfo {volume} {82}},\ \bibinfo {pages} {134526} (\bibinfo {year}
  {2010})}\BibitemShut {NoStop}%
\bibitem [{\citenamefont {Durst}\ and\ \citenamefont
  {Lee}(2002)}]{durst2002microwave}%
  \BibitemOpen
  \bibfield  {author} {\bibinfo {author} {\bibfnamefont {A.~C.}\ \bibnamefont
  {Durst}}\ and\ \bibinfo {author} {\bibfnamefont {P.~A.}\ \bibnamefont
  {Lee}},\ }\href@noop {} {\bibfield  {journal} {\bibinfo  {journal} {Phys.
  Rev. B}\ }\textbf {\bibinfo {volume} {65}},\ \bibinfo {pages} {094501}
  (\bibinfo {year} {2002})}\BibitemShut {NoStop}%
\bibitem [{\citenamefont {Schubert}\ \emph {et~al.}(2010)\citenamefont
  {Schubert}, \citenamefont {Schleede}, \citenamefont {Byczuk}, \citenamefont
  {Fehske},\ and\ \citenamefont {Vollhardt}}]{schubert2010distribution}%
  \BibitemOpen
  \bibfield  {author} {\bibinfo {author} {\bibfnamefont {G.}~\bibnamefont
  {Schubert}}, \bibinfo {author} {\bibfnamefont {J.}~\bibnamefont {Schleede}},
  \bibinfo {author} {\bibfnamefont {K.}~\bibnamefont {Byczuk}}, \bibinfo
  {author} {\bibfnamefont {H.}~\bibnamefont {Fehske}}, \ and\ \bibinfo {author}
  {\bibfnamefont {D.}~\bibnamefont {Vollhardt}},\ }\href@noop {} {\bibfield
  {journal} {\bibinfo  {journal} {Phys. Rev. B}\ }\textbf {\bibinfo {volume}
  {81}},\ \bibinfo {pages} {155106} (\bibinfo {year} {2010})}\BibitemShut
  {NoStop}%
\bibitem [{\citenamefont {Ouazi}\ \emph {et~al.}(2006)\citenamefont {Ouazi},
  \citenamefont {Bobroff}, \citenamefont {Alloul}, \citenamefont {Le~Tacon},
  \citenamefont {Blanchard}, \citenamefont {Collin}, \citenamefont {Julien},
  \citenamefont {Horvati{\'c}},\ and\ \citenamefont
  {Berthier}}]{ouazi2006impurity}%
  \BibitemOpen
  \bibfield  {author} {\bibinfo {author} {\bibfnamefont {S.}~\bibnamefont
  {Ouazi}}, \bibinfo {author} {\bibfnamefont {J.}~\bibnamefont {Bobroff}},
  \bibinfo {author} {\bibfnamefont {H.}~\bibnamefont {Alloul}}, \bibinfo
  {author} {\bibfnamefont {M.}~\bibnamefont {Le~Tacon}}, \bibinfo {author}
  {\bibfnamefont {N.}~\bibnamefont {Blanchard}}, \bibinfo {author}
  {\bibfnamefont {G.}~\bibnamefont {Collin}}, \bibinfo {author} {\bibfnamefont
  {M.~H.}\ \bibnamefont {Julien}}, \bibinfo {author} {\bibfnamefont
  {M.}~\bibnamefont {Horvati{\'c}}}, \ and\ \bibinfo {author} {\bibfnamefont
  {C.}~\bibnamefont {Berthier}},\ }\href@noop {} {\bibfield  {journal}
  {\bibinfo  {journal} {Phys. Rev. Lett.}\ }\textbf {\bibinfo {volume} {96}},\
  \bibinfo {pages} {127005} (\bibinfo {year} {2006})}\BibitemShut {NoStop}%
\bibitem [{\citenamefont {Zhou}\ \emph
  {et~al.}(2017{\natexlab{a}})\citenamefont {Zhou}, \citenamefont {Hirata},
  \citenamefont {Wu}, \citenamefont {Vinograd}, \citenamefont {Mayaffre},
  \citenamefont {Kr{\"a}mer}, \citenamefont {Horvati{\'c}}, \citenamefont
  {Berthier}, \citenamefont {Reyes}, \citenamefont {Kuhns}, \citenamefont
  {Liang}, \citenamefont {Hardy}, \citenamefont {Bonn},\ and\ \citenamefont
  {Julien}}]{zhou2017quasiparticle}%
  \BibitemOpen
  \bibfield  {author} {\bibinfo {author} {\bibfnamefont {R.}~\bibnamefont
  {Zhou}}, \bibinfo {author} {\bibfnamefont {M.}~\bibnamefont {Hirata}},
  \bibinfo {author} {\bibfnamefont {T.}~\bibnamefont {Wu}}, \bibinfo {author}
  {\bibfnamefont {I.}~\bibnamefont {Vinograd}}, \bibinfo {author}
  {\bibfnamefont {H.}~\bibnamefont {Mayaffre}}, \bibinfo {author}
  {\bibfnamefont {S.}~\bibnamefont {Kr{\"a}mer}}, \bibinfo {author}
  {\bibfnamefont {M.}~\bibnamefont {Horvati{\'c}}}, \bibinfo {author}
  {\bibfnamefont {C.}~\bibnamefont {Berthier}}, \bibinfo {author}
  {\bibfnamefont {A.~P.}\ \bibnamefont {Reyes}}, \bibinfo {author}
  {\bibfnamefont {P.~L.}\ \bibnamefont {Kuhns}}, \bibinfo {author}
  {\bibfnamefont {R.}~\bibnamefont {Liang}}, \bibinfo {author} {\bibfnamefont
  {W.~N.}\ \bibnamefont {Hardy}}, \bibinfo {author} {\bibfnamefont {D.~A.}\
  \bibnamefont {Bonn}}, \ and\ \bibinfo {author} {\bibfnamefont {M.-H.}\
  \bibnamefont {Julien}},\ }\href@noop {} {\bibfield  {journal} {\bibinfo
  {journal} {Phys. Rev. Lett.}\ }\textbf {\bibinfo {volume} {118}},\ \bibinfo
  {pages} {017001} (\bibinfo {year} {2017}{\natexlab{a}})}\BibitemShut
  {NoStop}%
\bibitem [{\citenamefont {Zhou}\ \emph
  {et~al.}(2017{\natexlab{b}})\citenamefont {Zhou}, \citenamefont {Hirata},
  \citenamefont {Wu}, \citenamefont {Vinograd}, \citenamefont {Mayaffre},
  \citenamefont {Kr{\"a}mer}, \citenamefont {Reyes}, \citenamefont {Kuhns},
  \citenamefont {Liang}, \citenamefont {Hardy}, \citenamefont {Bonn},\ and\
  \citenamefont {Julien}}]{zhou2017spin}%
  \BibitemOpen
  \bibfield  {author} {\bibinfo {author} {\bibfnamefont {R.}~\bibnamefont
  {Zhou}}, \bibinfo {author} {\bibfnamefont {M.}~\bibnamefont {Hirata}},
  \bibinfo {author} {\bibfnamefont {T.}~\bibnamefont {Wu}}, \bibinfo {author}
  {\bibfnamefont {I.}~\bibnamefont {Vinograd}}, \bibinfo {author}
  {\bibfnamefont {H.}~\bibnamefont {Mayaffre}}, \bibinfo {author}
  {\bibfnamefont {S.}~\bibnamefont {Kr{\"a}mer}}, \bibinfo {author}
  {\bibfnamefont {A.~P.}\ \bibnamefont {Reyes}}, \bibinfo {author}
  {\bibfnamefont {P.~L.}\ \bibnamefont {Kuhns}}, \bibinfo {author}
  {\bibfnamefont {R.}~\bibnamefont {Liang}}, \bibinfo {author} {\bibfnamefont
  {W.~N.}\ \bibnamefont {Hardy}}, \bibinfo {author} {\bibfnamefont {D.~A.}\
  \bibnamefont {Bonn}}, \ and\ \bibinfo {author} {\bibfnamefont {M.-H.}\
  \bibnamefont {Julien}},\ }\href@noop {} {\bibfield  {journal} {\bibinfo
  {journal} {Proceedings of the National Academy of Sciences}\ ,\ \bibinfo
  {pages} {201711445}} (\bibinfo {year} {2017}{\natexlab{b}})}\BibitemShut
  {NoStop}%
\bibitem [{\citenamefont {Franz}\ \emph {et~al.}(1996)\citenamefont {Franz},
  \citenamefont {Kallin},\ and\ \citenamefont {Berlinsky}}]{franz1996impurity}%
  \BibitemOpen
  \bibfield  {author} {\bibinfo {author} {\bibfnamefont {M.}~\bibnamefont
  {Franz}}, \bibinfo {author} {\bibfnamefont {C.}~\bibnamefont {Kallin}}, \
  and\ \bibinfo {author} {\bibfnamefont {A.~J.}\ \bibnamefont {Berlinsky}},\
  }\href@noop {} {\bibfield  {journal} {\bibinfo  {journal} {Phys. Rev. B}\
  }\textbf {\bibinfo {volume} {54}},\ \bibinfo {pages} {R6897} (\bibinfo {year}
  {1996})}\BibitemShut {NoStop}%
\bibitem [{\citenamefont {Kaminski}\ \emph {et~al.}(2000)\citenamefont
  {Kaminski}, \citenamefont {Mesot}, \citenamefont {Fretwell}, \citenamefont
  {Campuzano}, \citenamefont {Norman}, \citenamefont {Randeria}, \citenamefont
  {Ding}, \citenamefont {Sato}, \citenamefont {Takahashi}, \citenamefont
  {Mochiku}, \citenamefont {Kadowaki},\ and\ \citenamefont
  {Hoechst}}]{kaminski2000quasiparticles}%
  \BibitemOpen
  \bibfield  {author} {\bibinfo {author} {\bibfnamefont {A.}~\bibnamefont
  {Kaminski}}, \bibinfo {author} {\bibfnamefont {J.}~\bibnamefont {Mesot}},
  \bibinfo {author} {\bibfnamefont {H.}~\bibnamefont {Fretwell}}, \bibinfo
  {author} {\bibfnamefont {J.~C.}\ \bibnamefont {Campuzano}}, \bibinfo {author}
  {\bibfnamefont {M.~R.}\ \bibnamefont {Norman}}, \bibinfo {author}
  {\bibfnamefont {M.}~\bibnamefont {Randeria}}, \bibinfo {author}
  {\bibfnamefont {H.}~\bibnamefont {Ding}}, \bibinfo {author} {\bibfnamefont
  {T.}~\bibnamefont {Sato}}, \bibinfo {author} {\bibfnamefont {T.}~\bibnamefont
  {Takahashi}}, \bibinfo {author} {\bibfnamefont {T.}~\bibnamefont {Mochiku}},
  \bibinfo {author} {\bibfnamefont {K.}~\bibnamefont {Kadowaki}}, \ and\
  \bibinfo {author} {\bibfnamefont {H.}~\bibnamefont {Hoechst}},\ }\href@noop
  {} {\bibfield  {journal} {\bibinfo  {journal} {Phys. Rev. Lett.}\ }\textbf
  {\bibinfo {volume} {84}},\ \bibinfo {pages} {1788} (\bibinfo {year}
  {2000})}\BibitemShut {NoStop}%
\bibitem [{\citenamefont {Valla}\ \emph {et~al.}(2006)\citenamefont {Valla},
  \citenamefont {Kidd}, \citenamefont {Rameau}, \citenamefont {Noh},
  \citenamefont {Gu}, \citenamefont {Johnson}, \citenamefont {Yang},\ and\
  \citenamefont {Ding}}]{valla2006fine}%
  \BibitemOpen
  \bibfield  {author} {\bibinfo {author} {\bibfnamefont {T.}~\bibnamefont
  {Valla}}, \bibinfo {author} {\bibfnamefont {T.~E.}\ \bibnamefont {Kidd}},
  \bibinfo {author} {\bibfnamefont {J.~D.}\ \bibnamefont {Rameau}}, \bibinfo
  {author} {\bibfnamefont {H.-J.}\ \bibnamefont {Noh}}, \bibinfo {author}
  {\bibfnamefont {G.~D.}\ \bibnamefont {Gu}}, \bibinfo {author} {\bibfnamefont
  {P.~D.}\ \bibnamefont {Johnson}}, \bibinfo {author} {\bibfnamefont {H.-B.}\
  \bibnamefont {Yang}}, \ and\ \bibinfo {author} {\bibfnamefont
  {H.}~\bibnamefont {Ding}},\ }\href@noop {} {\bibfield  {journal} {\bibinfo
  {journal} {Phys. Rev. B}\ }\textbf {\bibinfo {volume} {73}},\ \bibinfo
  {pages} {184518} (\bibinfo {year} {2006})}\BibitemShut {NoStop}%
\bibitem [{\citenamefont {Koralek}\ \emph {et~al.}(2006)\citenamefont
  {Koralek}, \citenamefont {Douglas}, \citenamefont {Plumb}, \citenamefont
  {Sun}, \citenamefont {Federov}, \citenamefont {Murnane}, \citenamefont
  {Kapteyn}, \citenamefont {Cundiff}, \citenamefont {Aiura}, \citenamefont
  {Oka}, \citenamefont {Eisaki},\ and\ \citenamefont
  {Dessau}}]{koralek2006laser}%
  \BibitemOpen
  \bibfield  {author} {\bibinfo {author} {\bibfnamefont {J.~D.}\ \bibnamefont
  {Koralek}}, \bibinfo {author} {\bibfnamefont {J.~F.}\ \bibnamefont
  {Douglas}}, \bibinfo {author} {\bibfnamefont {N.~C.}\ \bibnamefont {Plumb}},
  \bibinfo {author} {\bibfnamefont {Z.}~\bibnamefont {Sun}}, \bibinfo {author}
  {\bibfnamefont {A.~V.}\ \bibnamefont {Federov}}, \bibinfo {author}
  {\bibfnamefont {M.~M.}\ \bibnamefont {Murnane}}, \bibinfo {author}
  {\bibfnamefont {H.~C.}\ \bibnamefont {Kapteyn}}, \bibinfo {author}
  {\bibfnamefont {S.~T.}\ \bibnamefont {Cundiff}}, \bibinfo {author}
  {\bibfnamefont {Y.}~\bibnamefont {Aiura}}, \bibinfo {author} {\bibfnamefont
  {K.}~\bibnamefont {Oka}}, \bibinfo {author} {\bibfnamefont {H.}~\bibnamefont
  {Eisaki}}, \ and\ \bibinfo {author} {\bibfnamefont {D.~S.}\ \bibnamefont
  {Dessau}},\ }\href@noop {} {\bibfield  {journal} {\bibinfo  {journal} {Phys.
  Rev. Lett.}\ }\textbf {\bibinfo {volume} {96}},\ \bibinfo {pages} {017005}
  (\bibinfo {year} {2006})}\BibitemShut {NoStop}%
\bibitem [{\citenamefont {Scalapino}\ \emph {et~al.}(1992)\citenamefont
  {Scalapino}, \citenamefont {White},\ and\ \citenamefont
  {Zhang}}]{scalapino1992superfluid}%
  \BibitemOpen
  \bibfield  {author} {\bibinfo {author} {\bibfnamefont {D.~J.}\ \bibnamefont
  {Scalapino}}, \bibinfo {author} {\bibfnamefont {S.~R.}\ \bibnamefont
  {White}}, \ and\ \bibinfo {author} {\bibfnamefont {S.~C.}\ \bibnamefont
  {Zhang}},\ }\href@noop {} {\bibfield  {journal} {\bibinfo  {journal} {Phys.
  Rev. Lett.}\ }\textbf {\bibinfo {volume} {68}},\ \bibinfo {pages} {2830}
  (\bibinfo {year} {1992})}\BibitemShut {NoStop}%
\bibitem [{\citenamefont {Scalapino}\ \emph {et~al.}(1993)\citenamefont
  {Scalapino}, \citenamefont {White},\ and\ \citenamefont
  {Zhang}}]{scalapino1993insulator}%
  \BibitemOpen
  \bibfield  {author} {\bibinfo {author} {\bibfnamefont {D.~J.}\ \bibnamefont
  {Scalapino}}, \bibinfo {author} {\bibfnamefont {S.~R.}\ \bibnamefont
  {White}}, \ and\ \bibinfo {author} {\bibfnamefont {S.}~\bibnamefont
  {Zhang}},\ }\href@noop {} {\bibfield  {journal} {\bibinfo  {journal} {Phys.
  Rev. B}\ }\textbf {\bibinfo {volume} {47}},\ \bibinfo {pages} {7995}
  (\bibinfo {year} {1993})}\BibitemShut {NoStop}%
\end{thebibliography}%

\end{document}